\documentstyle[12pt]{article}
\newcommand{\seqnoll}{\setcounter{equation}{0}}

\newcommand{\eq}[1]{\mbox{Eq.~(\ref{#1})}}
\def\figcap#1{\refstepcounter{figure}\par\small\bf Figure \thefigure: \sl#1}

\newcommand{\bort}[1]{}
\def\id{\leavevmode\hbox{\small1\kern-3.3pt\normalsize1}}
\newcommand{\goto}{\rightarrow}

\newcommand{\non}{\nonumber}
\newcommand{\nn}{\nonumber\\}
\newcommand{\abs}[1]{\left|#1\right|}

\newcommand{\inv}[1]{\frac{1}{#1}}
\newcommand{\cF}{{\cal F}}

\def\simleq{\; \raise0.3ex\hbox{$<$\kern-0.75em
      \raise-1.1ex\hbox{$\sim$}}\; }
\def\simgeq{\; \raise0.3ex\hbox{$>$\kern-0.75em
      \raise-1.1ex\hbox{$\sim$}}\; }
\def\void{}
\def\labelmark{}

\newenvironment{formula}[1]%
{\def\labelname{#1}
\ifx\void\labelname\begin{equation}
\else\labelmark\begin{equation}\label{\labelname}\fi}%
{\ifx\void\labelname\end{equation}\else\end{equation}\fi}

\newenvironment{formulas}[1]%
{\def\labelname{#1}
\ifx\void\labelname\begin{equation}
\else\labelmark\begin{equation}\label{\labelname}\fi
\begin{array}{lllllll}}%
{\end{array}\ifx\void\labelname\end{equation}\else
\end{equation}\fi}
\def\mus{$\mu$s}
\def\erf{{\rm erf}}
\def\O{{\cal O}}
\def\.{~~.}
\def\simleq{\; \raise0.3ex\hbox{$<$\kern-0.75em
      \raise-1.1ex\hbox{$\sim$}}\; }
\def\simgeq{\; \raise0.3ex\hbox{$>$\kern-0.75em
      \raise-1.1ex\hbox{$\sim$}}\; }

\def\noi{\noindent}
\def\ie{{\rm i.e.}}
\def\apriori{{\it a priori}}

\def\({\left(}
\def\){\right)}

\def\>{\rangle}
\def\<{\langle}

\def\Tr{{\rm Tr}}

\def\P{{\cal P}}
\def\n{{\hat n}}

\def\bfor#1{\begin{formula}{#1}}
\def\efor{\end{formula}}

\begin{document}
\vspace*{-2cm}
\vspace{1cm}
\begin{center}
{\Huge\bf Topics in Modern Quantum Optics}\\
\vspace{10mm}
\noi
{\large Lectures presented at {\it  The 17th Symposium on Theoretical
  Physics
- { \bf APPLIED FIELD THEORY}},  Seoul National University, Seoul,
Korea, 1998. \\[10mm]}
{\large 
 Bo-Sture Skagerstam\footnote{email:
boskag@phys.ntnu.no. Research supported in part by the
Research Council of Norway.}\\[10mm]
}
{
\large
Department of Physics, The Norwegian University of Science and
Technology,
N-7491 Trondheim, Norway\\
}
\end{center}
{\abstract{ Recent experimental developments in electronic and optical
    technology have made it possible to experimentally realize in
    space and time  well localized {\it single  photon} quantum-mechanical
    states.
    In these lectures we will first remind ourselves about some basic
    quantum mechanics and then discuss in what sense
    quantum-mechanical 
    single-photon interference has been observed experimentally. A
    relativistic quantum-mechanical description of single-photon
    states will then be outlined. Within such a single-photon
    scheme a derivation of the Berry-phase for photons will given. 
    In the second set of
    lectures we will discuss the highly idealized system of a single
    two-level atom interacting with a single-mode of the second
    quantized electro-magnetic field as e.g. realized in terms of the
    micromaser system.
    This system  possesses a variety of dynamical phase transitions  parameterized
    by the flux of atoms and the time-of-flight of the atom within the
    cavity as well as other parameters of the system. 
These phases may be revealed to  an  observer  outside  the  cavity
    using the long-time correlation length in the atomic beam. It is explained
    that some  of  the  phase
    transitions are not reflected in the average excitation level of  the  outgoing
    atom, which is one of the commonly used observable. The
    correlation length 
    is directly related to the leading eigenvalue of a certain
    probability conserving  time-evolution
    operator,  which  one can study in order to elucidate the phase
    structure. 
    It is found that as a function  of the time-of-flight the
    transition from  the  thermal  to  the  maser  phase  is
    characterized  by  a  sharp  peak  in  the  correlation  length.
    For   longer times-of-flight there is a transition to a phase
    where the  correlation  length grows exponentially  with  the atomic
    flux.  Finally, we present  a  detailed  numerical  and analytical
    treatment of the different phases and  discuss  the  
    physics  behind them in terms of the physical parameters at 
    hand. }}
\thispagestyle{empty}
\newpage
\vspace*{-3cm}
\tableofcontents
\newpage
\setcounter{page}{1}
\setcounter{footnote}{0}
%
\newcommand{\ibid}[3]{{\sl ibid.} {{\bf #1} {(#2)} {#3}}}
\newcommand{\Eqref}[1]{Eq.(\ref{#1})}
\newcommand{\ad}[1]{{#1}^{*}}
\newcommand{\fet}[1]{{\bf #1}}
\newcommand{\mean}[1]{\langle #1\rangle} 
\newcommand{\rv}[1]{|#1\rangle}
\newcommand{\lv}[1]{\langle #1|}
\newcommand{\cH}{{\cal H}}
\newcommand{\bea}{\begin{eqnarray}}
\newcommand{\eea}{\end{eqnarray}}
\newcommand{\ba}{\begin{array}}
\newcommand{\ea}{\end{array}}
\newcommand{\be}{\begin{equation}}
\newcommand{\ee}{\end{equation}}
\newcommand{\bk}[2]{\left\langle #1,#2 \right\rangle}
\newcommand{\ket}[1]{\left|\,#1\right\rangle}
\newcommand{\braket}[2]{\langle #1|#2\rangle}
\newcommand{\jdg}[3]{{ Journ. Diff. Geom.} {{\bf #1} {(#2)} {#3}}}
\newcommand{\pw}[3]{{ Particle World} {{\bf #1} {(#2)} {#3}}}
\newcommand{\fdp}[3]{{ Fortschr. Phys.} {{\bf #1} {(#2)} {#3}}}
\newcommand{\tmp}[3]{{ Theor. Math. Phys.}{{\bf #1} {(#2)} {#3}}}
\newcommand{\ps}[3]{{ Physica Scripta} {{\bf #1} {(#2)} {#3}}}
\newcommand{\cpam}[3]{{ Commun. Pure and Appl. Math.}{{\bf #1} {(#2)} {#3}}}
\newcommand{\pr}[3]{{ Phys. Rev.} {{\bf #1} {(#2)} {#3}}}
\newcommand{\pra}[3]{{  Phys. Rev. A} {{\bf #1} {(#2)} {#3}}}
\newcommand{\prb}[3]{{  Phys. Rev. B} {{{\bf #1} {(#2)} {#3}}}}
\newcommand{\prc}[3]{{  Phys. Rev. C} {{\bf #1} {(#2)} {#3}}}
\newcommand{\prd}[3]{{  Phys. Rev. D} {{\bf #1} {(#2)} {#3}}}
\newcommand{\prl}[3]{ { Phys. Rev. Lett.} {{\bf #1} {(#2)} {#3}}}
\newcommand{\ijmp}[3]{{ Int. J. Mod. Phys.} {{\bf #1} {(#2)} {#3}}}
\newcommand{\jmp}[3]{{  J. Math. Phys.} {{\bf #1} {(#2)} {#3}}}
\newcommand{\rmp}[3]{{  Rev. Mod. Phys.} {{\bf #1} {(#2)} {#3}}}
\newcommand{\cmp}[3]{{  Comm. Math. Phys.} {{\bf #1} {(#2)} {#3}}}
\newcommand{\cqg}[3]{{  Class. Quant. Grav.} {{\bf #1} {(#2)} {#3}}}
\newcommand{\np}[3]{{  Nucl. Phys.} {{\bf #1} {(#2)} {#3}}}
\newcommand{\nature}[3]{{  Nature} {{\bf #1} {(#2)} {#3}}}
\newcommand{\pl}[3]{{  Phys. Lett.} {{\bf #1} {(#2)} {#3}}}
\newcommand{\prep}[3]{{ Phys. Rep.} {{\bf #1} {(#2)} {#3}}}
\newcommand{\ptp}[3]{{  Prog. Theor. Phys.} {{\bf #1} {(#2)} {#3}}}
\newcommand{\ptps}[3]{{  Prog. Theor. Phys. suppl.} {{\bf #1} {(#2)} {#3}}}
\newcommand{\epl}[3]{{  Europhys. Lett.} {{\bf #1} {(#2)} {#3}}}
\newcommand{\ajp}[3]{{  Am. J. Phys.} {{\bf #1} {(#2)} {#3}}}
\newcommand{\apj}[3]{{  Astrophys. Jour.} {{\bf #1} {(#2)} {#3}}}
\newcommand{\apjl}[3]{{  Astrophys. Jour. Lett.} {{\bf #1} {(#2)} {#3}}}
\newcommand{\annp}[3]{{  Ann. Phys. (N.Y.)} {{\bf #1} {(#2)} {#3}}}
\newcommand{\jpa}[3]{{  J. Phys. A} {{\bf #1} {(#2)} {#3}}}
\newcommand{\jpc}[3]{{  J. Phys. C} {{\bf #1} {(#2)} {#3}}}
\newcommand{\apl}[3]{{  Appl. Phys. Lett.} {{\bf #1} {(#2)} {#3}}}  
\newcommand{\phys}[3]{{  Physica} {{\bf #1} {(#2)} {#3}}} 
\newcommand{\jap}[3]{{ J. Appl. Phys.} {{\bf #1} {(#2)} {#3}}}
\newcommand{\spu}[3]{{  Sov. Phys. Usp.} {{\bf #1} {(#2)} {#3}}}
\newcommand{\zfp}[3]{{  Z. Physik} {{\bf #1} {(#2)} {#3}}}
\newcommand{\jmo}[3]{{  J. Mod. Opt.} {{\bf #1} {(#2)} {#3}}}
\section{Introduction}
\label{sec-intro}
\seqnoll
\begin{flushright}
``{\sl Truth and clarity are complementary.}"\\
N. Bohr
\end{flushright}

 In the first part of these lectures we will focus our attention on 
some aspects of the notion of a photon in 
modern quantum optics and a relativistic description of single,
localized, photons.
In the second part we will discuss in great detail the 
``standard model" of quantum optics, i.e. the Jaynes-Cummings
model describing the 
interaction of a two-mode system with a single mode of the
second-quantized electro-magnetic 
field and its realization in resonant cavities in  
 terms of in particular the micro-maser system. Most of the material 
presented in these
lectures has appeared in one form or another elsewhere. Material for
the first set of lectures can be found in
Refs.\cite{skagerstam92,Skagerstam94} and for the second part of the
lectures we refer to Refs.\cite{Elmforsetal95,Elmforsetal95b}.

The lectures are organized  as  follows.  
In Section~\ref{sec-quantum}
we discuss some basic quantum mechanics and the notion of coherent and
semi-coherent  states. Elements form the photon-detection theory of
Glauber is discussed in Section~\ref{sec-pdt} as well as the experimental
verification of quantum-mechanical
single-photon interference. Some applications of the ideas of
photon-detection theory
in high-energy physics are also briefly mentioned. In  
Section~\ref{sec-photons} we outline a relativistic and
quantum-mechanical theory of single photons. The Berry phase for
single photons is then derived within such a quantum-mechanical
scheme. We also discuss
properties of single-photon wave-packets which by construction have
positive energy.
In  Section~\ref{basic}  we  present  the
standard theoretical framework  for  the  micromaser  and  introduce
the notion of a correlation length in the outgoing atomic beam as
was first introduced  in Refs.\cite{Elmforsetal95,Elmforsetal95b}.
  A  general  discussion  of  long-time  correlations  is
given   in Section~\ref{correlations}, where  we  also  show how one 
can determine  the  correlation length
numerically. Before entering the analytic investigation of the phase  structure
we introduce some useful concepts in  Section~\ref{analytic}  and  discuss  the
eigenvalue problem for the correlation length. In
Section~\ref{Phasestructure}  
 details  of  the  different   phases   are
analyzed. In Section~\ref{Spread} we discuss effects related to the finite spread
in atomic velocities. The phase boundaries are defined in the limit of an
infinite
flux of atoms, but there are several interesting effects related to finite
fluxes
as well. We discuss these issues in Section~\ref{finite}. Final
remarks and a summary
is given  in Section~\ref{conclusions}.

\section{Basic Quantum Mechanics}
\label{sec-quantum}
\seqnoll
%
\begin{flushright}
``{\sl Quantum mechanics, that mysterious, confusing \\ discipline, 
which none of us really understands,\\
but which we know how to use}"\\
M. Gell-Mann
\end{flushright}

Quantum
mechanics, we believe, is the fundamental framework  for the 
description of all known natural physical phenomena. Still we are, however,
often very often puzzled about the role of concepts from the domain of classical
physics within the quantum-mechanical language. 
The interpretation of the theoretical framework
of quantum mechanics is, of course, directly connected to the ``classical
picture" of physical phenomena. We often talk about quantization of the {\it
classical observables} in particular so with regard to classical dynamical
systems in the Hamiltonian formulation as has so beautifully  been discussed by
Dirac \cite{dirac50} and others 
(see e.g. Ref.\cite{hanson}).  
\subsection{Coherent States}
\label{sec-coherentstates}
\seqnoll

The concept of coherent states is very useful  
in trying to orient the inquiring mind in this
jungle of conceptually difficult issues when connecting classical pictures of
physical phenomena with the fundamental notion of quantum-mechanical
probability-amplitudes
and probabilities. 
We will
not try to make a general enough definition of the concept of coherent states
(for such an attempt see e.g. the introduction of
Ref.\cite{klauderskagerstam85}).  There
are, however, many  excellent text-books
 \cite{glauber64,klaudersudarshan68,pere85}, recent reviews
\cite{zfgilmore90} and other expositions of the subject
\cite{klauderskagerstam85} to which we will refer to for details and/or
other aspects of the subject under consideration. 
To our knowledge,  the modern notion of coherent states actually 
goes back to the pioneering work by Lee, Low 
and Pines in 1953 \cite{leelowpines53}  on a quantum-mechanical
variational principle. 
These authors studied electrons in
low-lying conduction bands. This is a strong-coupling problem due to 
interactions with
the longitudinal optical modes of lattice vibrations and in
Ref.\cite{leelowpines53}
a variational calculation was performed using coherent states. The
concept of coherent states as we use in the context of quantum optics
goes back Klauder \cite{klauder&60}, Glauber \cite{glauber63} and Sudarshan
\cite{sudarshan63}.
We will refer to these states as Glauber-Klauder coherent states.

As is well-known, coherent states appear in a very natural way when 
considering the 
classical limit or the infrared properties of quantum field theories
like quantum electrodynamics
(QED)\cite{kee68}-\cite{grecorossi67}
or in analysis of the infrared
properties of quantum gravity \cite{weinberg65,ceh78}. In the
conventional and extremely successful application of perturbative quantum
field theory in the description of elementary processes in Nature when gravitons
are not taken into account,
the number-operator Fock-space representation is
the natural Hilbert space.  The realization of the canonical 
commutation relations of
the quantum fields leads, of course, in general to mathematical
difficulties when interactions are taken into account. Over the years we have,
however, in practice learned how to deal with some of these mathematical
difficulties.

In presenting the theory of the second-quantized electro-magnetic
field on an elementary level, it is tempting to exhibit
an apparent ``paradox" of Erhenfest theorem in quantum mechanics and the
existence of the classical Maxwell's equations: any average of  the
electro-magnetic field-strengths in the physically natural number-operator basis
is zero and hence these averages will not obey the classical equations of
motion. The solution of this apparent paradox is, as is by now well established:
  the classical fields in Maxwell's equations corresponds to {\it
quantum states}
 with
an arbitrary number of photons. In classical physics, we may neglect the
quantum structure of the
 charged
sources. Let $\fet{j}(\fet{x},t)$ be such a classical current, like the
classical current in a coil, and $\fet{A}(\fet{x},t)$ the second-quantized
radiation field (in e.g. the radiation gauge). In the long wave-length
limit of the radiation 
field a classical current should be an appropriate approximation at
least for theories like 
quantum electrodynamics. The interaction Hamiltonian
$\cH_{I}(t)$  then takes the form
 \be 
\label{cc1}
\cH_{I}(t) = - \int d^3 x~\fet{j}(\fet{x},t)\cdot \fet{A}(\fet{x},t)~~~,
 \ee
and the quantum states in the interaction picture, $\rv{t}_{I}$, obey the
time-dependent Schr\"{o}dinger equation, i.e. using natural units 
($\hbar = c = 1$)
 \be
\label{se1}
i\frac{d}{dt}\rv{t}_{I} = \cH_{I}(t)\rv{t}_{I}~~~.
\ee
For reasons of simplicity, we will consider only one specific mode of the
electro-magnetic field described in terms of a canonical creation operator
($\ad{a}$) and an annihilation operator ($a$). The general case then easily
follows by considering a system of such independent modes
(see e.g. Ref.\cite{ebss78}). It is therefore
sufficient to consider the following single-mode interaction
Hamiltonian: 
 \be
\cH_{I}(t) = - f(t)\biggl(a\exp[-i\omega t] + \ad{a}
\exp[i\omega t]\biggr) ~~~, 
\ee
where the real-valued function $f(t)$ describes the  in general
time-dependent classical current.
The ``free" part $\cH_{0}$ of the  total Hamiltonian in natural units
then is
\be
\cH_{0} = \omega(\ad{a}a +1/2)~~~. 
\ee
In terms of canonical ``momentum" ($p$) and ``position" ($x$) field-quadrature
degrees of freedom defined by 
\bea
a &=&\sqrt{\frac{\omega}{2}}x + i\frac{1}{\sqrt{2\omega}}p~~~,\nonumber\\
\ad{a} &=&\sqrt{\frac{\omega}{2}}x -i\frac{1}{\sqrt{2\omega}}p~~~,
\eea
we therefore see that we are formally considering an harmonic
oscillator in the presence of a time-dependent external force. 
The explicit solution to  \Eqref{se1}
is easily found. We can write
\be
\label{sol1}
\rv{t}_{I} =
T\exp\left(-i\int_{t_{0}}^{t}H_{I}(t')dt'\right)\rv{t_{0}}_{I} =
\exp[i\phi(t)] 
\exp[iA(t)] \rv{t_{0}}_{I}~~~,
\ee
where the non-trivial time-ordering procedure is expressed in terms of
\be
A(t)  =  -\int_{t_{0}}^{t}dt'\cH_{I}(t')~~~,
\ee
and  the c-number phase $\phi(t)$ as given by
\be
\phi(t) = \frac{i}{2}\int_{t_{0}}^{t}dt'[A(t'),\cH_{I}(t')]~~~.
\ee
The form of this solution is valid for {\it any } interaction Hamiltonian which is
at most linear in
creation and annihilation operators (see e.g. Ref.\cite{letz77}).
 We now define the unitary operator
\be
U(z) = \exp[z\ad{a}-\ad{z}a]~~~.
\ee
Canonical coherent states $\rv{z;\phi_{0}}$, depending on the (complex)
parameter $z$ and the fiducial
 normalized state number-operator eigenstate
$\rv{\phi_{0}}$, are defined by
\be
\rv{z;\phi_{0}} = U(z)\rv{\phi_{0}}~~~,
\ee
such that
\be
1 = \int \frac{d^{2}z}{\pi}\rv{z}\lv{z} = \int
\frac{d^{2}z}{\pi}\rv{z;\phi_{0}}\lv{z;\phi_{0}}~~~. 
\ee
Here the canonical coherent-state $\rv{z}$ corresponds to the choice 
$\rv{z;0}$, i.e. to an initial Fock vacuum state. We then see that, up to a phase, the
solution \Eqref{sol1} is a canonical coherent-state if the initial
state is the vacuum state. It can be verified that
the expectation value of the second-quantized electro-magnetic field in the
state $\rv{t}_{I}$ obeys the classical Maxwell equations of motion for {\it any} fiducial
Fock-space state $\rv{t_{0}}_{I}=\rv{\phi_{0}}$. Therefore the corresponding
complex, and in general time-dependent, parameters $z$ constitute an explicit
mapping between  classical phase-space dynamical variables and a pure
quantum-mechanical state. In more general terms, quantum-mechanical models can
actually be constructed which demonstrates that by the process of
phase-decoherence one is naturally lead to such a correspondence between points
in classical phase-space and coherent states (see
e.g. Ref.\cite{zurek93}).
\subsection{Semi-Coherent or Displaced Coherent States}
\label{sec-semicoherent}

If the fiducial state $\rv{\phi_{0}}$ is a number operator eigenstate $\rv{m}$,
where $m$  is an integer,
the corresponding coherent-state $\rv{z;m}$  have
recently been  discussed
in detail in the literature  and is referred to as a {\it semi-coherent state}
\cite{bagrovetal76,frad91} or a {\it displaced number-operator state}
\cite{knight90}. 
For some recent considerations see e.g. Refs.\cite{dahl&96,nieto&97}
and in the context of 
resonant micro-cavities see Refs.\cite{carmichael&96,walther&98}.
We will now argue that a classical current can be used to amplify the
information contained in the pure fiducial vector $\rv{\phi_{0}}$. 
In Section \ref{basic} we will give further discussions on this topic.
For a given
initial fiducial Fock-state vector $\rv{m}$, it is a rather
trivial exercise to calculate the
probability $P(n)$ to find $n$ photons in the final state, i.e. 
(see e.g. Ref.\cite{carruthersnieto65}) 
\be P(n) = \lim_{t \rightarrow
\infty}|\braket{n}{t}_{I}|^{2}~~~, 
\ee
\noindent which then depends on the Fourier transform  
 $z = f(\omega) = \int^{\infty}_{-\infty}dtf(t)\exp(-i\omega
t)$. 
\begin{figure}[htb]
\unitlength=1mm
\begin{picture}(140,100)(-40,-80)
\includegraphics{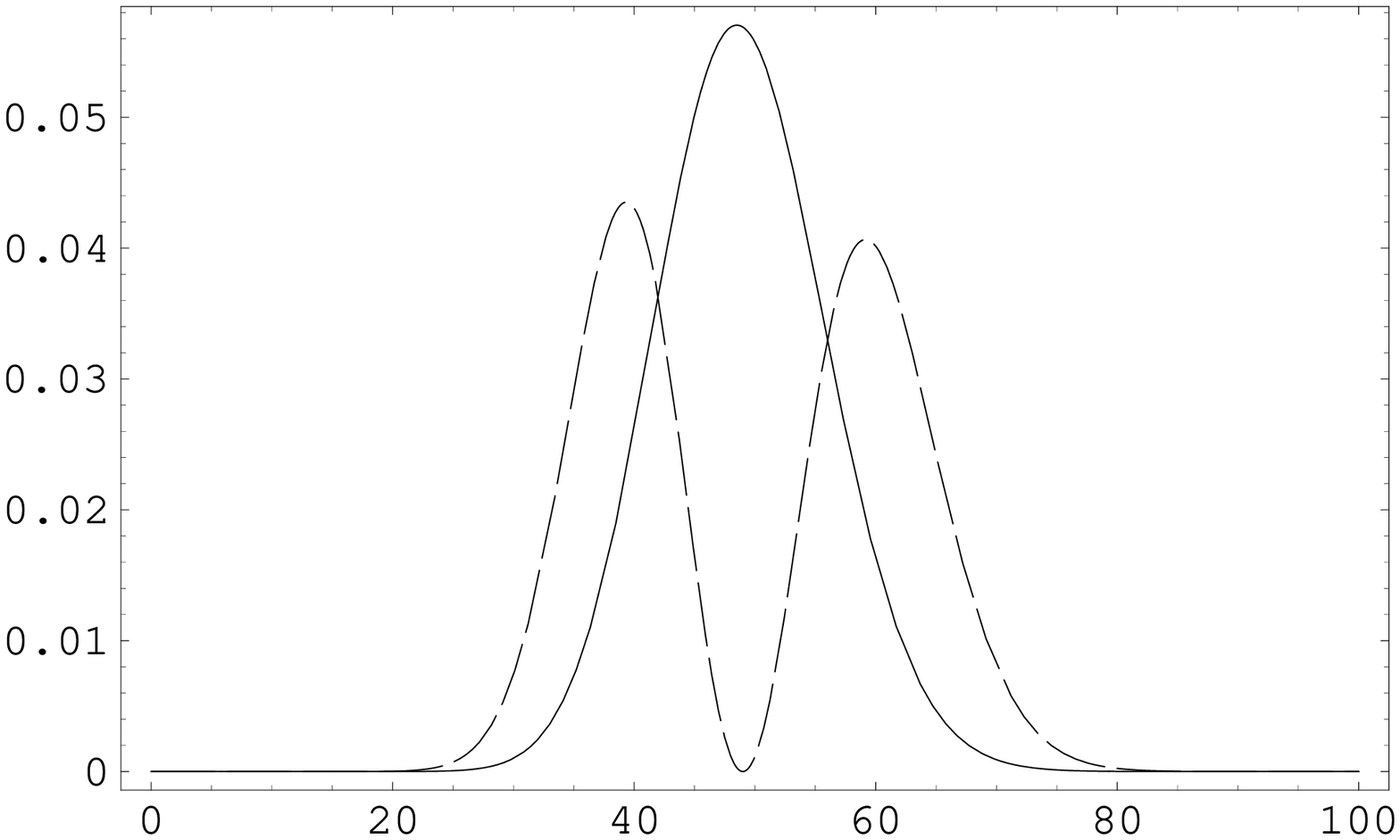}
\put(50,-5){\small $<n>~=~|z|^2~=~49$}
\put(-43,-29){\large $P(n)$}
\put(38,-70){$n$}
\end{picture}
\figcap{Photon number distribution of coherent (with an initial vacuum
  state $\rv{t=0}=\rv{0}$ - solid curve) and semi-coherent
  states (with an initial one-photon state $\rv{t=0}=\rv{1}$ - dashed curve).}
\label{FigSemiCoh}
\end{figure}
\noindent In Figure
\ref{FigSemiCoh}, 
the solid curve gives $P(n)$ for $\rv{\phi_{0}}= \rv{0}$, where we, for the
purpose of illustration, have chosen the Fourier transform of $f(t)$ such that
the mean value of the Poisson number-distribution of photons is 
$|f(\omega)|^2=49$. The
distribution $P(n)$ then characterize a {\it classical state} of the radiation
field. The dashed curve in Figure \ref{FigSemiCoh} 
corresponds to $\rv{\phi_{0}} = \rv{1}$,
and we observe the characteristic oscillations. It may be a slight surprise
that the minor change of the initial state by {\it one} photon 
completely change the
final distribution $P(n)$ of photons, i.e. {\it one} photon among a large
number of photons (in the present case $49$) makes a difference. 
If $\rv{\phi_{0}} = \rv{m}$ one finds in
the same way that the  $P(n)$-distribution will have $m$ zeros. If we sum the
distribution $P(n)$ over the initial-state quantum number $m$ we, of course,
obtain unity as a consequence of the unitarity of the time-evolution. Unitarity
is actually the simple quantum-mechanical reason why oscillations in $P(n)$
must be present. We also observe that two canonical coherent states
$\rv{t}_{I}$ are orthogonal if the initial-state fiducial vectors are
orthogonal. It is in the sense of  oscillations in $P(n)$, as described above,
that a classical current can amplify a quantum-mechanical pure state 
$\rv{\phi_{0}}$ to a coherent-state with a large number of coherent photons.
This effect is, of course, due to the boson character of photons.

It has, furthermore, been shown that one-photon states localized in space and
time can be generated in the laboratory (see
e.g. \cite{mandel86}-\cite{strekalov&98}).
%
It would be
interesting if such a state  could be
amplified  by means of a classical source in resonance with
the typical frequency of the photon. It has been argued by Knight  et al.
\cite{knight90} that an imperfect photon-detection by allowing for dissipation
of field-energy does not necessarily destroy the appearance of the oscillations
in the probability distribution $P(n)$ of photons in the displaced
number-operator eigenstates. It would, of course, be an interesting and
striking verification of quantum coherence  if the oscillations in the
$P(n)$-distribution could be observed experimentally.
 %
%
\section{Photon-Detection Theory} \label{sec-pdt}
\seqnoll
%
\begin{flushright}
``{\sl If it was so, it might be;
And if it were so, \\
it would 
be. But as it isn't, it ain't.~}''\\
Lewis Carrol
\end{flushright}

The quantum-mechanical description of optical coherence was developed
in a series of
beautiful papers by  Glauber \cite{glauber63}. Here we will only touch upon some
elementary considerations of photo-detection theory. Consider an
experimental situation where a beam of
particles, in our case a beam of photons, hits an ideal  beam-splitter. 
Two photon-multipliers
measures the corresponding intensities at times $t$ and $t + \tau $ of
the two beams
generated by the beam-splitter. The quantum state describing the detection of one
photon at  time $t$ and another one at time $t + \tau $ is then of the form $E^{+}(t +
\tau)E^{+}(t) \rv{i} $, where $\rv{i}$ describes the initial state and where $E^{+}(t)$
denotes a positive-frequency component of the second-quantized
electric field.
The quantum-mechanical amplitude for the detection of a final state $\rv{f}$
then is $\lv{f}E^{+}(t +\tau)E^{+}(t) \rv{i}$. The total
detection-probability, obtained by summing over all final states, is then proportional
to the {\it  second-order correlation function} $g^{(2)}(\tau ) $
given by 
\be 
g^{(2)}(\tau ) = \sum_{f}\frac{|\lv{f}E^{+}(t +\tau)E^{+}(t)\rv{i}|^2}
{(\lv{i}E^{-}(t)E^{+}(t)\rv{i})^{2})}=
\frac{\lv{i}E^{-}(t)E^{-}(t + \tau) E^{+}(t + \tau
)E^{+}(t) \rv{i}}{(\lv{i}E^{-}(t)E^{+}(t)\rv{i})^{2}}~~~.
 \ee
Here the normalization factor is just proportional to the intensity
of the source, 
i.e. $\sum_{f}|\lv{f}E^{+}(t) \rv{i}|^2 = (\lv{i}E^{-}(t)E^{+}(t)\rv{i})^{2}$.
A  classical treatment of the radiation field would then lead to 
\be
g^{(2)}(0) = 1 + \frac{1}{\mean{I}^{2}}\int dI P(I)(I -\mean{I})^{2}~~~,
\ee
where $I$ is the intensity of the radiation field and $P(I)$ is a 
quasi-probability distribution
(i.e. not in general an apriori positive definite function). What
we call  classical coherent light can then be described in terms of
Glauber-Klauder coherent states.
These states leads to $P(I) = \delta (I - \mean{I})$. As long as $P(I)$ is a positive
definite function, there is a complete equivalence 
between the classical theory of optical
coherence and the quantum field-theoretical description \cite{sudarshan63}. 
 Incoherent light, as thermal light, leads to a second-order correlation function
$g^{(2)}(\tau )$ which is larger than one. This feature is referred to
as {\it photon bunching} (see e.g. Ref.\cite{Tei88}). 
{\sl Quantum-mechanical light} is, however, described by a second-order
correlation function which may be smaller than one. If the beam consists of $N$
photons, all with the same quantum numbers, we easily find that  
\be 
g^{(2)}(0) = 1 - \frac{1}{N} < 1~~~. 
\ee
Another way to express this form of  
{\it photon anti-bunching } is to say that in this
case the quasi-probability $P(I)$ distribution cannot be positive, i.e. it cannot be
interpreted as a
probability (for an account of the early history of anti-bunching see e.g.
Ref.\cite{walls79,smirnov87}). 
\subsection{Quantum Interference of Single Photons}
\label{sec-interference}
\seqnoll

 A one-photon beam must, in particular, have the property
that $g^{(2)}(0) = 0$, which simply corresponds to maximal photon anti-bunching. One
would, perhaps, expect that a sufficiently attenuated classical source of radiation,
like the light from a pulsed photo-diode or a laser, would exhibit
photon maximal anti-bunching in a beam
splitter. This sort of reasoning is, in one way or another, explicitly assumed in many
of the beautiful tests of ``single-photon'' interference in quantum
mechanics.  It has, however,
been argued by Aspect and Grangier \cite{aspectgrangier87} that this
reasoning is {\it incorrect}. Aspect and Grangier actually measured 
the second-order correlation function $g^{(2)}(\tau)$ by making use of
a beam-splitter and found  this to be greater or equal to
one even for an attenuation of a classical light source below the one-photon level. The
conclusion, we guess, is that the radiation  emitted from e.g. a
monochromatic laser always behaves in  classical manner, 
i.e. even  for such a strongly attenuated source below the one-photon
flux limit the corresponding radiation has no non-classical features
(under certain circumstances one can, of course, arrange for such an attenuated
light source with a very low probability for more than one-photon at
a time  (see e.g. Refs.\cite{rieke&98,kwiat&98}) but, nevertheless,
the source can
still  be described in terms of classical electro-magnetic fields). As already
mentioned in the introduction, it is, however, possible to generate photon beams which
exhibit complete photon anti-bunching.
This has first been shown in the beautiful experimental work by Aspect and Grangier
\cite{aspectgrangier87} and   by Mandel and collaborators 
\cite{mandel86}. Roger, Grangier
and Aspect  in their beautiful study also verified that the one-photon 
states obtained exhibit one-photon interference in accordance with 
the rules of quantum mechanics as we, of course,
expect. 
In the experiment by e.g. the Rochester group \cite{mandel86} beams of one-photon
states, localized in {\it both} space and time, were generated. A quantum-mechanical
description of such relativistic one-photon states will now be the
subject for Chapter \ref{sec-photons}. 
 %
\subsection{Applications in High-Energy Physics}
\label{sec-highenergy}
\seqnoll

Many of the concepts from photon-detection theory has applications in
the context of 
high-energy physics. 
The use of photon-detection theory as mentioned in Section \ref{sec-pdt} goes
historically back to Hanbury-Brown and Twiss \cite{hbt57} in which case the
second-order correlation function was used in order to extract information on the size
of distant stars.  The same idea has been applied in high-energy  physics. The
two-particle correlation function $C_{2}(\fet{p}_{1},\fet{p}_{2})$,  where
$\fet{p}_{1}$ and $\fet{p}_{2}$ are three-momenta of the (boson) particles considered,
is in this case given by the ratio of two-particle probabilities
$P(\fet{p}_{1},\fet{p}_{2})$ and the product of the one-particle probabilities
$P(\fet{p}_{1})$ and $P(\fet{p}_{2})$, i.e. $C_{2}(\fet{p}_{1},\fet{p}_{2}) =
P(\fet{p}_{1},\fet{p}_{2})/P(\fet{p}_{1})P(\fet{p}_{2})$. For a  source of pions where
any phase-coherence is averaged out, corresponding to what is called a chaotic source,
there is an enhanced emission probability as compared to a non-chaotic source over a
range of momenta such that $R|\fet{p}_{1} - \fet{p}_{2}| \simeq 1$, where $R$
represents an average of the size of the pion source. For pions formed in a
 coherent-state one finds  that $C_{2}(\fet{p}_{1},\fet{p}_{2}) = 1$.  The width of the
experimentally determined correlation function of pions with different momenta, i.e.
$C_{2}(\fet{p}_{1},\fet{p}_{2})$,  can therefore give information about the size of the
pion-source.  A lot of experimental data has been compiled over the years and the
subject has recently been discussed in detail by e.g. Boal et al. \cite{boaletal90}. A
recent experimental analysis has been considered by the OPAL collaboration in the case
of like-sign charged track pairs at a center-off-mass energy close to the $Z^{0}$ peak.
146624 multi-hadronic $Z^{0}$ candidates were used leading to an estimate of the radius
of the pion source to be close to one fermi \cite{opal91}. Similarly the $NA44$
experiment at CERN have studied $\pi^{+}\pi^{+}$-correlations from 227000 reconstructed
pairs in $S + Pb$ collisions at $200~GeV/c$ per nucleon leading to a space-time
averaged pion-source radius of the order of a few fermi \cite{na4492}. The impressive
experimental data and its interpretation has been confronted by simulations using
relativistic molecular dynamics \cite{sullivanetal93}. In heavy-ion physics the
measurement of the second-order correlation function of pions is of
special interest since
it can give us information about the spatial extent of the quark-gluon plasma phase, if
it is formed. It has been suggested that one may make use of photons instead
of pions when studying possible signals from the quark-gluon plasma. In particular, it
has been suggested \cite{kaputsa93} that the correlation of high transverse-momentum
photons is sensitive to the details of the space-time evolution of the high density
quark-gluon plasma.
%
%
\section{Relativistic Quantum Mechanics of Single Photons}
\label{sec-photons}
\seqnoll
%
%
%
\begin{flushright}
``{\sl  Because the word photon is used in so many ways, \\ it is a
  source of much confusion. 
The reader always\\ has to figure out what the writer has in mind.}"\\
P. Meystre and M. Sargent III
\end{flushright}

The concept of a photon has a long and
intriguing history in physics. It is, e.g., in this context interesting to notice a
remark by A. Einstein; ``{\it All these fifty years of pondering have
  not brought me any closer to answering the question: What are light
  quanta?~}'' \cite{pais&82}. Linguistic considerations do not
appear to enlighten our conceptual understanding of this fundamental
concept either \cite{klyshko&94}.
Recently, it has even been suggested that
one should not make use of the concept of a photon at all
\cite{lamb&95}. As we have remarked above, {\it single} photons can,
however, be generated in the laboratory and the wave-function of
single photons can actually be measured \cite{smithey&93}. The decay
of a {\it single} photon quantum-mechanical state
in a resonant cavity has also recently been
studied experimentally \cite{maitre&etal&97}.

A related concept is that of localization of relativistic elementary
systems, which also has a long and
intriguing history 
(see e.g. Refs. \cite{newton}-\cite{adlard&97}).
Observations of physical phenomena takes place in space and time.
The notion of {\it localizability} of particles, elementary or not, then refers to the
empirical fact that particles, at a given instance of time, appear to be localizable in
the physical space. 

In the realm of non-relativistic quantum mechanics the concept of
localizability of particles is built into the theory at a very
fundamental  level and is expressed in terms of the fundamental
canonical commutation relation
between a position operator and the corresponding generator of translations,
i.e. the canonical momentum of a particle. In relativistic theories the
concept of localizability of physical systems is  deeply connected to our
notion of space-time, the arena of physical phenomena, as a 4-dimensional
continuum. In the context of the classical  theory of general relativity
 the localization of light rays in space-time is e.g. a fundamental
ingredient. In fact, it has been argued \cite{ehlers} that the Riemannian
metric is basically determined by basic properties of light propagation.

In a fundamental paper by Newton and Wigner \cite{newton} it
 was argued that in the context of relativistic quantum mechanics a
 notion of point-like 
localization of a single particle 
can be, uniquely, determined by kinematics. Wightman
\cite{wightman} extended this notion to localization to finite domains of
space and it was, rigorously, shown that massive particles are always
localizable if they are elementary, 
i.e. if they are described in terms of
irreducible representations of the Poincar\'{e} group \cite{wigner}. 
Massless
elementary systems with non-zero helicity, like a gluon, graviton, 
neutrino or a photon, are {\it not} localizable in the sense of Wightman. The
axioms used by Wightman can, of course, be weakened. It was actually shown
by Jauch, Piron and Amrein \cite{jpa} that in such a sense the photon is
{\it weakly localizable}. As will be argued below, 
the notion of weak localizability essentially corresponds
to allowing for non-commuting observables in order to characterize the
localization of massless and spinning particles in general.

Localization of relativistic particles, at a fixed
 time, as alluded to above, has been shown to be
incompatible with a natural notion of (Einstein-) causality \cite{heger}. 
If relativistic  elementary system has an exponentially small tail outside a
finite domain of localization at $t=0$, then, according to the hypothesis of
a weaker form of causality, this should remain true at later times, i.e. the
tail should only be shifted further out to infinity. As was shown by
Hegerfeldt \cite{heger2}, even this notion of causality is incompatible with
the notion of a positive and bounded observable whose expectation value gives
the probability to a find a particle inside a finite domain of space at a
given instant of time. It has been argued that the use of local
 observables in the context of relativistic quantum field theories
 does not lead to such apparent difficulties with Einstein
 causality \cite{buch&yng&94}.

We will now reconsider some of these
questions related to the concept of localizability in terms of a 
quantum mechanical description of a
massless particle with given helicity $\lambda $
\cite{baletal,atre,albs1} (for a related construction see
Ref.\cite{iwo&87}). The one-particle states we are considering are, of
course, nothing else than the positive energy
one-particle states of quantum field
theory. We simply endow such states with a set of appropriately 
defined quantum-mechanical observables and, in terms of these, we
 construct the generators of
the Poincar\'{e} group.
We will then show how one can extend this description to include both
positive and negative
helicities, i.e. including reducible representations of the
Poincar\'{e} group. We are then in the position  to e.g. study the motion of a
linearly polarized photon
in the framework of relativistic quantum mechanics and the appearance
of non-trivial phases of wave-functions.
%
%
\subsection{Position Operators for Massless Particles}
\label{sec-positionoperators}
\seqnoll
%
%
It is easy to show that the components of the position operators for a
massless particle must
be non-commuting\footnote{This argument has, as far as we  know, first been
suggested by N. Mukunda.}  if the helicity $\lambda \neq 0$. 
If $J_{k}$ are the generators
of rotations and $p_{k}$ the diagonal momentum operators, $k = 1,2,3$, then 
 we should have ${\bf J}\cdot {\bf p} = \pm \lambda$ for a massless
 particle like the photon (see e.g. Ref.\cite{wigner&57}).
Here ${\bf J } = (J_{1},J_{2},J_{3})$ and
${\bf p } = (p_{1},p_{2},p_{3})$. In terms of natural units 
($\hbar = c = 1$) we then have that
\begin{equation}
[J_{k},p_{l}] = i\epsilon _{klm} p_{m}~~~~.
\end{equation}
If a canonical position operator ${\bf x }$ exists with components
$x_{k}$  such that 
\begin{eqnarray}
&[x_{k},x_{l}] &= 0~~~~, \\
&[x_{k},p_{l}] &= i\delta _{kl}~~~~, 
\label{ccr} \\
&[J_{k},x_{l}] &= i\epsilon _{klm} x_{m}~~~~,
\end{eqnarray}
then we can define generators of orbital angular momentum in the conventional
way, i.e. 
\begin{equation}
L_{k} = \epsilon _{klm} x_{l}p_{m}~~~~.~
\end{equation}
Generators of spin 
are then defined by
\begin{equation}
 S_{k} = J_{k} - L_{k}~~~~.
\end{equation}
They fulfill the correct algebra, i.e.
\begin{equation}
[S_{k}, S_{l}] = i\epsilon _{klm} S_{m}~~~~,
\end{equation}
and they, furthermore, commute with ${\bf x}$ and ${\bf p}$.
 Then, however, the spectrum
of ${\bf S}\cdot {\bf p}$ is $\lambda ,\lambda -1 ,..., -\lambda $, which
contradicts the requirement ${\bf J}\cdot {\bf p} = \pm \lambda$
 since, by construction,
${\bf J}\cdot {\bf p} ={\bf S}\cdot {\bf p}$.
 
 As has been discussed in detail
in the literature, the non-zero commutator of the components of the 
position operator  for a
massless particle primarily emerges due to the non-trivial topology of the
momentum space \cite{baletal,atre,albs1}. The irreducible representations of the
Poincar\'{e} group for massless particles
 \cite{wigner} can be constructed from a knowledge of the little group
 $G$ of a light-like
momentum four-vector $p = (p^{0}, {\bf p})$ . This group is the Euclidean group
$E(2)$. Physically, we are interested in possible finite-dimensional
representations of the covering of this little group.
We therefore restrict ourselves to the compact subgroup, i.e. we
represent the 
$E(2)$-translations trivially and
consider $G=SO(2)=U(1)$. Since the origin in the momentum space is
excluded for massless particles  one is therefore led to consider appropriate
$G$-bundles over $S^2$ since the energy of the particle can be kept
fixed. Such $G$-bundles are classified by mappings from the equator to
G, i.e. by the first homotopy group $\Pi_{1} (U(1))$={\bf Z}, 
where it turns out that each integer
corresponds to twice the helicity of the particle. A massless particle
with helicity $\lambda $ and {\it sharp momentum} is thus  described in
terms of a non-trivial line bundle characterized by $\Pi _{1}(U(1)) = \{
2\lambda \} $ \cite{nagel}.

This consideration can easily be extended to higher space-time
dimensions \cite{albs1}. 
If $D$ is the number of space-time
dimensions, the corresponding $G$-bundles are classified by the 
homotopy groups $\Pi _{D-3}
(Spin(D-2))$. These homotopy groups are in general non-trivial. It is a
remarkable fact that the only trivial homotopy groups  of this form in
higher space-time
dimensions correspond to $D=5$ and $D=9$  due to the existence of quaternions
and the Cayley  numbers (see e.g. Ref. \cite{white}).  In these space-time
dimensions, and for $D=3$, it then turns that one can explicitly
construct 
canonical {\it and}
commuting position operators for massless particles \cite{albs1}. 
The mathematical fact that the spheres $S^{1}$, $S^{3}$ and $S^{7}$
are parallelizable can then be expressed in terms of the existence of
canonical {\it and} commuting position operators for massless spinning particles
in  $D=3$, $D=5$ and $D=9$ space-time dimensions.

In terms of a
canonical momentum $p_{i}$ and coordinates $x_{j}$ satisfying the 
canonical commutation relation Eq.(\ref{ccr}) we can easily derive
the commutator of two components of the position operator ${\bf x}$ by making use of a 
simple consistency argument as follows.  If the massless
particle has a given helicity $\lambda$, then the generators of angular
momentum is given by:
 \begin{equation}
J_{k} = \epsilon _{klm}x_{l}p_{m} +
\lambda \frac{p_{k}}{|{\bf p}|}~~~.\label{eqn2}
\end{equation}
The canonical momentum then transforms as a vector under rotations,
i.e.
\begin{equation}
[J_{k},p_{l}] = i\epsilon _{klm}p_{m}~~~,
\end{equation}
without any condition on the commutator of two components of the 
position operator {\bf x}. The
 position operator will, however, not transform like a vector
unless the following commutator is postulated
\begin{equation}
i[x_{k},x_{l}] = \lambda \epsilon _{klm}\frac{p_{m}}{|{\bf p}|^{3}}~~~,
\label {eqn4} 
 \end{equation}
where we notice that commutator formally corresponds to a point-like Dirac magnetic
monopole \cite{dirac} localized at the origin in momentum space with strength $4\pi
\lambda$. The energy $p^{0}$ of the massless particle is, of course, given by
 $\omega = |{\bf p}|$. In terms of a singular $U(1)$ connection ${\cal
   A}_{l}\equiv {\cal A}_{l}({\bf p})$ we can write
\begin{equation}
x_k = i\partial _k - {\cal A }_{k}~~~,
\label{dcon}
\end{equation}
where $\partial _{k} =\partial /\partial p_{k} $ and 
\begin{equation}
\partial _{k}{\cal A}_{l} - \partial _{l}{\cal A}_{k} = 
\lambda \epsilon _{klm}\frac{p_{m}}{|{\bf p }|^{3}}~~~.
\end{equation}
Out of the observables $x_{k}$ and the energy $\omega$ one can
easily construct the generators (at time $t=0$) of Lorentz boots, i.e.
\begin{equation} 
K_{m}= (x_{m}\omega + \omega x_{m})/2~~~~,
\label{boost}
\end{equation}
and verify that $J_{l}$ and $K_{m}$ lead to a
realization of the Lie algebra of the Lorentz group, i.e.
\begin{eqnarray}
&[J_{k},J_{l}]&=\, i\epsilon _{klm}J_{m}~~~~,\\
&[J_{k},K_{l}]&=\, i\epsilon _{klm}K_{m}~~~~,\\
&[K_{k},K_{l}]&=\, -i\epsilon_{klm}J_{m}~~~~.
\end{eqnarray} 
The components of the Pauli-Plebanski operator $W_\mu$ are given by
\begin{equation}
W^{\mu} = (W^{0},{\bf W}) = ({\bf J}\cdot {\bf p}, {\bf J}p^{0} +
{\bf K}\times {\bf p}) = \lambda p^{\mu}~~~,
\end{equation}
i.e. we also obtain an irreducible representation of the Poincar\'{e}
group. 
The additional non-zero commutators are
\begin{eqnarray}
& [K_{k},\omega ] &= ip_{k}~~~~, \\
& [K_{k},p_{l}] &=i\delta _{kl}\omega ~~~~.
\end{eqnarray}
At $t\equiv x^{0}(\tau )\neq 0$ the Lorentz boost generators $K_{m}$ as
given by Eq.(\ref{boost}) are extended to
\begin{equation}
K_{m}= (x_{m}\omega + \omega x_{m})/2 - tp_{m}~~~~.
\label{Kt}
\end{equation} 
In the Heisenberg picture, the quantum equation of motion of an
observable ${{\cal O}(t)}$ is obtained by using
\begin{equation}
\frac{d{\cal O}(t)}{dt} = \frac{\partial {\cal O}(t)}{\partial t}
+ i[H,{\cal O}(t)]~~~~,
\end{equation}
where the Hamiltonian $H$ is given by the $\omega $. One then finds that
all generators of the Poincar\'{e} group are conserved as they
should. 
The equation of motion for {\bf x}(t) is
\begin{equation}
\frac{d}{dt} {\bf x}(t) = \frac{{\bf p}}{\omega}~~~~,
\end{equation}
which is an expected equation of motion for a massless particle.

The  non-commuting components $x_{k}$ of the position operator ${\bf x}$ transform
as the components of a vector under spatial rotations. Under Lorentz boost we find in
addition that
\begin{equation}
i[K_{k},x_{l}] = \frac{1}{2}\left( x_{k}\frac{p_{l}}{\omega}
+ \frac{p_{l}}{\omega}x_{k} \right) - t\delta _{kl} +
\lambda \epsilon _{klm}\frac{p_m}{|{\bf p }|^2}~~~~.
\label{kx} 
\end{equation}
The first two terms in Eq.(\ref{kx}) corresponds to the correct limit for
$\lambda = 0$ since the proper-time condition $x^{0}(\tau ) \approx
\tau $ is 
not Lorentz invariant (see e.g.  \cite{hanson}, Section 2-9). 
The last term in Eq.(\ref{kx}) is
due to the 
non-zero commutator Eq.(\ref{eqn4}). This anomalous term can be dealt
with by 
introducing an appropriate two-cocycle for finite transformations
consisting of translations generated by the position operator ${\bf
  x}$, rotations generated by ${\bf J}$ and Lorentz boost 
generated by ${\bf K}$. For pure translations this two-cocycle will be
explicitly 
constructed in Section~\ref{sec-berry}.

The
algebra discussed above can be extended in a rather straightforward manner
to incorporate both positive and negative helicities  needed in order to
describe linearly polarized light. As we now will see this extension
corresponds to a replacement of the Dirac monopole at the origin
in momentum space with
a $SU(2)$ Wu-Yang \cite{wuyang} monopole. The procedure below follows
a rather 
standard method of imbedding the singular $U(1)$ connection ${\cal A}_{l}$ into 
a {\it regular} $SU(2)$ connection.
 Let us specifically consider a massless, spin-one particle. 
The Hilbert space, ${\cal H}$, of  one-particle transverse 
wave-functions $\phi_{\alpha }({\bf p}), \alpha = 1,2,3$ is defined in
terms of a scalar product
\begin{equation} 
(\phi,\psi) = \int d^{3}p \phi ^{*}_{\alpha }({\bf p})\psi _{\alpha}({\bf p})~~~, 
\end{equation}
where $\phi ^{*}_{\alpha }({\bf p})$ denotes the complex conjugated
 $\phi _{\alpha}({\bf p})$.
In terms of a Wu-Yang connection ${\cal A}^{a}_{k }\equiv {\cal
 A}^{a}_{k }({\bf p})$, i.e.
\begin{equation}
{\cal A}^{a}_{k }({\bf p})= \epsilon _{alk}\frac{p_{l}}{|{\bf
p}|^{2}}~~~, 
\end{equation}
Eq.(\ref{dcon}) is extended to
\begin{equation}
x_k = i\partial _k - {\cal A }_{k}^{a}({\bf p})S_{a}~~~,
\label{wupos}
\end{equation}
where
\begin{equation}
(S_{a})_{kl}= -i\epsilon _{akl}
\end{equation}
 are the spin-one generators. By means of a singular
 gauge-transformation the 
Wu-Yang connection can be transformed into the singular
 $U(1)$-connection ${\cal A}_{l}$ 
times the third component of the spin generators $S_{3}$ 
(see e.g. Ref.\cite{baletal91}).
This position operator defined by Eq.(\ref{wupos}) is compatible with
 the transversality 
condition
on the one-particle wave-functions, i.e. $x_{k}\phi _{\alpha}({\bf p})$ is
transverse. With suitable conditions on the one-particle
 wave-functions, the 
position operator ${\bf x}$ therefore has a well-defined action on
 ${\cal H }$. Furthermore, 
\begin{equation}
i[x_{k},x_{l}] = {\cal
F}^{a}_{kl}S^{a} =  \epsilon _{klm}\frac{p_{m}}{|{\bf p}|^{3}}
\hat{{\bf p}}\cdot {\bf S}~~~,
\label{eq:gen}
 \end{equation}
where
\begin{equation}
{\cal F}^{a}_{kl} = \partial _{k}{\cal A}^{a}_{l}
 - \partial _{l}{\cal A}^{a}_{k} 
-\epsilon _{abc}{\cal A}^{b}_{k}{\cal
A}^{c}_{l} = \epsilon _{klm}\frac{p_{m}p_{a}}{|{\bf p }|^{4}}~~~,
\end{equation}
is the non-Abelian $SU(2)$ field strength tensor and $\hat{{\bf p}}$
is a unit vector in the 
direction of the particle
momentum ${\bf p}$. The generators of angular momentum are now defined
as follows
 \begin{equation}
J_{k} = \epsilon _{klm}x_{l}p_{m} + 
\frac{p_{k}}{|{\bf p}|}\hat{{\bf p}}\cdot {\bf S}~~~.
\label{eq:ang}
\end{equation}
The helicity operator $ \Sigma \equiv \hat{{\bf p}}\cdot {\bf S}$ is
covariantly constant, i.e.
\begin{equation}
\partial _{k} \Sigma + i\left[A_{k},\Sigma \right] = 0~~~,
\label{ccon}
\end{equation}
where $A_{k} \equiv {\cal A}_{k}^{a}({\bf p})S_{a}$. The position
 operator ${\bf x}$ therefore
 commutes with  $\hat{{\bf p}}\cdot {\bf S}$. One can therefore verify
 in a straightforward 
manner that the observables $p_{k}, \omega  , J_{l}$ and $K_{m} =
 (x_{m}\omega +\omega x_{m})/2$ 
close to the Poincar\'{e} group. At $t \neq 0$ the Lorentz boost
 generators $K_{m}$ are 
defined as in Eq.(\ref{Kt}) and Eq.(\ref{kx}) is  extended to
\begin{equation}
i[K_{k},x_{l}] = \frac{1}{2}\left( x_{k}\frac{p_{l}}{\omega}
+ \frac{p_{l}}{\omega}x_{k} \right) - t\delta _{kl} +
i\omega[x_{k},x_{l}]~~~~.
\label{kx2}
\end{equation}

For helicities $\hat{{\bf p}}\cdot {\bf S}= \pm \lambda$ one 
extends the previous considerations
 by considering ${\bf S}$ in the spin $|\lambda |$-representation.
Eqs.(\ref{eq:gen}), (\ref{eq:ang}) and (\ref{kx2}) are then valid in
general. A reducible 
representation for the generators
of the Poincar\'{e} group for an arbitrary spin has therefore been
constructed for a massless particle. We observe that the helicity 
operator $\Sigma$ can be interpreted as a
generalized ``magnetic charge'', and since $\Sigma$ is covariantly 
conserved one can use the general theory of topological quantum 
numbers \cite{goddard} and derive the quantization condition
\begin{equation}
\exp (i4\pi \Sigma ) =1~~~~,
\end{equation}
i.e. the helicity is properly quantized. In the next section we will
present an alternative way to derive helicity quantization.

\subsection{Wess-Zumino Actions and Topological Spin}

Coadjoint orbits on a group $G$  
has a geometrical structure which naturally admits
a symplectic two-form (see e.g.
\cite{kir,yaffe82,woodhouse&92}) which can be used 
to construct topological Lagrangians, i.e.
Lagrangians constructed by means of Wess-Zumino terms \cite{wesszumino} (for a
general account see e.g. Refs.\cite{baletal91,baletal82}).
Let us
illustrate the basic ideas for a non-relativistic spin and $G = SU(2)$.
Let ${\cal K}$ be an element of the Lie algebra ${\cal G}$ of $G$ in
the fundamental representation.
Without loss of generality we can write 
${\cal K} = \lambda _{\alpha}\sigma _{\alpha} = \lambda \sigma _3$, 
where $\sigma _{\alpha}, \alpha = 1,2,3$ denotes the three Pauli spin
matrices.  Let $H$ be the little
group  of ${\cal K}$. Then the coset space $G/H$ is isomorphic to
$S^2$ and defines an adjoint
orbit (for semi-simple Lie groups adjoint and coadjoint representations are
equivalent due to the existence of the non-degenerate Cartan-Killing form).
The action for the spin degrees of freedom is then expressed in terms 
of the group $G$ itself, i.e.
\begin{equation}
S_{P} = -i\int \bk{{\cal K}}{g^{-1}(\tau )dg(\tau )/d\tau} d\tau ~~~~,
\end{equation}
where $\bk{A}{B}$ denotes the trace-operation of two
Lie-algebra elements $A$ and $B$ in ${\cal G}$ and where 
\begin{equation}
g(\tau )= \exp (i\sigma
_{\alpha} \xi _{\alpha}(\tau ))
\end{equation}
 defines the (proper-)time dependent
dynamical group element.
 We observe that $S_{P}$ has
a gauge-invariance, i.e. the transformation 
\begin{equation}
g(\tau ) \longrightarrow g(\tau ) \exp \left( i\theta (\tau )\sigma _3
\right) \label{uone}
\end{equation}
only change the Lagrangian density 
$\bk{{\cal K }}{g^{-1}(\tau )dg(\tau )/d\tau }$ by a total time
derivative. The gauge-invariant components of spin, $S_{k}(\tau )$,
 are defined in terms of ${\cal K}$
by the relation
 \begin{equation}
 S(\tau ) \equiv S_{k}(\tau )\sigma _{k} = \lambda
g(\tau )\sigma _{3} g^{-1}(\tau ) ~~~,
\label{spindef}
 \end{equation}
such that
\begin{equation}
S^2 \equiv S_{k}(\tau )S_{k}(\tau ) = \lambda ^{2}~~~.
\end{equation}
By adding a non-relativistic particle kinetic term as well as  a 
conventional magnetic moment 
interaction term to the action $S_{P}$, one can verify that the
components  $S_{k}(\tau )$  obey the correct
classical equations of motion for spin-precession \cite{baletal,baletal91}.
 
Let $M = \{ \sigma ,\tau | \sigma \in [0,1] \}$ and
 $ (\sigma ,\tau ) \rightarrow g(\sigma ,\tau )$ parameterize $\tau
-$dependent paths in $G$ such that $g(0 ,\tau ) = g_{0}$ is an arbitrary
reference element and $g(1,\tau ) = g(\tau )$. The Wess-Zumino term in this
case is given by 
\begin{equation}
 \omega _{WZ} = -id \bk{{\cal
K}}{g^{-1}(\sigma, \tau )dg(\sigma, \tau )} =
 i\bk{{\cal K}}{(g^{-1}(\sigma, \tau )dg(\sigma, \tau ))^{2}}~~~,
  \end{equation} 
where
$d$ denotes exterior differentiation and where now 
\begin{equation}
g(\sigma, \tau )= \exp (i\sigma
_{\alpha} \xi _{\alpha}(\sigma, \tau ))~~~~.
\end{equation}
 Apart from boundary terms which do not contribute to the equations of
motion, we then have that 
\begin{equation}
S_{P} = S_{WZ} \equiv  \int _{M}\omega _{WZ} =
-i\int _{\partial M} \bk{{\cal K}}{g^{-1}(\tau )dg(\tau )}~~~,
\label{wzaction}
\end{equation} 
where the one-dimensional boundary $\partial M $ of $M$ , parameterized 
by $\tau $, can play the role of (proper-) time. $\omega _{WZ}$ is now 
gauge-invariant under a larger $U(1)$ symmetry, i.e. Eq.(\ref{uone}) 
is now extended to
\begin{equation}
g(\sigma,\tau ) \longrightarrow g(\sigma,\tau ) 
\exp \left( i\theta (\sigma,\tau )\sigma _3 \right)~~~~.
\end{equation}
$\omega _{WZ}$ is therefore a closed but not exact two-form defined on the coset space $G/H$.
A canonical analysis then shows that there are no gauge-invariant 
dynamical degrees of freedom in the interior of $M$.
The Wess-Zumino action Eq.(\ref{wzaction}) is the
topological action for spin degrees of freedom.

As for the quantization of the theory described by the action
Eq.(\ref{wzaction}), one may use methods from geometrical quantization and
especially the Borel-Weil-Bott theory of representations of compact Lie
groups \cite{kir,baletal91}. One then finds that $\lambda$ is half an
integer, i.e. $|\lambda |$ corresponds to the spin. 
This quantization of $\lambda $ also
naturally emerges by demanding that the action Eq.(\ref{wzaction}) is well-defined in
quantum mechanics for  periodic motion as recently was discussed by e.g.
Klauder \cite{klauder}, i.e.
\begin{equation}
4\pi \lambda = \int _{S^{2}} \omega _{WZ} = 2\pi n ~~~~,
\end{equation}
where $n$ is an integer. The symplectic two-form $\omega _{WZ}$ must
then belong to an 
integer class cohomology.
 This geometrical approach is in principal
straightforward, but it requires explicit coordinates on $G/H$. An
alternative approach, as used in \cite{baletal,baletal91}, is a
canonical Dirac analysis and quantization \cite{hanson}. This procedure leads to the
condition $\lambda ^{2} = s(s+1)$, where $s$ is half an integer. The
fact that one can arrive at different answers for $\lambda $ illustrates
a certain lack of uniqueness in the quantization procedure of the action
Eq.(\ref{wzaction}). The quantum theories obtained describes, however, the
same physical system namely one irreducible representation of the group
$G$.

The action Eq.(\ref{wzaction}) was first proposed in \cite{bor}. The action can
be derived quite naturally in terms of a coherent state path integral (for a
review see e.g. Ref.\cite{klauderskagerstam85}) using spin coherent states. It is
interesting to notice that structure of the action Eq.(\ref{wzaction})
actually appears 
in such a language already in a
paper by Klauder on continuous representation theory \cite{klauderaction}.

A classical action which after quantization leads to a description of a
massless particle in terms of an irreducible representations of the Poincar\'{e}
group can be constructed in a similar fashion \cite{baletal}. Since the
Poincar\'{e} group is non-compact the geometrical analysis referred to above for
non-relativistic spin must be extended and one should consider coadjoint
orbits instead of adjoint orbits (D=3 appears to be an 
exceptional case due to the existence of a non-degenerate bilinear form
on the D=3 Poincar\'{e} group Lie algebra \cite{witten}. In this case there
is a topological action for irreducible representations of the form
Eq.(\ref{wzaction}) \cite{topq}). The point-particle action in D=4 
then takes the form
\begin{equation} 
 S= \int d\tau \left( p_{\mu}(\tau )\dot{x}^{\mu}(\tau
) + \frac{i}{2} \mbox{Tr} [{\cal K } \Lambda ^{-1}(\tau ) 
\frac{d}{d\tau}\Lambda (\tau )]
 \right)~~~.
\label{paction}
\end{equation}
Here $[\sigma _{\alpha \beta}]_{\mu \nu} = -i(\eta _{\alpha \mu}\eta _{\beta \nu}
-\eta _{\alpha \nu } \eta_{\beta \mu})$ are the Lorentz group
generators in the spin-one 
representation and $\eta _{\mu \nu} = (-1,1,1,1)$ is the Minkowski 
metric. The trace operation has a conventional meaning, 
i.e. $\mbox{Tr} [{\cal M}] = {\cal M}^{\alpha}_{~~\alpha}$.
The Lorentz group Lie-algebra element ${\cal K }$ is here chosen to
be $\lambda \sigma _{12}$. 
The $\tau $-dependence of the Lorentz group element $\Lambda _{\mu
  \nu}(\tau )$ is defined by 
\begin{equation}
\Lambda _{\mu \nu}(\tau ) = \left[ \exp \left(i\sigma _{\alpha \beta} 
\xi ^{\alpha \beta }(\tau ) \right)\right]_{\mu \nu}~~~~.
\end{equation}
 The momentum variable $p_{\mu}(\tau )$ is defined by
\begin{equation}
p_{\mu }(\tau ) = \Lambda _{\mu \nu} (\tau ) k^{\nu }~~~,
\end{equation}
where the constant reference momentum $ k^{\nu }$ is given by 
\begin{equation}
k^{\nu }
 = (\omega ,0,0,|{\bf k }|)~~~,
\end{equation}
where $\omega = |{\bf k }|$. The momentum $p_{\mu }(\tau )$ is then
light-like by construction. 
The action Eq.(\ref{paction}) leads to the equations of motion
\begin{equation}
\frac{d}{d\tau}p_{\mu}(\tau ) = 0~~~,
\end{equation}
and
\begin{equation}
\frac{d}{d\tau } \biggl \{ x_{\mu }(\tau )p_{\nu}(\tau ) - 
x_{\nu}(\tau )p_{\mu}(\tau ) + S_{\mu \nu }(\tau )  \biggr\} = 0 ~~~.
\end{equation}
Here we have defined gauge-invariant spin degrees of freedom $ S_{\mu
  \nu }(\tau ) $ by
\begin{equation}
 S_{\mu \nu }(\tau )  = 
\frac{1}{2} \mbox{Tr}[\Lambda (\tau ){\cal K }\Lambda ^{-1}(\tau )\sigma _{\mu\nu}]
\end{equation}
in analogy with Eq.(\ref{spindef}). These spin degrees of freedom satisfy
the relations
\begin{equation}
p_{\mu} (\tau ) S^{\mu \nu }(\tau ) = 0~~~,
\label{CON1}
\end{equation} 
and
\begin{equation}
\frac{1}{2} S_{\mu \nu }(\tau ) S^{\mu \nu }(\tau ) = \lambda ^{2}~~~.
\label{CON2}
\end{equation}
Inclusion of external electro-magnetic and gravitational fields 
leads to the classical Bargmann-Michel-Telegdi \cite{telegdi&etal&59}
 and Papapetrou \cite{papape&51} equations of
motion respectively \cite{baletal}. Since the equations derived are
expressed in terms of {\it bosonic} variables these equations of motion
admit a straightforward classical interpretation. (An alternative 
{\it bosonic} or {\it fermionic}
treatment of internal degrees of freedom which also leads to Wongs
equations of motion \cite{wong70} in the presence of in general
non-Abelian external gauge fields can be found in Ref.\cite{bssw}.)

Canonical quantization of the system described by {\it bosonic}
degrees of freedom and 
the action Eq.(\ref{paction}) leads to a realization of the
Poincar\'{e}
 Lie algebra with generators $p_{\mu}$ and $J_{\mu \nu}$ where
\begin{equation}
J_{\mu \nu} = x_{\mu}p_{\nu} -  x_{\nu}p_{\mu} + S_{\mu \nu}~~~~.
\end{equation}
The four vectors $x_{\mu}$ and $p_{\nu}$ commute with the spin
generators $S_{\mu \nu}$ 
and are canonical, i.e.
\begin{eqnarray}
&[x_{\mu},x_{\nu}] &=\,  [p_{\mu},p_{\nu}] = 0~~~~, \\
&[x_{\mu},p_{\nu}] &=\, i\eta _{\mu \nu}~~~~.
\end{eqnarray}
The spin generators $S_{\mu \nu}$ fulfill the conventional algebra
\begin{equation}
[S_{\mu \nu},S_{\lambda \rho}] = i(\eta _{\mu \lambda}S_{\nu \rho} +
\eta _{\nu \rho}S_{\mu \lambda} - \eta _{\mu \rho}S_{\nu \lambda}
-\eta _{\nu \lambda}S_{\mu \rho})~~~~.
\end{equation}
The mass-shell condition $p^{2}=0$ as well as the constraints
Eq.(\ref{CON1}) and Eq.(\ref{CON2}) are all first-class constraints \cite{hanson}. 
In the proper-time gauge $x^{0}(\tau ) \approx \tau $ one obtains the 
system described in Section \ref{sec-positionoperators}, i.e. we obtain an irreducible
representation of the Poincar\'{e} group with helicity $\lambda $ \cite{baletal}. 
For half-integer helicity, i.e. for fermions, one can verify in a 
straightforward manner that the wave-functions obtained change with a
minus-sign under a $2\pi $
rotation \cite{baletal,albs1,baletal91} as they should.
%
%
\subsection{The Berry Phase for Single Photons}
\label{sec-berry}
%
%
We have constructed a set of $O(3)$-covariant position  operators of massless
particles and a  reducible representation of the
Poincar\'{e} group corresponding 
to a combination of positive and negative helicities. 
It is interesting to notice that the
construction above leads to observable effects. Let us specifically consider
photons and the motion of photons along e.g. an optical fibre. Berry has argued
\cite{berry} that a spin in an adiabatically changing magnetic field leads
to the appearance of an observable phase factor, called the Berry phase. It
was suggested in Ref.\cite{wu} that a similar geometric phase could appear for
photons. We will now,  within the framework of the {\it relativistic quantum
mechanics} of a single massless particle as discussed above, 
give a {\it derivation} of this geometrical phase. The Berry phase
for a {\it single }
photon can e.g. be obtained as follows. We consider the motion of a
photon with fixed energy moving e.g. along an optical fibre. We assume that as  the
photon moves in the fibre, the momentum vector traces out a closed loop in
momentum space on the constant energy surface, i.e. on a two-sphere
$S^2$. This simply means that the initial and final momentum vectors of the
photon are the same.  We
therefore consider a wave-function $\ket{\bf p}$ which is diagonal in
momentum. We also define the translation operator  $U({\bf a}) = \exp (i{\bf
a}\cdot {\bf x})$. It is straightforward to show, using  Eq.(\ref{eqn4}), that
 \begin{equation} U({\bf a})U({\bf b}) \ket{\bf p} = \exp
(i\gamma [{\bf a},{\bf b};{\bf p}]) \ket{{\bf p}+{\bf a}+{\bf b}}~~~~,
\end{equation}
where the two-cocycle phase $\gamma [{\bf a},{\bf b};{\bf p}]$ is equal
to the flux of the magnetic monopole in momentum space through the simplex
spanned by the vectors ${\bf a}$ and ${\bf b}$ localized at  the point ${\bf
p}$, i.e.
 \begin{equation} \gamma [{\bf a},{\bf b};{\bf p}]
 = \lambda \int_{0}^{1} \int_{0}^{1} d\xi _{1} d\xi _{2} a_{k}b_{l}
\epsilon _{lkm} B_{m}({\bf p} + \xi _{1} {\bf a} + \xi _{2} {\bf b})~~~~,
\end{equation}
where $B_{m}({\bf p}) = p_{m}/|{\bf p}|^{3}$. The non-trivial phase appears
because the second de Rham cohomology group of $S^{2}$ is
non-trivial. The two-cocycle 
phase $\gamma [{\bf a},{\bf b};{\bf p}]$ is therefore not a coboundary
and 
hence it cannot be removed by a redefinition of $U({\bf a})$. This result 
has a close analogy in the theory of
magnetic monopoles \cite{jackiw}. The anomalous commutator
Eq.(\ref{eqn4}) 
therefore leads to a ray-representation of the translations in momentum space.

A closed loop
in momentum space, starting
and ending at ${\bf p}$, can then be obtained by using a sequence of
infinitesimal translations
 $U(\delta {\bf a})\ket{{\bf p}} = \ket{{\bf p}+ \delta{\bf a}}$ such that
$\delta {\bf a}$ is orthogonal to argument of the wave-function on which it
acts (this defines the adiabatic transport of the system). The momentum
vector ${\bf p}$ then traces out a closed curve on the constant energy
surface $S^2$ in momentum space.  The total phase of these translations then
gives a phase $\gamma $ which is the $\lambda $ times the solid angle of the
closed curve the momentum vector traces out on the constant energy surface.
This phase does not depend on Plancks constant.
This is precisely the Berry phase for the photon with a given helicity
$\lambda$. In the original experiment by Tomita and Chiao 
\cite{chiao} one considers a beam of
linearly polarized photons (a single-photon experiment is considered
in Ref.\cite{kwiat&chiao&91}). The same line of arguments above but making
use Eq.(\ref{eq:gen}) 
instead of  Eq.(\ref{eqn4}) leads to the desired change of
polarization as the photon moves 
along the optical fibre. 

A somewhat alternative derivation of the Berry phase
for photons is based 
on observation that the covariantly conserved helicity operator
$\Sigma$ can be 
interpreted as a generalized ``magnetic charge''. Let $\Gamma$ denote
a closed 
path in momentum space parameterized by $\sigma \in [0,1]$ such 
that ${\bf p}(\sigma =0) = {\bf p}(\sigma = 1) = {\bf p}_{0}$ is
fixed. 
The parallel transport of a one-particle state $\phi _{\alpha}({\bf p})$ 
along the path $\Gamma$ is then determined by a path-ordered
exponential, i.e.
\begin{equation}
\phi _{\alpha}({\bf p}_{0}) \longrightarrow  
\left[ P\exp \left(i\int _{\Gamma}A_{k}({\bf p}(\sigma ))
\frac{dp_{k}(\sigma )}
{d\sigma}d\sigma \right) \right]_{\alpha \beta}\phi _{\beta} ({\bf p}_{0})~~~~,
\end{equation}
where $A_{k}({\bf p}(\sigma))  \equiv {\cal A}_{k}^{a}({\bf p}(\sigma ))S_{a}$.
By making use of a non-Abelian version of Stokes theorem
\cite{goddard} one can then show that
\begin{equation}
\label{eq:helicity}
 P\exp \left(i\int _{\Gamma}A_{k}({\bf p}(\sigma ))\frac{dp_{k}(\sigma
 )}{d\sigma}d\sigma \right)
 = \exp\left(i\Sigma \Omega [\Gamma ]\right)~~~~,
\end{equation}
where $\Omega [\Gamma ]$ is the solid angle subtended by the path
$\Gamma$ on 
the two-sphere $S^{2}$. This result leads again to the desired change
of linear 
polarization as the photon moves along the path described by
$\Gamma$. Eq.(\ref{eq:helicity}) also directly leads to
helicity quantization, as alluded to already in 
Section \ref{sec-positionoperators}, 
 by considering a sequence of loops  which
converges to a point and at the same time has covered a solid angle of $4\pi$.  
This derivation does not require that $|{\bf p}(\sigma )|$ is constant along the path.

In the experiment of 
Ref.\cite{chiao}
the photon flux is large. In
order to strictly 
apply our results under such conditions one can consider a second
quantized version 
of the theory we have presented following e.g. the discussion of
Amrein \cite{jpa}. 
By making use of coherent states of the electro-magnetic field in a
standard and 
straightforward manner (see e.g. Ref.\cite{klauderskagerstam85}) one then
realize that our  considerations survive. This is so since the coherent states are
parameterized in terms of the one-particle states. By construction the
coherent states 
then inherits the transformation properties of the one-particle states
discussed above. It is, of course, of vital importance that the Berry
phase of single-photon states has experimentally been observed 
\cite{kwiat&chiao&91}.
%
\subsection{Localization of Single-Photon States}
\begin{figure}[htb]
\unitlength=1mm
\begin{picture}(140,100)(-40,-80)
\includegraphics{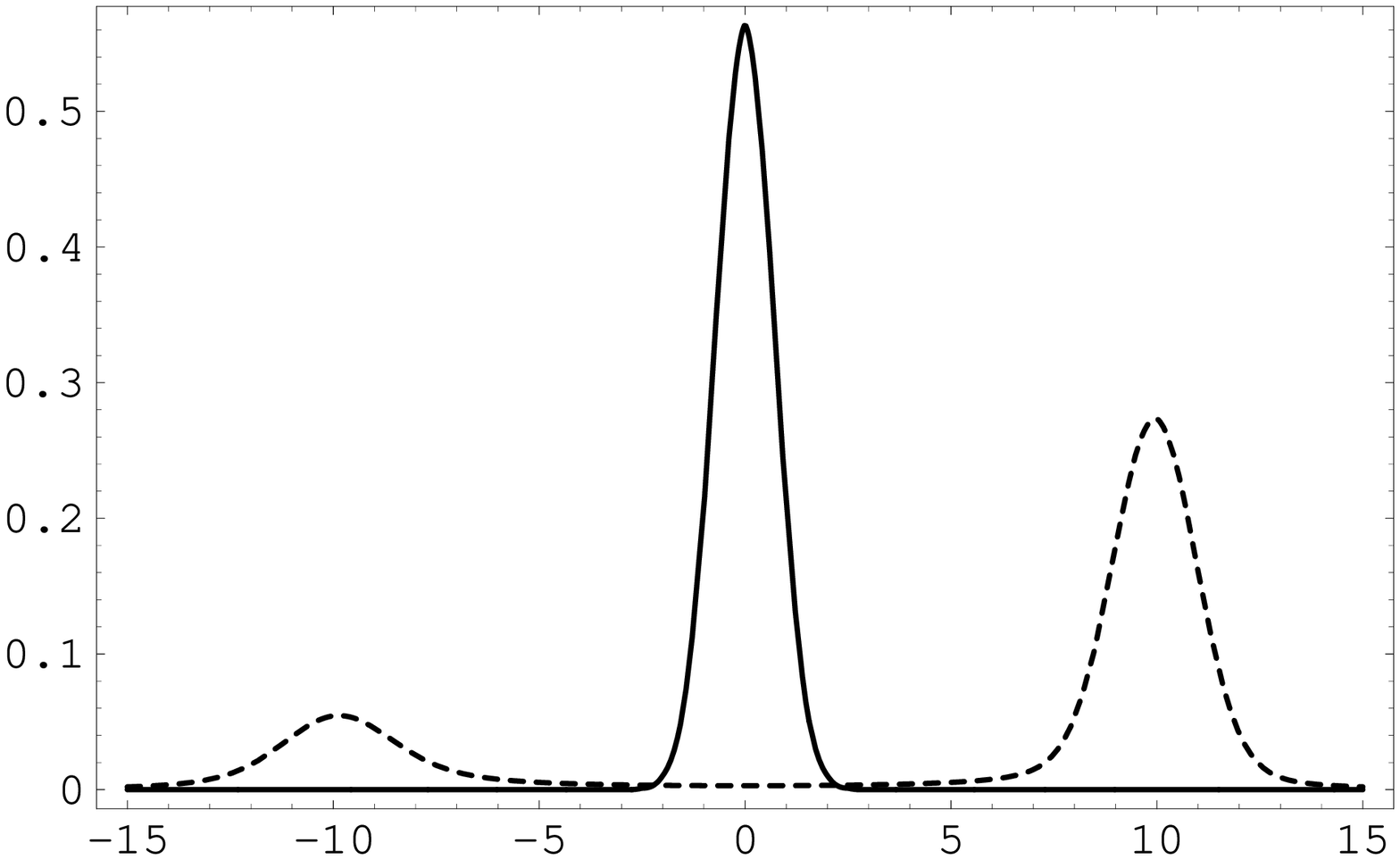}
\put(60,-5){\small $k_{0}a = 0.5$}
\put(-43,-29){\large $P(x)$}
\put(38,-70){$x/a$}
\end{picture}
\figcap{Probability density-distribution $P(x) = |\psi(x,t)|^2$ for a
one-dimensional Gaussian
  single-photon wave-packet with $k_{0}a=0.5$ at $t=0$ (solid curve)
  and at $t/a = 10$ (dashed curve).}
\label{LocalizedPhoton}
\end{figure}

In this section we will see that the fact that a one-photon state has
{\it positive} energy, generically makes a localized one-photon wave-packet
de-localized
in space in the course of its time-evolution. 
We will, for reasons of simplicity, restrict ourselves to a one-dimensional motion,
i.e. we have assume that the transverse dimensions of the propagating
 localized
one-photon
state  are much large than the longitudinal scale. We will also
neglect the effect of photon polarization. Details of a more
general treatment can be found in Ref.\cite{skagerstam99a}.
In one dimension
we have seen above that the conventional notion of a position operator
makes sense for a single photon. We can therefore consider
wave-packets not only in momentum space but also in the longitudinal
co-ordinate space in a conventional quantum-mechanical
 manner. One can easily address the same
issue in terms of photon-detection theory but in the end no essential
differences will emerge. In the Sch\"{o}dinger picture we are then
considering the following initial value problem ($c = \hbar = 1$)
\bea
\label{eq:photoneqm}
i\frac{\partial \psi(x,t)}{\partial t} &=& \sqrt{- \frac{d^2}{dx^2}}
\psi(x,t)~~~,\nonumber \\
\psi(x,0)&=& \exp(-x^2/2a^2 )\exp(ik_{0}x)~~~,
\eea
which describes the unitary time-evolution of a single-photon wave-packet
localized within the distance $a$ and with mean-momentum $<p>=k_{0}$.  The
{\it non-local} pseudo-differential operator $\sqrt{- d^2 / dx^2 }$ is 
defined in terms of Fourier-transform
techniques, i.e.
\be
\sqrt{- \frac{d^2}{dx^2}}\psi(x) =
\int_{-\infty}^{\infty}dyK(x-y)\psi(y) ~~~,
\ee
where the kernel $K(x)$ is given by
\be
K(x) = \frac{1}{2\pi}\int_{-\infty}^{\infty}dk|k|\exp(ikx)~~~.
\ee
The co-ordinate wave function at any finite time can now easily be
written down and the probability density is shown in
Figure \ref{LocalizedPhoton} in the case of a non-zero average momentum
of the photon. The form of the soliton-like peaks is preserved for
sufficiently large times. In the limit of $ak_0 =0$ one gets two
soliton-like 
identical peaks
propagating in opposite directions. The structure of these peaks are
actually very similar to the directed localized energy pulses in 
Maxwells theory \cite{ziolkowski&89} or to the pulse splitting processes in non-linear
dispersive media  (see e.g. Ref.\cite{ranka&96}) but the physics 
is, of course, completely different.

Since the wave-equation Eq.(\ref{eq:photoneqm}) leads to the
second-order wave-equation in one-dimension the physics so obtained
can, of course, be described in terms of solutions to the
one-dimensional d'Alembert wave-equation of Maxwells theory of electromagnetism.
The quantum-mechanical wave-function above in momentum space is then simply
 used to parameterize a coherent
state. The average of a  second-quantized 
electro-magnetic free field operator in such a coherent state
will then be a solution of this
wave-equation. The solution to the d'Alembertian wave-equation can
then be written in terms of the general d'Alembertian formula, i.e.
\be
\psi_{cl}(x,t) = \frac{1}{2}(\psi(x+t,0) + \psi(x-t,0)) +
\frac{1}{2i}\int_{x-t}^{x+t}dy\int_{-\infty}^{\infty}dzK(y-z)\psi(z,0)~~~,
\ee
where the last term corresponds to the initial value of the
time-derivative of the classical
electro-magnetic field. The fact this term is non-locally connected to
the initial value of the classical electro-magnetic field is perhaps somewhat
unusual. By construction $ \psi_{cl}(x,t)= \psi(x,t)$. 
The physical interpretation of the two functions $\psi(x,t)$ and
$\psi_{cl}(x,t)$ are, of course, very different. In the quantum-mechanical case the
detection of the photon destroys the coherence properties of the
wave-packet $\psi(x,t)$ entirely. In the classical case the detection of a single photon
can still preserve the coherence properties of the classical field $\psi_{cl}(x,t)$
since there are infinitely many photons present in the corresponding
coherent state.
\subsection{Various Comments}
\seqnoll
%

In the analysis of Wightman, corresponding to commuting position 
variables, the natural mathematical tool turned out to be systems of
imprimitivity for the representations of the three-dimensional Euclidean
group. In the case of non-commuting position operators we have also seen
that notions from differential geometry are important. It is interesting
to see that such a broad range of mathematical methods enters into the
study of the notion of localizability of physical systems.

 We have, in particular, argued that
Abelian as well as non-Abelian magnetic monopole field configurations
reveal themselves 
in a description of localizability of
massless spinning particles.
  Concerning the physical existence of magnetic monopoles
 Dirac remarked in 1981 \cite{craige} that ``{\it I am inclined now
to believe that monopoles do not exist. So many years have gone by without
any encouragement from the experimental side}''. The
``monopoles'' we are considering are, however, only mathematical objects
in the momentum space of the massless particles. Their existence, we have
argued, is then only indirectly revealed to us by the properties of e.g.
the photons moving along optical fibres.

Localized states of massless particles will necessarily develop
non-exponential tails in space as a consequence of the Hegerfeldts theorem
\cite{heger2}. Various  number operators representing the number of
massless, spinning particles localized in a finite volume $V$ at time $t$
has been discussed in the literature. The non-commuting position
observables we have discussed for photons correspond to the point-like
limit of the weak localizability of Jauch, Piron and Amrein \cite{jpa}.
This is so since our construction, as we have seen in 
Section~\ref{sec-positionoperators},
corresponds to an explicit enforcement of the transversality condition of
the one-particle wave-functions.

In a finite volume, photon number operators appropriate for weak
localization \cite{jpa} do not agree with the photon number operator
introduced by Mandel \cite{mandel1} for sufficiently small wavelengths as
compared to the linear dimension of the localization volume. It would be
interesting to see if there are measurable differences. A necessary
ingredient in answering such a question would be the experimental
realization of a localized one-photon state. It is interesting to notice
that such states can be generated in the laboratory 
\cite{mandel86}-\cite{strekalov&98}.

As a final remark of this first set of lectures we recall a statement of
Wightman which, to a large extent, still is true \cite{wightman}: ``{\it Whether, in
  fact, the position of such particle is observable in the sense of
  quantum theory is, of course, a much deeper problem that probably
  can only be be decided within the context of a specific consequent
  dynamical theory of particles. All investigations
  of localizability for relativistic particles up to now, including
  the present one, must be regarded as preliminary from this point of
  view:
They construct position observables consistent with a given
transformation law. It remains to construct complete dynamical
theories consistent with a given transformation law and then to investigate
whether the position observables are indeed observable with the
apparatus that the dynamical theories themselves predict~}''. This is,
indeed, an ambitious programme to which we have not added very much in
these lectures.
%
%
%

\newpage
\section{Resonant Cavities  and the Micromaser System}
\seqnoll
\begin{flushright}
``{\sl The interaction of a single dipole with a monochromatic radiation\\
 field presents an important problem in electrodynamics. It is an\\
unrealistic problem in the sense that experiments are not
done\\
with single atoms or single-mode fields.}"\\
L. Allen and J.H. Eberly
\end{flushright}
The  highly  idealized  physical  system  of  a  single  two-level  atom  in  a
super-conducting   cavity,   interacting   with    a    quantized    single-mode
electro-magnetic field, has  been  experimentally  realized  in  the  micromaser
\cite{Goy83}--\cite{Walther88} and microlaser systems \cite{An94}. It
is interesting to consider this remarkable experimental development in
view of the quotation above.
Details  and a limited set of
references  to  the   literature   can   be   found   in   e.g.   the   reviews
\cite{Haroche92}--\cite{Milonni91}. In the absence of dissipation (and  in  the
rotating wave approximation) the two-level atom and its  interaction  with  the
radiation field is well  described  by  the
Jaynes--Cummings  (JC)  Hamiltonian
\cite{Jaynes63}. Since  this  model  is  exactly  solvable  it  has  played  an
important role in the development  of  modern  quantum  optics  (for recent
accounts see e.g. Refs.~\cite{Barnett86,Milonni91}). The  JC model
predicts  non-classical
phenomena, such as revivals of the initial excited   state   of   the   atom
\cite{Meystre74}--\cite{Arroyo90}, experimental signs  of   which
have been reported for the micromaser system \cite{Rempe87}.

Correlation  phenomena  are  important  ingredients  in  the  experimental  and
theoretical investigation of physical systems. Intensity correlations of  light
was e.g. used by Hanbury--Brown
and Twiss \cite{hbt57} as a tool to determine  the
angular diameter of distant stars. The quantum theory of intensity correlations
of light was later developed by Glauber \cite{glauber63}.  These  methods  have
a wide range of physical applications including investigation of  the
space-time
evolution    of    high-energy     particle     and     nuclei     interactions
\cite{boaletal90,Skagerstam94}. In the case of  the  micromaser it has
recently been
suggested \cite{Elmforsetal95,Elmforsetal95b} 
that correlation measurements on  atoms  leaving
the micromaser system can be used to infer properties of the quantum state  of
the
radiation field in the cavity.

We will now discuss in great  detail  the  role  of  long-time
correlations in the outgoing atomic beam and  their  relation  to  the  various
phases of the micromaser system. Fluctuations in the number  of  atoms  in  the
lower maser level for a fixed transit time $\tau $ is known to  be  related  to
the photon-number  statistics  \cite{Filipowicz86}--\cite{PaulRichter91}.  The
experimental  results  of  \cite{Rempe90}  are  clearly  consistent  with   the
appearance of non-classical, sub-Poissonian statistics of the radiation  field,
and exhibit the intricate correlation between the atomic beam and  the  quantum
state of the cavity. Related work on characteristic statistical  properties  of
the beam of  atoms  emerging  from  the  micromaser  cavity  may  be  found  in
Ref.~\cite{Brigeletal94,WagnerSW94,Herzog94}.

\begin{figure}[htb]
\unitlength=1mm
\begin{picture}(140,100)(0,0)
\includegraphics{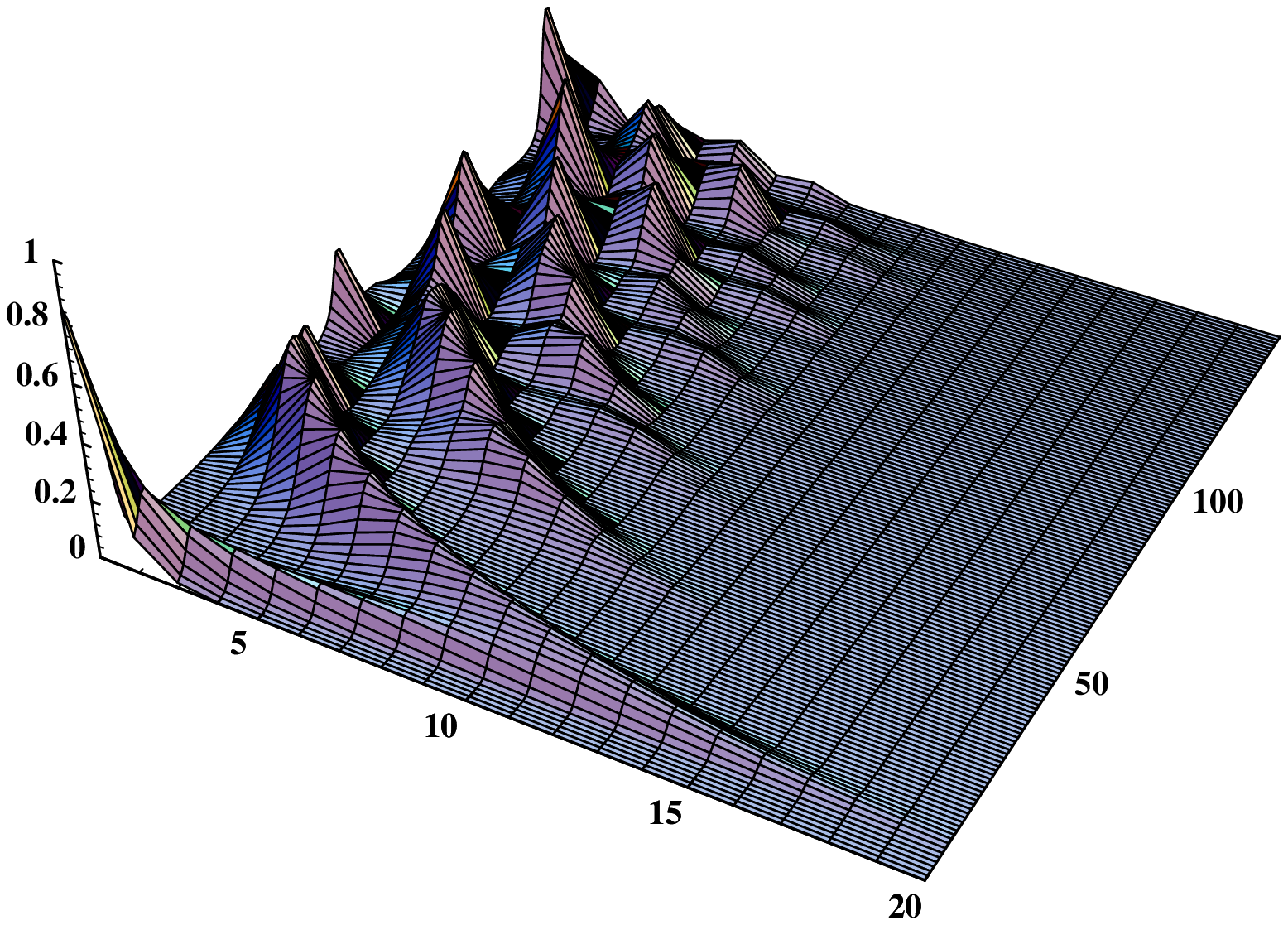}
\end{picture}
\put(-100,20){\small $n$}
\put(-5,35){\small $\tau~[\mu s]$}
\put(-155,65){\small $p_n(\tau)$}
\figcap{The rugged  landscape  of  the  photon  distribution  $p_n(\tau)$  in
\eq{Equilibrium} at equilibrium for the micromaser as a function of the  number
of photons in the cavity, $n$,  and  the  atomic  time-of-flight  $\tau$.  The
parameters correspond to a super-conducting niobium  maser,  cooled  down  to  a
temperature of ~$T=0.5\,$K, with an average thermal photon occupation number of
$n_b=0.15$, at the maser  frequency  of  $21.5\,$GHz.  The  single-photon  Rabi
frequency $\Omega $ is $44\,$kHz, the photon lifetime in the cavity is $T_{\rm
cav}=0.2~{\sl s}$, and the atomic beam intensity is $R=50/{\sl s}$. }
\label{FigLandscape}
\end{figure}

\newpage

\section{Basic Micromaser Theory}
\label{basic}
\seqnoll
\begin{flushright}
``{\sl It is the enormous progress in constructing super-conducting\\
cavities with high quality factors together with  the laser\\
preparation of highly exited atoms - Rydberg atoms - that \\have made
the realization of such a one-atom maser possible.}"\\
H. Walther
\end{flushright}
In the micromaser a beam of excited atoms is sent through  a  cavity  and  each
atom interacts with the cavity during a well-defined transit time  $\tau$.  The
theory     of     the      micromaser      has      been      developed      in
\cite{Filipowicz86,Filipowicz86a}, and in this section we  briefly  review  the
standard theory, generally following the notation of that paper. We assume that
excited atoms are injected into the cavity at an average rate $R$ and that  the
typical decay rate for photons in the cavity is $\gamma$. The number  of  atoms
passing the cavity  in  a  single  decay  time  $N=R/\gamma$  is  an  important
dimensionless parameter, effectively controlling the average number of  photons
stored in a high-quality cavity. We shall assume that the time $\tau$
during which the  atom
interacts with the cavity is so small that effectively only one atom  is  found
in the cavity at any  time,  \ie\  $R\tau\ll1$.  A  further  simplification  is
introduced by assuming that the cavity decay time  $1/\gamma$  is  much  longer
than the interaction time, \ie\ $\gamma\tau\ll1$, so that damping  effects  may
be ignored while the atom passes through the  cavity.  This  point  is  further
elucidated  in  Appendix~\ref{AppDamping}.  In  the   typical   experiment   of
Ref.~\cite{Rempe90}   these   quantities   are   given   the   values   $N=10$,
$R\tau=0.0025$ and $\gamma\tau=0.00025$.


\subsection{The Jaynes--Cummings Model}

The electro-magnetic interaction between a two-level atom with level  separation
$\omega_0$ and a single mode with frequency $\omega$ of the radiation field  in
a cavity is described,
in the rotating wave approximation, by the Jaynes--Cummings
(JC) Hamiltonian \cite{Jaynes63}
\begin{formula}{JCH}
H=\omega a^{*} a+\frac12\omega_0\sigma_z+g(a\sigma_++a^*\sigma_-)~~,
\end{formula}
\noi where the coupling constant $g$  is  proportional  to  the  dipole  matrix
element of the atomic transition\footnote{This coupling constant turns  out  to
be identical to the single photon Rabi frequency  for  the  case  of  vanishing
detuning, \ie\ $g=\Omega$. There is actually some confusion in  the  literature
about what is called the Rabi frequency \cite{Knight80}.  With  our
definition,
the  energy  separation  between   the   shifted   states   at   resonance   is
$2\Omega$.}. We use the Pauli matrices to describe the  two-level  atom  and
the   notation   $\sigma_\pm=(\sigma_x\pm   i\sigma_y)/2$.   The
second quantized single mode electro-magnetic field is described in a
conventional manner (see e.g. Ref.\cite{fermi&32}) by means of an
annihilation (creation) operator $a$ ($a^*$), where we have suppressed
the mode quantum numbers. For   $g=0$    the
atom-plus-field  states  $|n,s\>$  are  characterized  by  the  quantum  number
$n=0,1,\ldots$ of  the  oscillator  and  $s=\pm$  for  the  atomic  levels
(with $-$ denoting the ground state) with energies $
E_{n,-}= \omega n - \omega_0 /2$ and $E_{n,+}= \omega n +\omega_0 /2$.  At
resonance $\omega=\omega_0$ the levels $|n-1,+\>$ and $|n,-\>$  are  degenerate
for $n\ge1$ (excepting the ground state $n=0$), but this degeneracy  is  lifted
by  the  interaction.  For  arbitrary  coupling  $g$  and  detuning   parameter
$\Delta\omega= \omega_0-\omega$ the system reduces to a  $2\times2$  eigenvalue
problem,
which may be trivially solved. The result is that two  new  levels,
$|n,1\>$ and $|n,2\>$,
   are
formed as superpositions of the previously degenerate ones at
zero detuning according to
\begin{formulas}{JStates}
|n,1\> = \cos(\theta_n)|n+1,-\> +\sin(\theta_n)|n,+\>~~,\\
|n,2\> = -\sin(\theta_n)|n+1,-\> +\cos(\theta_n)|n,+\>~~,
\end{formulas}
\noindent with energies 
\begin{formulas}{Jenergies}
E_{n1} = \omega(n+1/2) + \sqrt{\Delta\omega^2/4+g^2(n+1)}~~,\\
E_{n2} = \omega(n+1/2) - \sqrt{\Delta\omega^2/4+g^2(n+1)}~~,
\end{formulas}
respectively. 
The ground-state of
the coupled system is given by $|0,-\>$ with 
energy $E_0 = -\omega_0 /2$. Here the mixing angle $\theta_n$ is given
by
\begin{formula}{Jmixing}
\tan(\theta_n) = \frac{2g\sqrt{n+1}}{\Delta\omega  +
  \sqrt{\Delta\omega^2 + 4g^2(n+1)}}~~.
\end{formula}
\noindent The interaction therefore leads to a separation in
energy  $\Delta E_n = \sqrt{\Delta\omega^2+4g^2(n+1)}$ for quantum number
$n$.  The  system  performs
Rabi oscillations with the corresponding
 frequency between the original, unperturbed  states
with transition probabilities \cite{Jaynes63,stenholm&73}
\begin{formulas}{TransEl}
|\<n,-|e^{-iH \tau}|n,-\>|^2&=&1-q_n(\tau)~~,\\
|\<n-1,+|e^{-iH \tau}|n,-\>|^2&=&q_n(\tau)~~,\\
|\<n,+|e^{-iH \tau}|n,+\>|^2&=&1-q_{n+1}(\tau)~~,\\
|\<n+1,-|e^{-iH \tau}|n,+\>|^2&=&q_{n+1}(\tau)~~.\\
\end{formulas}

\noi These are all expressed in terms of

\begin{formula}{}
q_n(\tau)=\frac{g^2n}{g^2n+\frac14\Delta\omega^2}
\sin^2\(\tau\sqrt{g^2n+{\textstyle\frac14}\Delta\omega^2}\) ~~.
\end{formula}

\noi Notice that for $\Delta\omega=0$ we have
$q_n=\sin^2 (g\tau\sqrt n )$. Even though
most of the following discussion will be limited to this case, the
equations
given below will often be valid in general.

Denoting the probability of finding $n$ photons in the cavity by $p_n$
we  find
a general expression for
the conditional probability that an excited atom decays to the ground state  in
the cavity to be

\begin{formula}{}
\label{eq:incomsum}
\P(-)=\<q_{n+1}\>=\sum_{n=0}^\infty q_{n+1} p_n ~~.
\end{formula}

\begin{figure}[p]
\unitlength=1mm
\begin{picture}(140,90)(-10,5)
\includegraphics{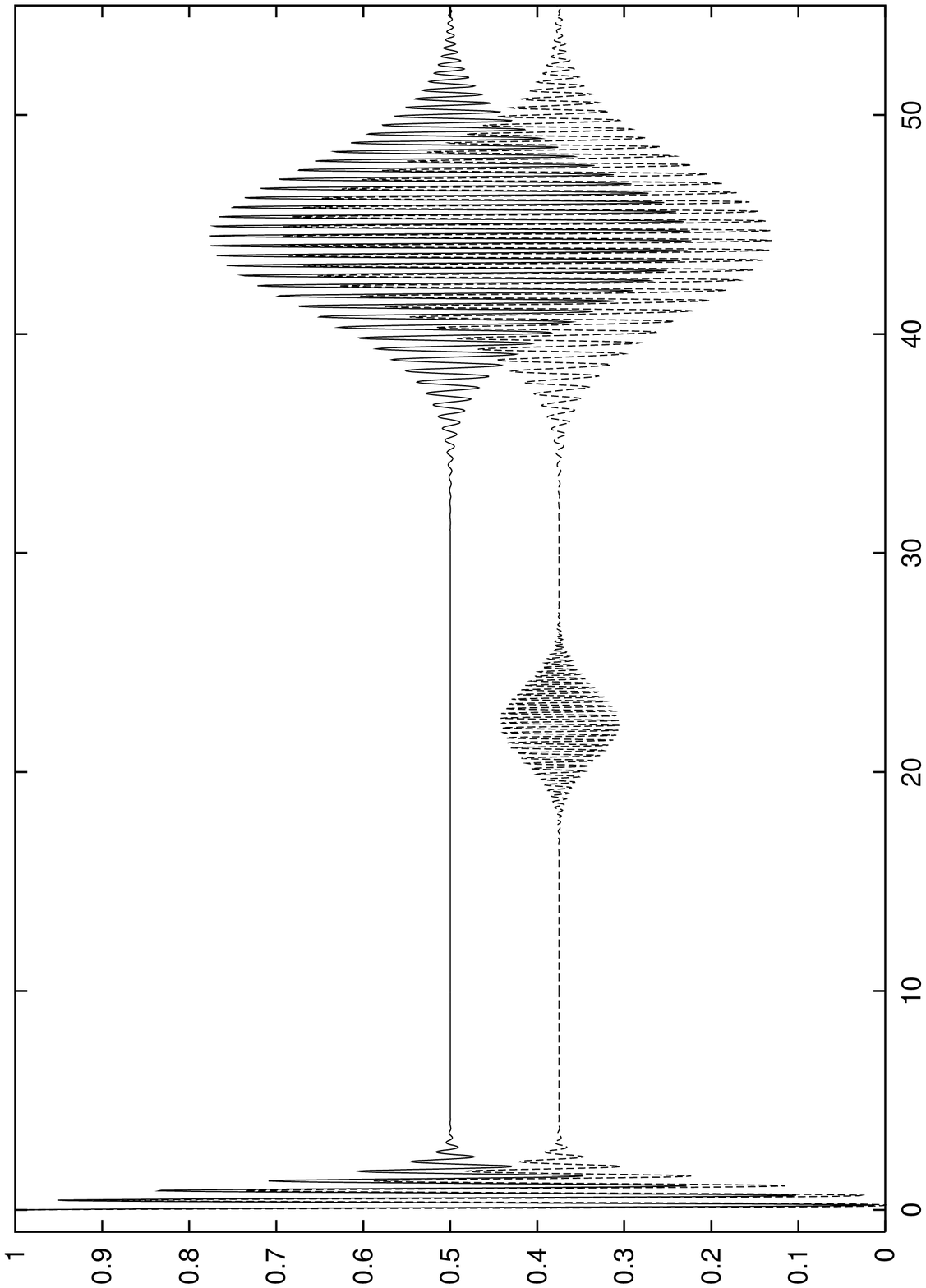}
\put(20,80){\small\parbox{40mm}{$\Delta\omega = 0\\z = 7\\
\tau = \tau_A = \tau_B$}}
\put(-18,53){$\P(+)$}
\put(-18,40){$\P(+,+)$}
\end{picture}

\begin{picture}(140,90)(-10,5)
\includegraphics{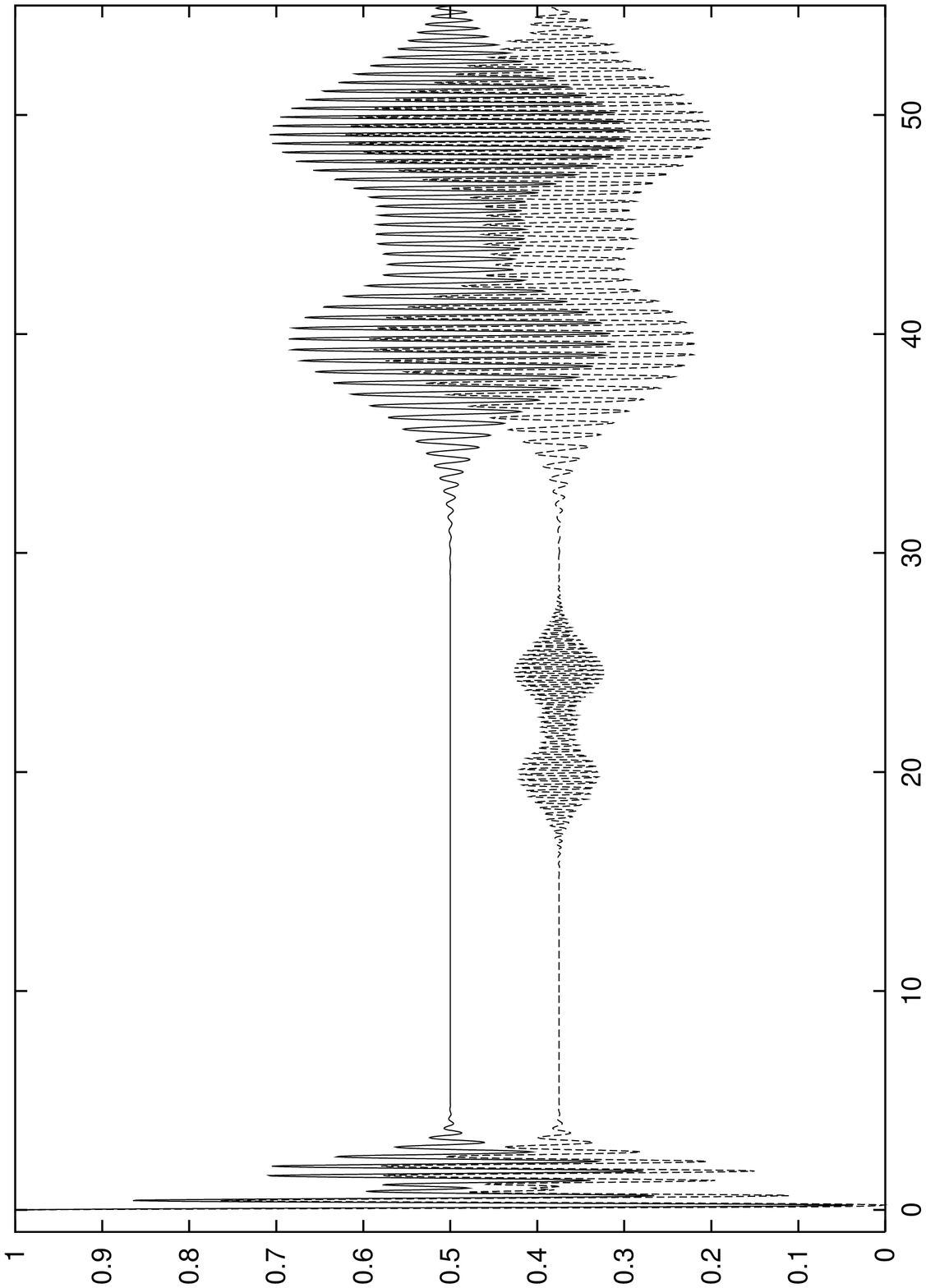}
\put(20,80){\small\parbox{40mm}{$\Delta\omega = 0\\z = 7\\
\tau = \tau_A = \tau_B$}}
\put(-18,53){$\P(+)$}
\put(-18,40){$\P(+,+)$}
\put(60,8){$g\tau$}
\end{picture}
\figcap{The upper figure shows the revival probabilities
 $\P(+)$ and $\P(+,+)$ for a coherent state $\rv{z}$ with a mean number
 $|z|^2 =49$ of photons as a
 function of the atomic passage time $g\tau $.
The lower figure shows the same revival probabilities for a displaced coherent
  state $\rv{z,1}$ with a mean value of $|z|^2 +1  =50$  photons.}
\label{JC1revivals}
\end{figure}

\begin{figure}[p]
\unitlength=1mm
\begin{picture}(140,90)(-10,5)
\includegraphics{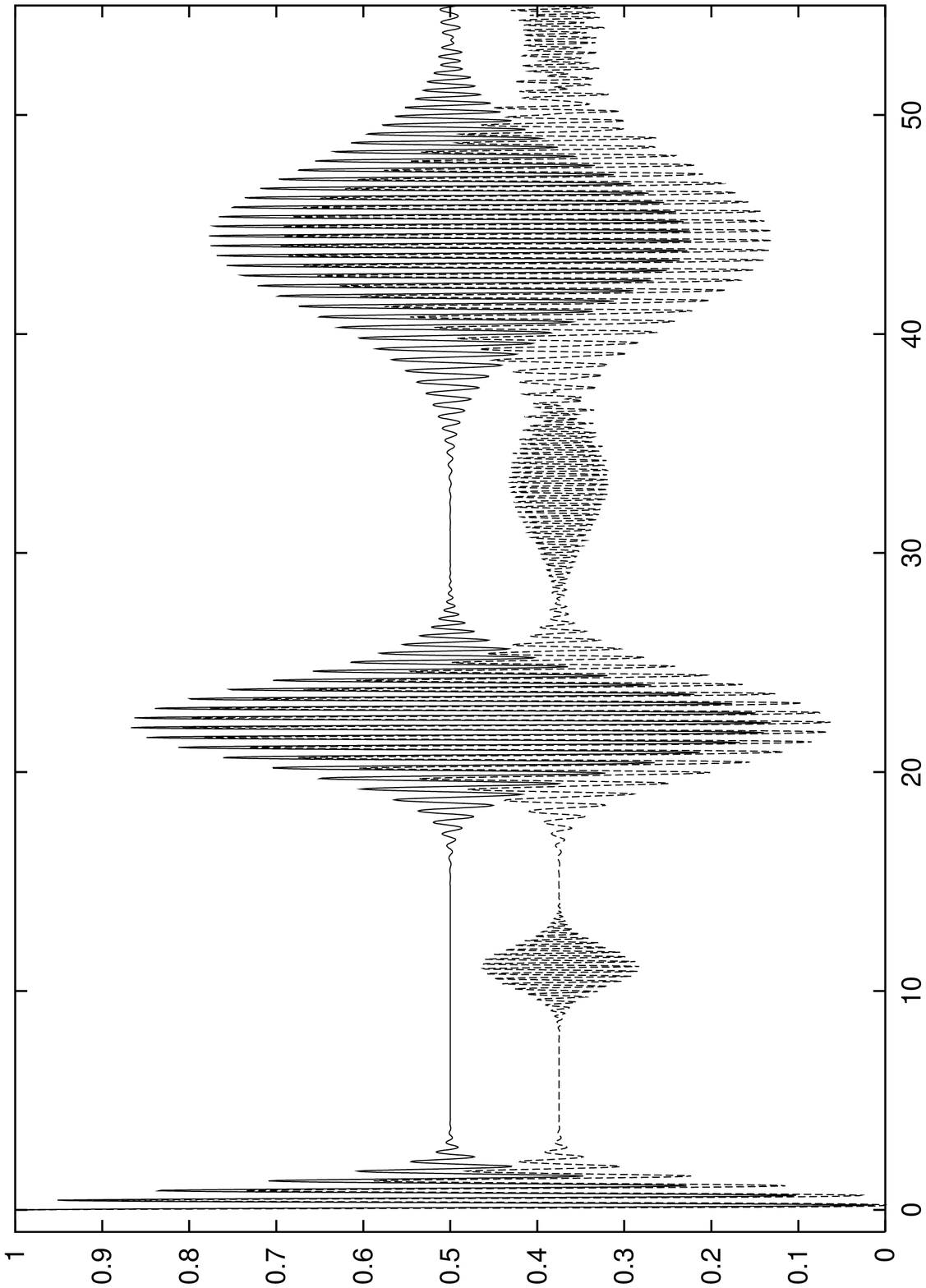}
\put(20,80){\small\parbox{40mm}{$\Delta\omega = 0\\z = 7\\
\tau = \tau_A = \tau_B$}}
\put(-18,53){$\P(+)$}
\put(-18,43){$\P(+,+)$}
\end{picture}
\begin{picture}(140,90)(-10,5)
\includegraphics{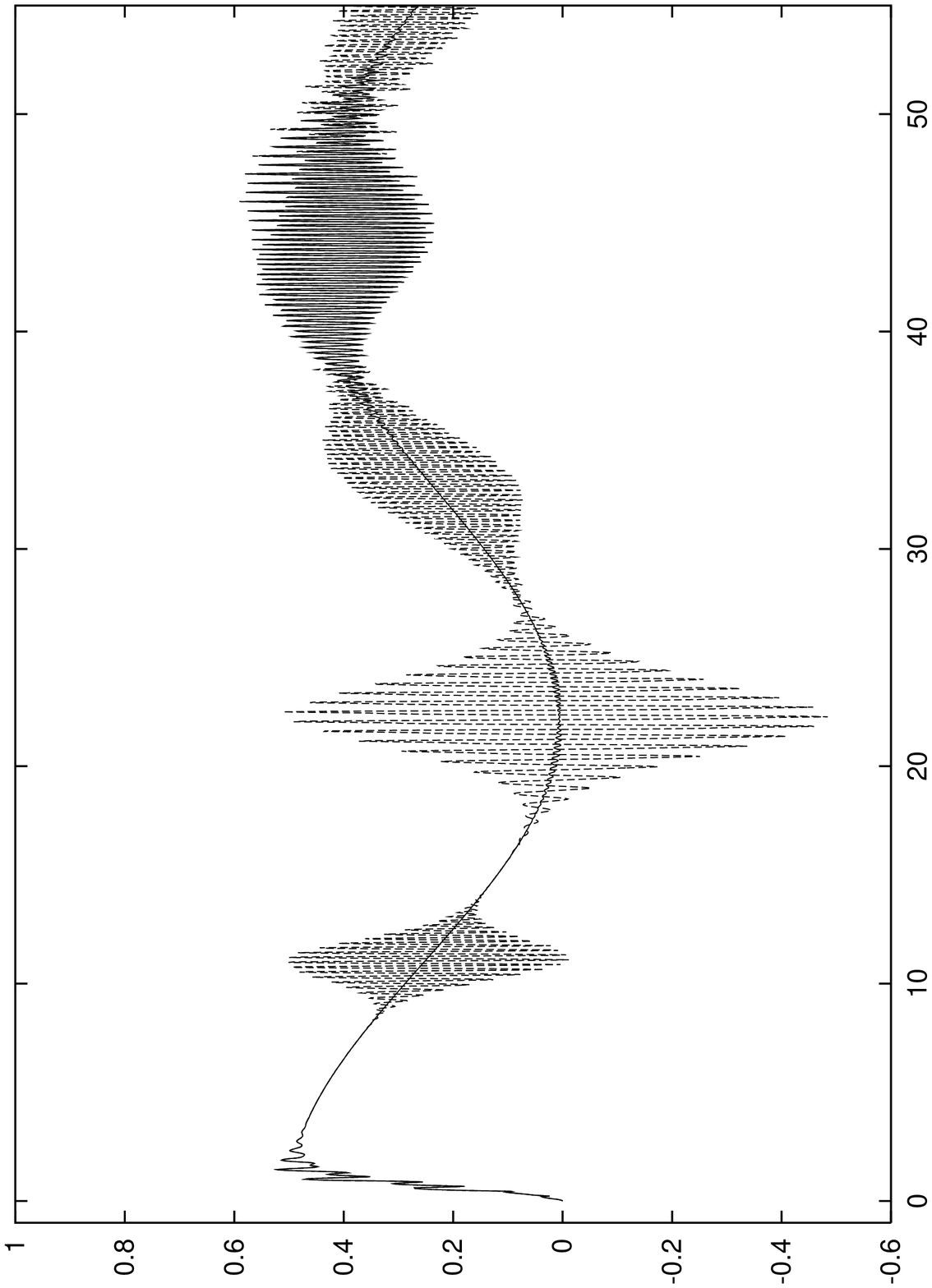}
\put(20,80){\small\parbox{40mm}{$\Delta\omega = 0\\z = 7\\
\tau = \tau_A = \tau_B$}}
\put(60,8){$g\tau$}
\put(-8,55){$\eta $}
\end{picture}
\figcap{The upper figure shows the revival probabilities
 $\P(+)$ and $\P(+,+)$ for a normalized Schr\"{o}dinger cat 
state as given by Eq.(\ref{eq:catstate}) with $z=7$
as a function of the atomic passage time $g\tau$.
 The lower figure shows the correlation coefficient $\eta $ for 
 a coherent state with $z=7$ (solid curve) and for the
  the same Schr\"{o}dinger cat state (dashed curve) as in the upper figure.
}
\label{JC2revivals}
\end{figure}

\noi From this equation we find $\P(+) = 1- \P(-)$, i.e. the
conditional probability that the atom remains excited. 
In a similar manner we may consider a situation when two atoms,
$A$ and $B$, have passed through the cavity with transit times
$\tau_A$ and $\tau_B$. Let  $\P(s_{1},s_{2})$ be the probability that the
second  atom $B$ is 
in the state $s_2 = \pm$
 {\it if} the first atom has been found in the state $s_1 = \pm$. Such
 expressions then contain information further information about the
 entanglement between the atoms and the state of the radiation field
 in the cavity. If damping of the resonant cavity is not taken into
 account than  $\P(+,-)$ and  $\P(-,+)$ are in general different.
It is such sums like in Eq.(\ref{eq:incomsum})
over the incommensurable frequencies $g\sqrt n$  that  is
the cause of some of the most important properties of the micromaser, such as
quantum collapse and revivals, to be discussed again in Section \ref{ss:prerev} 
(see e.g. Refs.\cite{Meystre74}-\cite{Arroyo90},
\cite{Averbukh89a}--\cite{Fleischhauer93}). If we are at resonance,
i.e. $\Delta \omega = 0$,  we in particular obtain the expressions
\be
\label{eq:singlerevivalsum}
\P(+)= \sum_{n=0}^\infty  p_n \cos^2(g\tau\sqrt{n+1})~~~,
\ee
for $\tau = \tau_A$ or $\tau_B$, 
and
\bea
\label{eq:revivalsums}
\P(+,+)= \sum_{n=0}^\infty  p_n \cos^2(g\tau_A\sqrt{n+1})\cdot
\cos^2(g\tau_B\sqrt{n+1})~~~,
\nonumber \\
\P(+,-)= \sum_{n=0}^\infty  p_n \cos^2(g\tau_A\sqrt{n+1})\cdot
\sin^2(g\tau_B\sqrt{n+1})~~~,
\\
\P(-,+)= \sum_{n=0}^\infty  p_n \sin^2(g\tau_A\sqrt{n+1})\cdot 
\cos^2(g\tau_B\sqrt{n+2})~~~,\nonumber \\
\P(-,-)= \sum_{n=0}^\infty  p_n \sin^2(g\tau_A\sqrt{n+1})\cdot
\sin^2(g\tau_B\sqrt{n+2})~~~.\nonumber
\eea
It is clear that these expressions obey the general conditions
$\P(+,+)+\P(+,-) = \P(+)$ and $\P(-,+)+\P(-,-) = \P(-)$. As a measure
of the coherence due to the entanglement of the state of an atom and
the state of the cavities radiation field one may consider the
difference of conditional probabilities
\cite{haroche&etal&96,haroche&gimo}, 
i.e.
\bea
\eta = && \frac{\P(+,+)}{\P(+,+)+\P(+,-)}-
\frac{\P(-,+)}{\P(-,+)+\P(-,-)}
 \nonumber \\
= && \frac{\P(+,+)}{\P(+)}-
\frac{\P(-,+)}{1-\P(+)}~~~. 
\eea
These  effects of quantum-mechanical revivals  are  most
easily displayed in the case that the cavity field is coherent  with  Poisson
distribution
\begin{equation}
\label{Poisson}
p_n=\frac{\<n\>^n}{n!}e^{-\<n\>} ~~.
\end{equation}
\noindent In Figure \ref{JC1revivals} we exhibit the well-known revivals in the
probability $\P(+)$ for a
coherent state. In the same figure we also notice the existence of
{\it prerevivals} \cite{Elmforsetal95,Elmforsetal95b} expressed
  in terms $\P(+,+)$. In Figure \ref{JC1revivals} we also consider the same
  probabilities for the semi-coherent state considered in 
Figure \ref{FigSemiCoh}. The presence of one additional photon
clearly manifests itself in the revival and prerevival structures.
For the purpose of illustrating the revival phenomena we also consider
a special from of Schr\"{o}dinger cat states (for an excellent review
see e.g. Ref.\cite{gerry&knight&97}) which is a superposition of the
coherent states $\rv{z}$ and  $\rv{-z}$ for a real parameter $z$, i.e.
\be
\label{eq:catstate}
\rv{z}_{sc} = \frac{1}{(2+2\exp(-2|z|^2))^{1/2}} (\rv{z} +
\rv{-z})~~~.
\ee
In Figure \ref{JC2revivals} we exhibit revivals and prerevivals for such
 Schr\"{o}dinger cat state with $z=7$ ($\tau = \tau_A = \tau_B$). 
When compared to the revivals
 and prerevivals of a coherent state with the same value of $z$ as in 
Figure \ref{JC1revivals} one
 observes that  Schr\"{o}dinger cat state revivals  occur much
 earlier. It is possible to view these earlier revivals as due to 
a quantum-mechanical interference effect. It is known
 \cite{gea&90} that the Jaynes-Cummings model has the
 property that with a coherent state of the radiation field
 one reaches a {\it pure} atomic state at time corresponding to
 approximatively one half of the
 first revival time {\it independent} of the initial atomic state. The states
$\rv{z}$ and $\rv{-z}$ in the construction of the Schr\"{o}dinger cat
 state are approximatively orthogonal. These two states will then
 approximatively behave as
 independent system. Since they lead to the same intermediate 
pure atomic state mentioned above, quantum-mechanical
 interferences will  occur. It can be verified
 \cite{skagerstam&99b}
 that that this
 interference effect will survive moderate damping corresponding to
present experimental cavity conditions. In Figure \ref{JC2revivals} we also
 exhibit the  $\eta $ for a coherent state with $z=7$ (solid curve) and the
same Schr\"{o}dinger cat as above. The  Schr\"{o}dinger cat state
 interferences are clearly revealed. It can again be shown that moderate
 damping effects do not change the  qualitative features of 
this picture \cite{skagerstam&99b} .

\begin{figure}[p]
\unitlength=1mm
\begin{picture}(140,80)(0,0)
\includegraphics{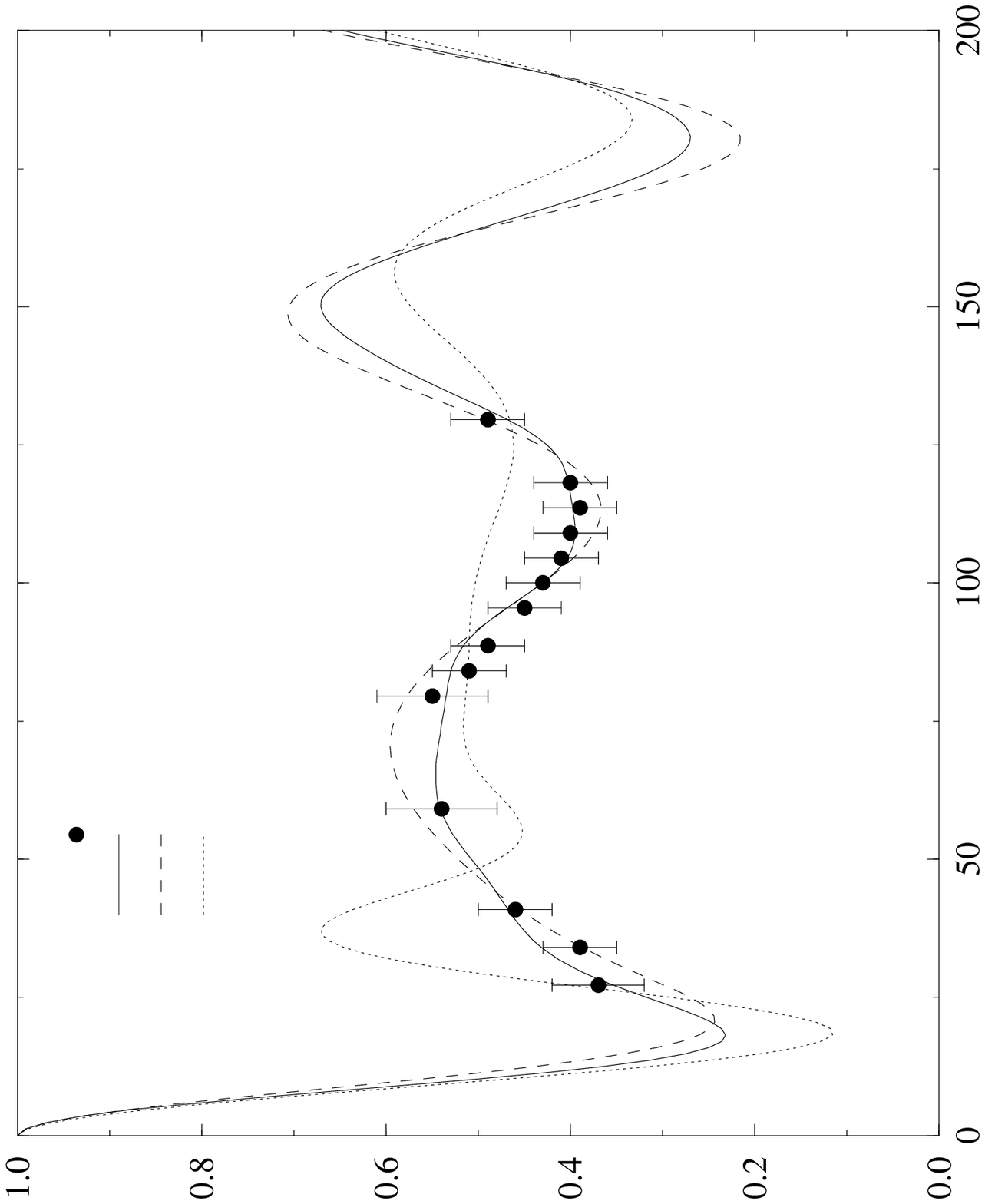}
\put(52,87){\small Experiment}
\put(52,83){\small Equilibrium: $n_b=2,~N=1$}
\put(52,79){\small Thermal: $n_b=2$}
\put(52,75){\small Poisson: $\<n\>=2.5$}
\put(70,0){\small$\tau~[\mu {\rm s}]$}
\put(0,50){\small$\P(+)$}
\put(60,23){\small$^{85}$Rb 63p$_{3/2}\leftrightarrow$ 61d$_{5/2}$}
\put(60,13){\small$R=500~{\rm s}^{-1}$,~$n_b=2$,~$\gamma=500~{\rm s}^{-1}$}
\end{picture}

\begin{picture}(140,100)(0,0)
\includegraphics{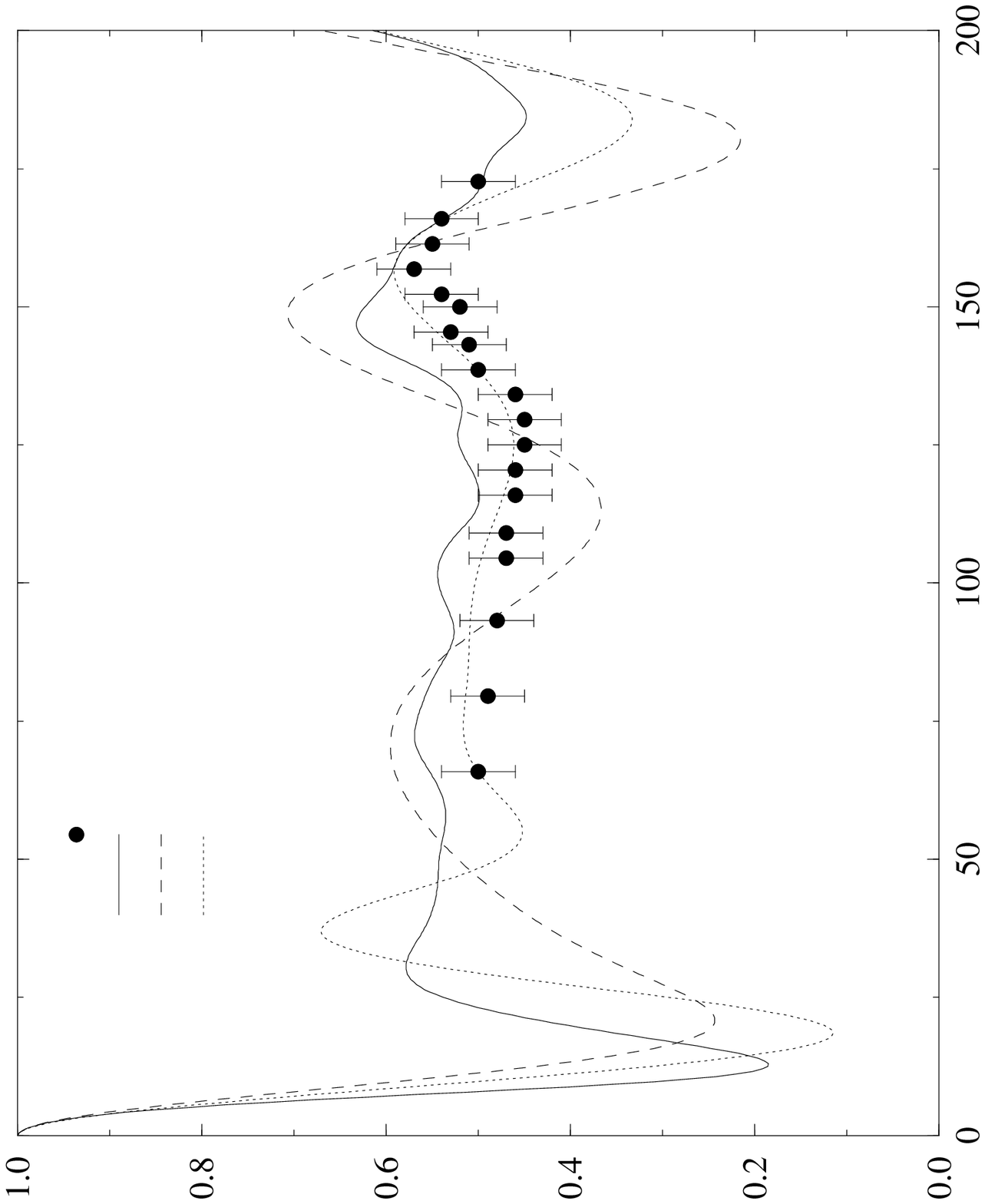}
\put(52,87){\small Experiment}
\put(52,83){\small Equilibrium: $n_b=2,~N=6$}
\put(52,79){\small Thermal: $n_b=2$}
\put(52,75){\small Poisson: $\<n\>=2.5$}
\put(70,0){\small$\tau~[\mu {\rm s}]$}
\put(0,50){\small$\P(+)$}
\put(60,23){\small$^{85}$Rb 63p$_{3/2}\leftrightarrow$ 61d$_{5/2}$}
\put(60,13){\small$R=3000~{\rm s}^{-1}$,~$n_b=2$,~$\gamma=500~{\rm s}^{-1}$}
\end{picture}
\figcap{Comparison of $\P(+)=1-\P(-)=1-\<q_{n+1}\>$ with experimental  data  of
Ref.~\cite{Rempe87}  for  various  probability  distributions.   The   Poisson
distribution   is   defined   in   Eq.~(\ref{Poisson}),    the    thermal    in
Eq.~(\ref{Thermal}),   and   the   micromaser   equilibrium   distribution   in
Eq.~(\ref{Equilibrium}). In  the  upper  figure  ($N=R/\gamma=1$)  the  thermal
distribution agrees well with the data and in the  lower  ($N=6$)  the  Poisson
distribution fits the data best. It is curious  that  the  data  systematically
seem to deviate from the micromaser equilibrium  distribution.
} \label{FigDataComparison}
\end{figure}

\noi In passing we notice that revival phenomena and the appearance of
Schr\"{o}dinger like cat states have  been studied and
observed in many other physical systems like in atomic systems 
\cite{Nauenberg}-\cite{Stroud&d}, in
ion-traps \cite{wineland,monroe} and recently also in the case of Bose-Einstein
condensates \cite{Pitaevskii}
(for a recent pedagogical account on revival phenomena see
e.g. Ref.\cite{bluhm}).

In the more realistic case, where the changes of the cavity field due to
the passing atoms is taken into account, a complicated statistical  state  of
the     cavity     arises     \cite{Filipowicz86},
\cite{Wright89}--\cite{Herzog95,BogarBH94}
\bort{\cite{Filipowicz86,Wright89,Guzman89,Herzog95}}
(see Figure \ref{FigLandscape}).
It is the  details  of  this  state  that  are
investigated in these lectures.


\subsection{Mixed States}
\label{mixed}

The above formalism is directly applicable when  the  atom  and  the  radiation
field are both in pure states initially. In general the  statistical  state  of
the system is described by an  initial  density  matrix  $\rho$,  which
evolves
according to the usual rule $\rho\to\rho(t)=\exp(-iHt)\rho\exp(iHt)$.
If we disregard, for
the moment, the decay of the cavity field due to interactions with the
environment, the evolution is governed by the JC Hamiltonian in \eq{JCH}. It is
natural to assume that the atom and the radiation field of the cavity initially
are completely uncorrelated so that the initial density matrix factories in  a
cavity part and a product of $k$ atoms as

\begin{formula}{rhofact}
    \rho=\rho_C\otimes\rho_{A_1}\otimes\rho_{A_2}\otimes \cdots \otimes
\rho_{A_k}~~.
\end{formula}

\noi When the first atom $A_1$ has passed through the cavity,  part  of  this
factorizability is destroyed by the interaction and the state has become

\begin{formula}{rhotau}
\rho(\tau)=\rho_{C,A_1}(\tau)   \otimes\rho_{A_2}\otimes \cdots \otimes
\rho_{A_k}~~.
\end{formula}

\noi   The   explicit   form   of   the   cavity-plus-atom   entangled    state
$\rho_{C,A_1}(\tau)$  is  analyzed  in  Appendix~\ref{AppDamping}.  After   the
interaction, the cavity decays, more atoms pass through and the  state  becomes
more and more entangled. If we decide never  to  measure  the  state  of  atoms
$A_1\ldots  A_i$  with  $i<k$,  we  should  calculate  the   trace   over   the
corresponding states and only the $\rho_{A_k}$-component remains.  Since  the  time
evolution  is  linear,  each  of  the   components   in   \eq{rhotau}   evolves
independently, and it does not matter when we calculate the trace. We can do
it
after each atom has passed the cavity, or at the end  of  the  experiment.  For
this we do not even have to assume that the  atoms  are  non-interacting  after
they leave the cavity, even though this simplifies the time evolution. If we do
perform a measurement of the state of an intermediate atom $A_i$, a correlation
can be observed between that result and a measurement of atom  $A_k$,  but  the
statistics of the unconditional measurement of  $A_k$  is  not  affected  by  a
measurement of $A_i$. In a real experiment also the efficiency of the measuring
apparatus should be taken into account when using  the  measured  results  from
atoms  $A_1,\ldots,A_i$  to  predict  the  probability  of  the  outcome  of  a
measurement of $A_k$ (see Ref. \cite{Brigeletal94} for a detailed investigation
of this case).

As a generic case let us assume that the initial  state  of  the  atom  is  a
diagonal mixture of excited and unexcited states

\begin{formula}{InitAtom}
\rho_A=\(\begin{array}{cc}a&0\\0&b\\\end{array}\)~~,
\end{formula}

\noi where, of course, $a,b\ge0$ and $a+b=1$. Using that both  preparation  and
observation are diagonal in the atomic states, it may  now  be  seen  from  the
transition elements in \eq{TransEl} that  the  time  evolution  of  the  cavity
density matrix does not mix different diagonals of this matrix.  Each  diagonal
so to speak ``lives its own life'' with respect to dynamics. This implies  that
if the initial cavity density matrix is diagonal, \ie\ of the form

\begin{formula}{}
\rho_C=\sum_{n=0}^\infty p_n |n\>\<n|~~,
\end{formula}

\noi with $p_n\ge0$ and  $\sum_{n=0}^\infty  p_n=1$,  then  it  stays  diagonal
during the interaction between atom and cavity and may always be described by a
probability distribution $p_n(t)$. In fact,  we  easily  find  that  after  the
interaction we have

\begin{formula}{Master}
p_n(\tau)=a q_n(\tau)p_{n-1}+bq_{n+1}(\tau)p_{n+1}
+(1-a q_{n+1}(\tau)-bq_{n}(\tau))p_n~~,
\end{formula}

\noi where the first term is the probability of decay for  the  excited  atomic
state, the second the probability of excitation for the  atomic  ground  state,
and the third is the probability  that  the  atom  is  left  unchanged  by  the
interaction. It is convenient to write this in matrix form
\cite{Herzog94}

\begin{formula}{}
p(\tau)=M(\tau)p~~,
\end{formula}

\noi with a transition matrix $M=M(+)+M(-)$ composed of two parts, representing
that the outgoing atom is either in the excited state (+) or  in  the  ground
state ($-$). Explicitly we have

\begin{formulas}{Pump}
M(+)_{nm}=&bq_{n+1}\delta_{n+1,m}
+a(1-q_{n+1})\delta_{n,m}~~,\\
M(-)_{nm}=&aq_n\delta_{n,m+1}
+b(1-q_{n})\delta_{n,m}~~.
\end{formulas}

\noi Notice that these formulas are completely classical and may  be  simulated
with a standard Markov  process.  The  statistical  properties  are  not
quantum
mechanical as long as the incoming atoms have a diagonal density matrix and  we
only measure elements in the diagonal. The only quantum-mechanical  feature  at
this stage is the discreteness  of  the  photon  states,  which  has  important
consequences for the correlation  length  (see
Section~\ref{LLCav}). The quantum-mechanical
discreteness of photon states in the cavity can actually be tested
experimentally \cite{bruneetal&96}.

 If  the
atomic
density matrix has off-diagonal elements, the above formalism breaks down.  The
reduced cavity density matrix will then  also  develop  off-diagonal  elements,
even if initially it is diagonal. We shall not go further  into  this  question
here (see for example Refs.~\cite{Krause86}--\cite{Zaheer89}).
\bort{\cite{Krause86,Lugiato87,Zaheer89}}


\subsection{The Lossless Cavity}\label{LLCav}

The above discrete master equation (\ref{Master}) describes the pumping of a
lossless cavity with a beam of atoms. After $k$ atoms have  passed  through
the cavity, its state has become  $M^kp$.  In  order  to  see  whether  this
process may reach statistical  equilibrium  for  $k\to\infty$  we  write
Eq.~(\ref{Master}) in the form

\begin{formula}{Current}
p_n(\tau)=p_n+J_{n+1}-J_n~~,
\end{formula}

\noi where $J_n=-aq_np_{n-1}+bq_np_n$. In statistical equilibrium we must  have
$J_{n+1}=J_n$, and the common value $J=J_n$ for all $n$ can only be zero  since
$p_n$, and therefore $J$, has to vanish for $n\to\infty$. It  follows  that
this
can only be the case for $a<b$ \ie\ $a<0.5$. There must thus be fewer than 50\%
excited atoms in the beam, otherwise the lossless cavity blows up. If $a<0.5$,
the cavity will reach an equilibrium distribution of  the  form  of  a  thermal
distribution  for   an   oscillator   $p_n=(1-a/b)(a/b)^n$.   The   statistical
equilibrium may be shown to be stable, \ie\ that all non-trivial eigenvalues of
the matrix $M$ are real and smaller than 1.


\subsection{The Dissipative Cavity}

A single oscillator interacting with an environment having a huge number of
degrees of freedom, for example a heat bath, dissipates energy
according to the well-known damping formula (see for example
\cite{Agarwal73,Walls95}):

\begin{formulas}{Damping}
{\displaystyle\frac{d\rho_C}{dt}}=&~~i[\rho_C,\omega a^*a]\\
&-\frac12\gamma(n_b+1)(a^*a\rho_C+\rho_C a^*a-2a\rho_C a^*)\\
&-\frac12\gamma n_b(aa^*\rho_C+\rho_C aa^*-2a^*\rho_C a)~~,\\
\end{formulas}

\noi where $n_b$  is  the  average  environment  occupation  number  at  the
oscillator frequency and $\gamma$ is the decay constant. This evolution also
conserves diagonality, so we have for any diagonal cavity state:

\begin{formula}{Dissipative}
\frac1\gamma\frac{dp_n}{dt}=-(n_b+1)(np_n-(n+1)p_{n+1})-n_b((n+1)p_n-np_{n-1})
{}~~,
\end{formula}

\noi which of course conserves probability. The right-hand  side  may  as  for
Eq.~(\ref{Current}) be written as $J_{n+1}-J_n$ with
$J_n=(n_b+1)np_n-n_bnp_{n-1}$
and the same arguments as above lead to a thermal equilibrium distribution with

\begin{formula}{Thermal}
\label{eq:thermalprob}
p_n=\frac1{1+n_b}\(\frac{n_b}{1+n_b}\)^n ~~.
\end{formula}


\subsection{The Discrete Master Equation}

We now take into account both pumping and damping. Let the next atom arrive  in
the cavity after a time $T\gg\tau$. During this interval the cavity damping  is
described by Eq.~(\ref{Dissipative}), which we shall write in the form

\begin{formula}{}
\frac{dp}{dt}=-\gamma  L_C  p~~,
\end{formula}

\noi where $L_C$ is the cavity decay matrix from above

\begin{formula}{LC}
(L_C)_{nm}=(n_b+1)(n\delta_{n,m}-(n+1)\delta_{n+1,m})
+n_b((n+1)\delta_{n,m}-n\delta_{n-1,m}) ~~.
\end{formula}
This decay matrix conserves probability, i.e. it is trace-preserving:
\be
\sum_{n=0}^{\infty}\left(L_C\right)_{nm} = 0~~~.
\ee
\noi The statistical state of the cavity when the next atom arrives is  thus
given by

\begin{formula}{stat1}
p(T)=e^{-\gamma L_CT}M(\tau)p ~~.
\end{formula}

\noi
In using the full interval $T$ and not $T-\tau$ we allow for
the decay  of
the cavity in the interaction time, although this decay is not properly
included  with  the  atomic  interaction  (for  a  more
 correct   treatment   see
Appendix~\ref{AppDamping}).

This would be the master equation describing the evolution of the cavity if the
atoms in the beam arrived with definite and known intervals. More commonly, the
time  intervals  $T$  between  atoms  are  Poisson-distributed   according   to
$d\P(T)=\exp(-RT)RdT$  with  an  average  time  interval  $1/R$  between  them.
Averaging the exponential in Eq.~(\ref{stat1}) we get

\begin{formula}{aveL}
\<p(T)\>_T= S p~~,
\end{formula}

\noi where

\begin{formula}{Discrete1}
S=\frac1{1+L_C/N}M~~,
\end{formula}

\noi and $N=R/\gamma$ is the dimensionless  pumping  rate  already  introduced.

Implicit in the above consideration is the lack  of  knowledge  of  the  actual
value of the atomic state after the interaction. If we know that the  state  of
the atom is $s=\pm$ after the  interaction,  then  the  average  operator  that
transforms the cavity state is  instead

\begin{formula}{Discrete2}
S(s)=(1+L_C/N)^{-1}M(s)~,
\end{formula}

\noi  with  $M(s)$ given by Eq.~(\ref{Pump}). This average operator
$S(s)$ is now by construction probability preserving, i.e.
\be
\sum_{n,m=0}^{\infty}S_{nm}(s)p_m = 1~~~.
\ee

Repeating the process for a sequence of $k$ unobserved atoms we find  that  the
initial probability distribution $p$ becomes $S^kp$. In the general  case  this
Markov process converges towards a  statistical  equilibrium  state  satisfying
$Sp=p$, which has the solution \cite{Filipowicz86,Lugiato87} for $n\ge1$

\begin{formula}{Equilibrium}
p_n=p_0\prod_{m=1}^n \frac{n_bm+Naq_m}{(1+n_b)m+Nbq_m}~~.
\end{formula}

\noi The overall constant $p_0$ is determined by $\sum_{n=0}^\infty
p_n=1$.
In passing we observe that if $a/b = n_b / (1+n_b)$ then this
statistical equilibrium distribution is equal to the thermal statistical
distribution Eq.(\ref{eq:thermalprob}) as it should.
The
photon landscape formed by this expression as a function of $n$ and  $\tau$  is
shown in Figure \ref{FigLandscape} for $a=1$ and $b=0$.  For  greater  values  of
$\tau$ it becomes very rugged.


\section{Statistical Correlations}
\label{correlations}
\seqnoll
\begin{flushright}
``{\sl Und was in schwankender Erscheinung schwebt,\\
 Befestiget mit dauernden Gedanken.}"\\
J.~W. von Goethe
\end{flushright}
After studying stationary single-time properties of the micromaser,  such  as
the average photon number in the cavity and the  average  excitation  of  the
outgoing atoms, we now proceed to dynamical properties. Correlations  between
outgoing atoms are not only determined by the equilibrium distribution in the
cavity but also by its approach to this equilibrium. Short-time  correlations,
such as the correlation between two  consecutive atoms 
\cite{PaulRichter91,Herzog94},
are difficult to determine experimentally,  because  they  require  efficient
observation of the states of atoms emerging from  the  cavity in
rapid  succession.
We propose instead to study and  measure  long-time  correlations,
which  do  not  impose  the  same  strict  experimental   conditions.   These
correlations turn out  to  have  a  surprisingly  rich  structure  (see  Figure
\ref{FigXi}) and reflect global properties of  the  photon  distribution.  In
this section we introduce the concept of long-time correlations  and  present
two ways of calculating them numerically. In the following sections we  study
the analytic properties of these correlations and elucidate their relation to
the dynamical phase structure, especially those aspects that are poorly
seen in the single-time observables or short-time correlations.


\subsection{Atomic Beam Observables}

Let us imagine that we know the state of all the  atoms  as  they  enter  the
cavity, for example that they are all  excited,  and  that  we  are  able  to
determine the state of each atom as it exits from the cavity. We shall assume
that the initial beam is statistically stationary, described by  the  density
matrix (\ref{InitAtom}), and that we have obtained an experimental record  of
the exit states of all the atoms after the  cavity  has  reached  statistical
equilibrium with the beam. 
The effect of non-perfect measuring efficiency
has been considered in several papers \cite{Brigeletal94,WagnerSW94, Herzog94}
but we ignore that complication since it is a purely experimental problem.
{}From this record  we  may  estimate  a  number  of
quantities, for example the probability  of  finding  the  atom  in  a  state
$s=\pm$ after the interaction, where we choose $+$ to represent  the  excited
state and $-$ the ground state. The  probability  may  be  expressed  in  the
matrix form

\begin{formula}{SingleP}
\P(s)={u^0}^\top M(s) p^0~~,
\end{formula}

\noi
where $M(s)$ is given  by  Eq.~(\ref{Pump})  and  $p^0$  is  the  equilibrium
distribution (\ref{Equilibrium}). The quantity $u^0$ is  a  vector  with  all
entries equal to 1, $u^0_n=1$, and represents the sum over all possible final
states of the cavity. In Figure \ref{FigDataComparison} we  have  compared  the
behavior of $\P(+)$ with some characteristic experiments.

Since $\P(+)+\P(-)=1$ it is sufficient to measure the average spin value
(see Figure \ref{FigMuQf}):

\begin{formula}{}
\label{eq:spinaverage}
\<s\>=\P(+)-\P(-)~~.
\end{formula}

\noi Since $s^2=1$ this quantity also determines the variance to be
$\<s^2\>-\<s\>^2=1-\<s\>^2$.

Correspondingly, we may define the joint probability for observing the states
of two atoms, $s_1$ followed $s_2$, with $k$ unobserved atoms between them,

\begin{formula}{kProb}
\P_k(s_1,s_2)={u^0}^\top S(s_2) S^k S(s_1) p^0~~,
\end{formula}

\noi  where  $S$  and  $S(s)$  are  defined  in  Eqs.  (\ref{Discrete1})  and
(\ref{Discrete2}).  
The joint probability of finding two consecutive excited outcoming
atoms, $\P_0(+,+)$, was calculated in \cite{PaulRichter91}.
It  is  worth  noticing  that  since  $S=S(+)+S(-)$   and
$Sp^0=p^0$ we have $\sum_{s_1} \P_k(s_1,s_2)=\P(s_2)$.  Since  we  also  have
${u^0}^\top  L=  {u^0}^\top  (M-1)=0$  we  find  likewise  that   ${u^0}^\top
S={u^0}^\top$ so that  $\sum_{s_2}  \P_k(s_1,s_2)=\P(s_1)$.  Combining  these
relations we derive that $\P_k(+,-)=\P_k(-,+)$, as  expected.  Due  to  these
relations there is  essentially  only  one  two-point  function,  namely  the
``spin--spin'' covariance function

\begin{formulas}{}
\<ss\>_k&=&\sum_{s_1,s_2} s_1s_2\P_k(s_1,s_2)\\
&=&\P_k(+,+)+ \P_k(-,-)- \P_k(+,-)- \P_k(-,+)\\
&=&1-4\P_k(+,-)~~.
\end{formulas}

\noi From this we derive the properly normalized correlation function

\begin{formula}{}
\gamma^A_k=\frac{\<ss\>_k-\<s\>^2}{1-\<s\>^2}~~,
\end{formula}

\noi which satisfies $-1\le\gamma^A_k\le1$.

At large times, when $k\to\infty$, the correlation function is  in  general
expected to decay exponentially, and we define the atomic beam  correlation
length $\xi_A$ by the asymptotic behavior for large $k\simeq Rt$

\begin{formula}{}
\gamma^A_k\sim\exp\(-\frac{k}{R\xi_A}\)~~.
\end{formula}

\noi Here we have scaled with $R$, the average  number  of  atoms  passing  the
cavity per unit of time, so that $\xi_A$ is the typical length of time that the
cavity remembers previous pumping events.


\subsection{Cavity Observables}

In the context  of  the  micromaser  cavity,  one  relevant  observable  is
the
instantaneous number of photons $n$, from which we may form the average $\<n\>$
and correlations in time. The quantum state of light in  the  cavity  is  often
characterized by  the  Fano--Mandel  quality  factor  \cite{Mandel79},  which
is
related to the fluctuations of $n$ through

\begin{formula}{}
Q_f=\frac{\<n^2\>-\<n\>^2}{\<n\>}-1~~.
\end{formula}

\begin{figure}[p]
\unitlength=1mm
\begin{picture}(140,90)(0,0)
\includegraphics{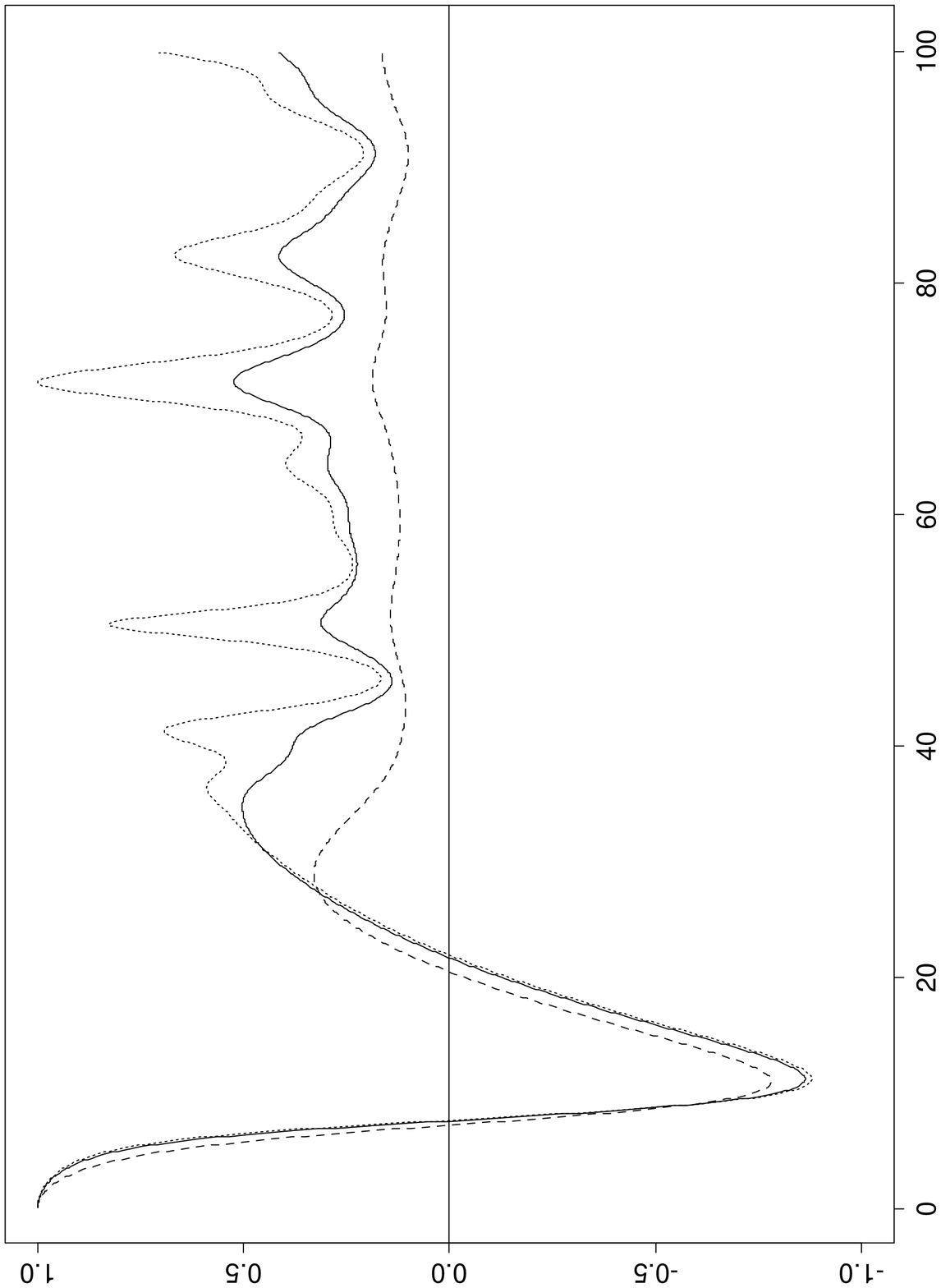}
\put(35,85){\small\parbox{40mm}{$N=10, \Omega=44 {\rm
{}~kHz}$\\$n_b=0.0,0.15,1.0$}}
\put(10,52){\small$\<s\>$}
\end{picture}

\begin{picture}(140,90)(0,0)
\includegraphics{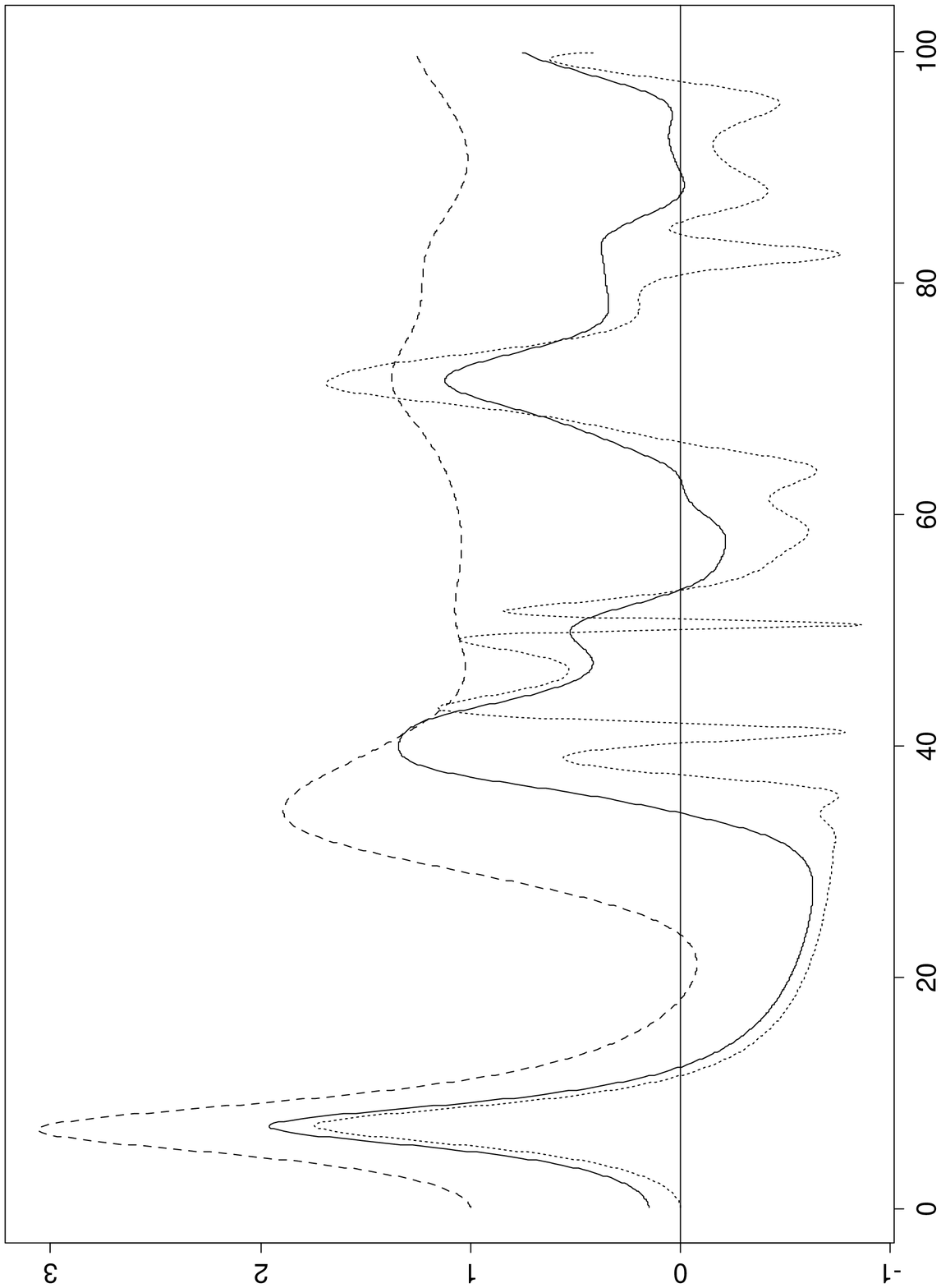}
\put(40,85){\small\parbox{40mm}{$N=10, \Omega=44 {\rm
{}~kHz}$\\$n_b=0.0,0.15,1.0$}}
\put(75,3){\small$\tau$ [\mus]}
\put(6,55){\small$Q_f$}
\end{picture}
\figcap{The upper figure shows the mean value  of  the  spin  variable  as  a
function of the atomic passage time for three different values of $n_b$.  The
dotted line $n_b=0$, the solid line $n_b=0.15$, and the dashed line  $n_b=1$.
The lower figure shows the Mandel quality factor in the same region  for  the
same values of $n_b$. The pronounced structures in the case of $n_b=0$ are
caused by trapping states (see Section \ref{Trapping}).
}
\label{FigMuQf}
\end{figure}

\noi This quantity vanishes for coherent (Poisson) light  and  is  positive
for classical light.
 (see Figure \ref{FigMuQf})

In equilibrium there is a relation between the average photon occupation number
and the spin average in the atomic beam, which is trivial to derive from the
equilibrium distribution (a=1)

\begin{formula}{}
\<n\>={u^0}^\top \hat{n} p^0=n_b+N\P(-)=n_b+N\frac{1-\<s\>}{2}~~,
\end{formula}

\noi where $\hat n$ is a diagonal matrix representing the quantum number $n$. A
similar but more uncertain relation  between  the  Mandel  quality  factor  and
fluctuations in the atomic beam may also be derived \cite{Rempe90a}.

The covariance between the values of the photon occupation number $k$ atoms
apart in equilibrium is easily seen to be given by

\begin{formula}{Relation}
\<n n\>_k={u^0}^\top \hat n S^k \hat n p^0~~,
\end{formula}

\noi and again a normalized correlation function may be defined

\begin{formula}{}
\gamma^C_k=\frac{\<n n\>_k-\<n\>^2}{\<n^2\>-\<n\>^2}~~.
\end{formula}

\noi The cavity correlation length $\xi_C$ is defined by

\begin{formula}{}
\gamma^C_k\sim \exp\(-\frac{k}{R\xi_C}\)~~.
\end{formula}

\noi Since the same power of the matrix $S$  is  involved,  both  correlation
lengths are determined by  the  same  eigenvalue,  and  the  two  correlation
lengths are  therefore  identical  $\xi_A=\xi_C=\xi$  and  we  shall  no longer
distinguish between them.


\subsection{Monte Carlo Determination of Correlation Lengths}

Since the statistical behaviour of the micromaser is a classical Markov process
it is possible to simulate it by means of Monte Carlo methods using the  cavity
occupation number $n$ as stochastic variable.

A sequence of excited atoms is generated at Poisson-distributed times and are
allowed  to  act  on  $n$   according   to   the   probabilities   given   by
Eq.~(\ref{TransEl}). In these simulations we have for simplicity chosen $a=1$
and $b=0$. After the interaction the cavity is allowed to  decay  during  the
waiting time until the next atom arrives. The action of this process  on  the
cavity variable $n$ is simulated by means  of  the  transition  probabilities
read off from the dissipative master  equation  (\ref{Dissipative})  using  a
suitably small time step $dt$. The states  of  the  atoms  in  the  beam  are
determined by the pumping transitions and the atomic correlation function may
be determined from this sequence of spin values $\{s_i\}$ by making  suitable  averages
after the system has reached equilibrium. Observables are then
measured in a standard manner. For  the averge spin
Eq.(\ref{eq:spinaverage}) we e.g. write
\be
\<s\> = \lim_{L \rightarrow \infty} \frac{1}{L}\sum_{i=1}^{L}s_i~~~,
\ee
and for the joint probability Eq.(\ref{kProb}) we write
\be
\P_k(s_1,s_2) =
\lim_{L \rightarrow \infty} \frac{1}{L}\sum_{i=1}^{L}s_i s_{i+k}~~~,
\ee
Finally we can then  extract  the correlation lengths
 numerically from the Monte Carlo data.

\begin{figure}[htb]
\unitlength=1mm
\begin{picture}(140,100)(0,0)
\includegraphics{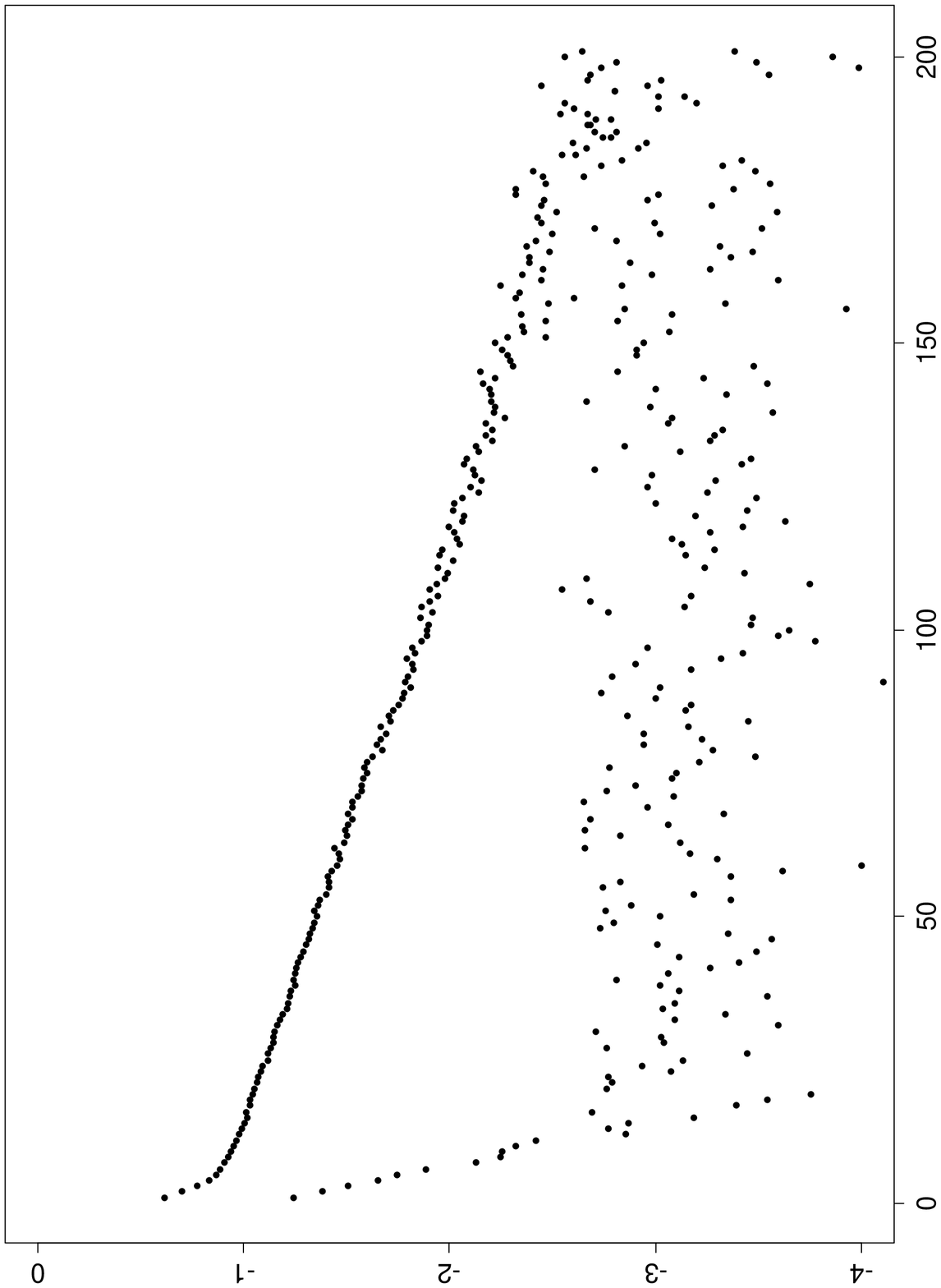}
\put(75,5){\small$k\simeq Rt$}
\put(0,53){\small$\log_{10}\gamma ^{A}_k$}
\put(80,90){\small$^{85}$Rb 63p$_{3/2}\leftrightarrow$ 61d$_{5/2}$}
\put(80,85){\small$R=50~{\rm s}^{-1}$}
\put(80,80){$n_b=0.15$}
\put(80,75){$\gamma=5~{\rm s}^{-1}$}
\end{picture}
\figcap{Monte Carlo data (with $10^6$ simulated atoms) for the correlation as a
function of the separation $k\simeq Rt$ between the  atoms  in  the  beam  for
$\tau=25~\mu$s (lower data points) and $\tau=50~\mu$s (upper data  points).
In
the latter case the exponential  decay  at  large  times  is  clearly  visible,
whereas it is hidden in the noise in the former. The parameters  are  those  of
the experiment described in Ref. \cite{Rempe90}.
}
\label{FigFit}
\end{figure}

This extraction is, however, limited by noise due to the finite  sample  size
which in our simulation is $10^6$ atoms. In  regions  where  the  correlation
length is large,  it  is  fairly  easy  to  extract  it  by  fitting  to  the
exponential decay, whereas it is more difficult in the regions  where  it  is
small (see Figure \ref{FigFit}). This accounts for the differences between  the
exact numerical calculations and the Monte Carlo data in Figure \ref{FigXi}. It
is expected that real experiments will face the  same  type  of  problems  in
extracting the correlation lengths from real data.


\subsection{Numerical Calculation of Correlation Lengths}

\begin{figure}[htb]
\unitlength=1mm
\begin{picture}(140,100)(0,0)
\includegraphics{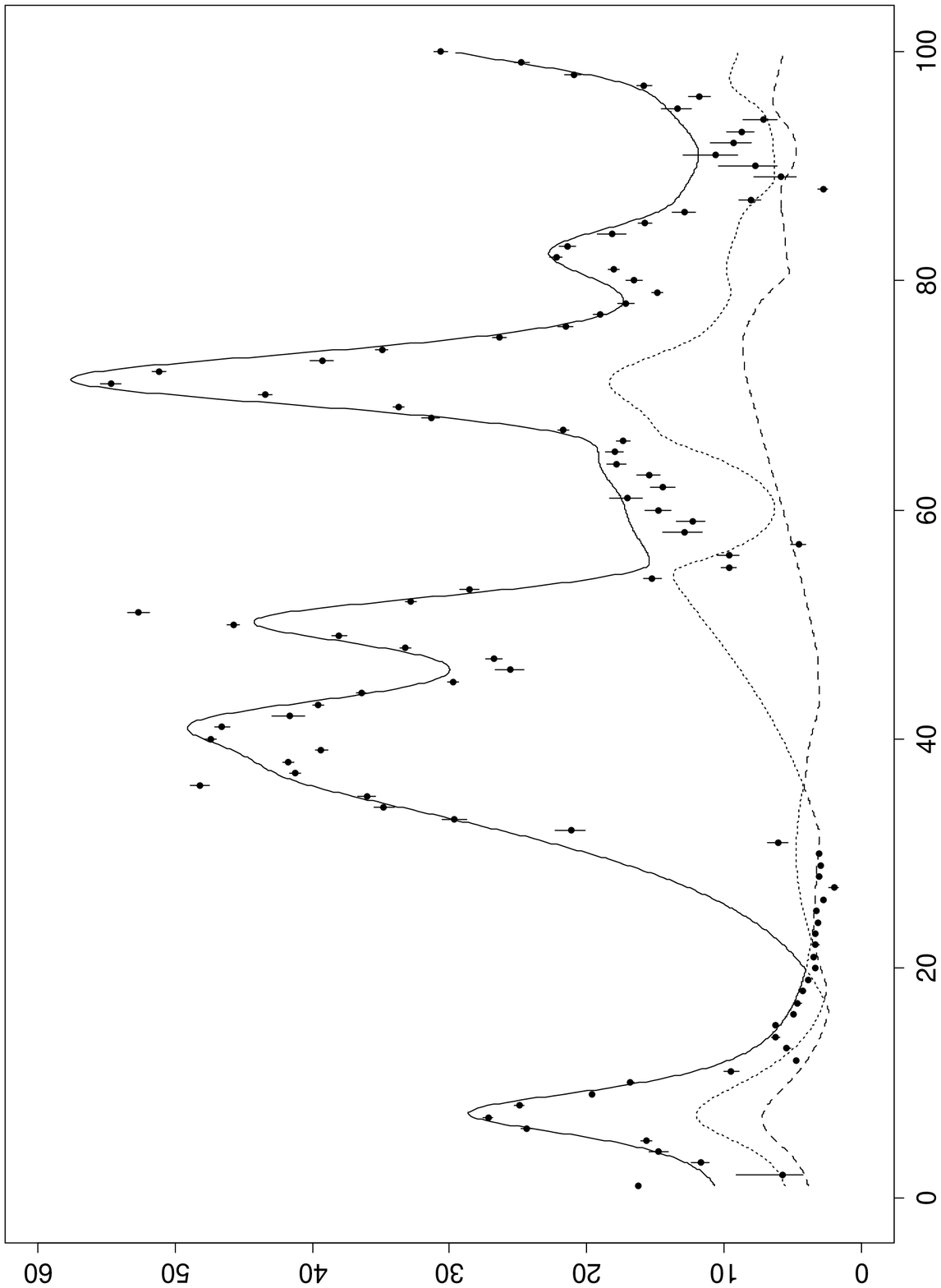}
\put(75,5){\small$\tau~[\mu {\rm s}]$}
\put(0,53){\small$R\xi$}
\put(20,90){\small$^{85}$Rb 63p$_{3/2}\leftrightarrow$ 61d$_{5/2}$}
\put(20,80){\small$R=50~{\rm s}^{-1}$}
\put(20,70){$n_b=0.15$}
\put(20,60){$\gamma=5~{\rm s}^{-1}$}
\end{picture}
\figcap{Comparison of theory  (solid  curve)  and  MC  data  (dots)  for  the
correlation length $R\xi$ (sample size $10^6$ atoms). The dotted  and  dashed
curves correspond to sub-leading eigenvalues ($\kappa_{2,3}$) of  the  matrix
$S$. The parameters are those of the experiment in Ref.~\cite{Rempe90}. }
\label{FigXi}
\end{figure}

The micromaser equilibrium distribution is the solution of $Sp=p$,  where  $S$
is the one-atom propagation matrix (\ref{Discrete1}), so  that  $p^0$  is  an
eigenvector  of  $S$  from  the  right  with  eigenvalue  $\kappa_0=1$.   The
corresponding eigenvector  from  the  left  is  $u^0$  and  normalization  of
probabilities is expressed as  ${u^0}^\top  p^0=1$.  The  general  eigenvalue
problem concerns solutions to $Sp=\kappa p$ from the right and  $u^\top  S  =
\kappa u^\top $ from the left. It is shown below  that  the  eigenvalues  are
non-degenerate, which implies that there exists a spectral resolution  of  the
form

\begin{formula}{}
S= \sum_{\ell=0}^\infty \kappa_\ell p^\ell {u^\ell}^\top~~,
\end{formula}

\noi with eigenvalues $\kappa_\ell$ and eigenvectors  $p^\ell$  and  $u^\ell$
from right and left respectively. The long-time behavior of the  correlation
function is governed by the next-to-leading eigenvalue  $\kappa_1<1$, and  we
see that

\begin{formula}{}
R\xi=-\frac1{\log\kappa_1}~~.
\end{formula}

The   eigenvalues   are   determined   by   the    characteristic    equation
$\det\{S-\kappa\}=0$, which may be solved  numerically.  This  procedure  is,
however, not well-defined for the  infinite-dimensional  matrix  $S$, and  in
order to evaluate the determinant we have truncated the matrix to a large and
finite-size $K\times K$ with typical $K\simeq 100$. The explicit form of $S$
in Eq.~(\ref{Discrete1}) is used, which reduces the problem to the calculation
of the determinant for a Jacobi matrix. Such a matrix  vanishes  outside  the
main diagonal and the two sub-leading diagonals on each side. It is  shown  in
Section~\ref{EigenvalueProblem} that the eigenvalues found from this equation
are indeed non-degenerate, real, positive and less than unity.

The next-to-leading eigenvalue is shown in Figure \ref{FigXi} and  agrees  very
well with the Monte Carlo calculations. This figure shows a surprising amount
of structure and part of the effort in the following will  be  to  understand
this structure in detail.

\begin{figure}[htb]
\unitlength=1mm
\begin{picture}(140,100)(-10,5)
\includegraphics{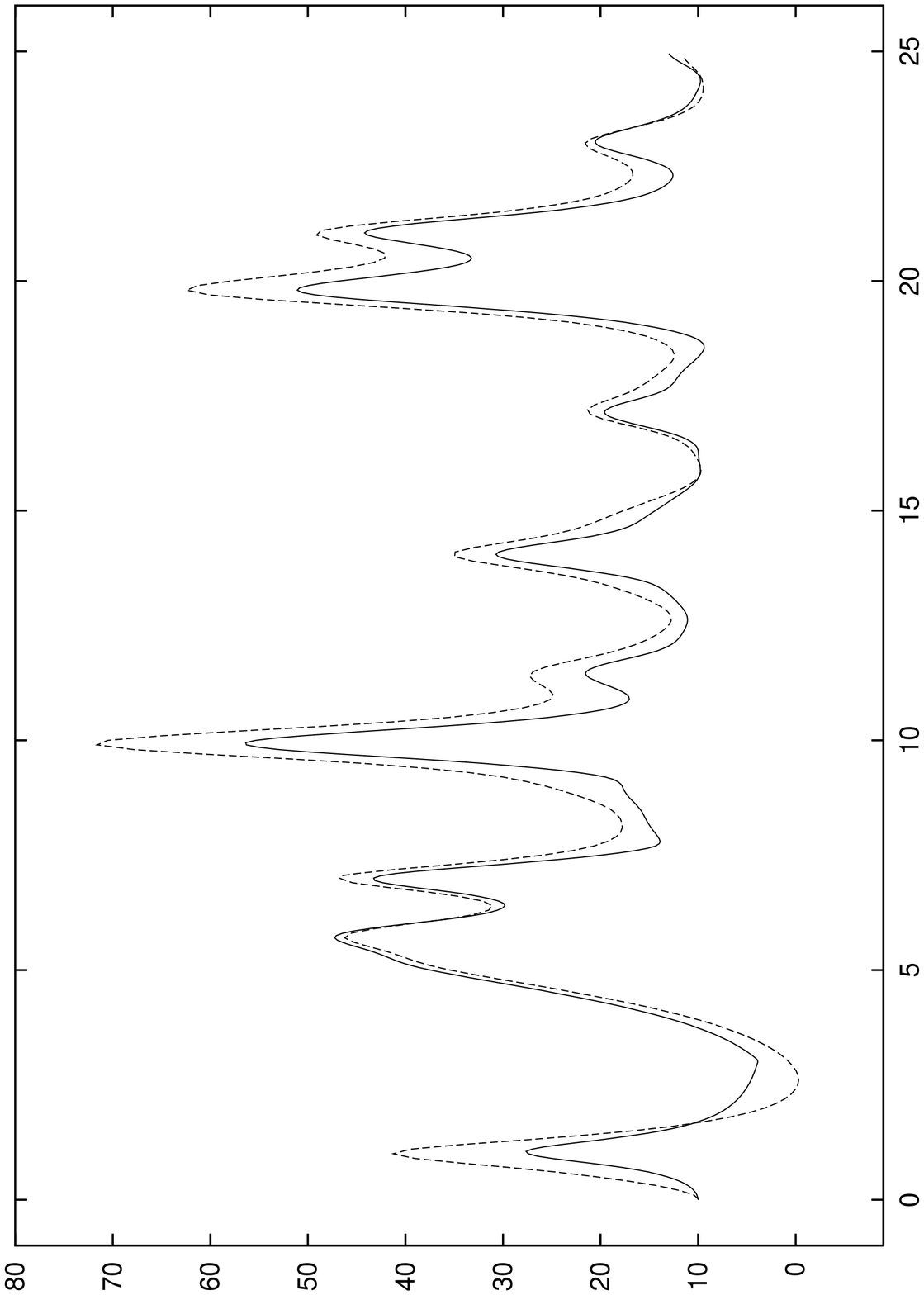}
\put(65,8){\small$\theta$}
\put(-10,54){\small$R\xi$}
\put(10,88){\small$^{85}$Rb 63p$_{3/2}\leftrightarrow$ 61d$_{5/2}$}
\put(10,83){\small$R=50~{\rm s}^{-1}$}
\put(10,78){$n_b=0.15$}
\put(10,73){$\gamma=5~{\rm s}^{-1}$}
\end{picture}
\figcap{Comparison  of  the  sum  in  Eq.  (\ref{SubSumRule})  over  reciprocal
eigenvalues (dotted  curve)  with  numerically  determined  correlation  length
(solid curve) for the same parameters as in Figure  \ref{FigXi} as a
function of $\theta = g\tau \sqrt{N}$, where $N=R/\gamma$.  The  difference
between the curves is entirely due to the sub-dominant eigenvalues that have not
been taken into account in Eq.~(\ref{SubSumRule}).}
\label{FigSumRule}
\end{figure}

It is possible to derive an exact sum rule for the reciprocal eigenvalues  (see
Appendix \ref{AppSumRule}), which yields the approximate expression:

\begin{formula}{SubSumRule}
\gamma\xi\simeq1+
\sum_{n=1}^\infty\(\frac{P_n(1-P_n)}{((1+n_b)n + Nbq_n )p_n}
-\frac{1-[n_b/(1+n_b)]^n}n\)~~,
\end{formula}

\noi when the sub-dominant eigenvalues may  be  ignored. This formula
for the correlation length is numerically rapidly converging.
Here $p_n$ is the
equilibrium distribution Eq.  (\ref{Equilibrium})  and  $P_n=\sum_{m=0}^{n-1}
p_m$ is the cumulative probability. In Figure \ref{FigSumRule} we  compare  the
exact numerical calculation and the result of the sum rule in the case
when $a=1$, which is much  less
time-consuming to compute\footnote{Notice that we have corrected for a numerical error in
 Figure 4  of Ref.\cite{Elmforsetal95b}.}.

\begin{figure}[htb]
\unitlength=1mm
\begin{picture}(140,100)(-10,5)
\includegraphics{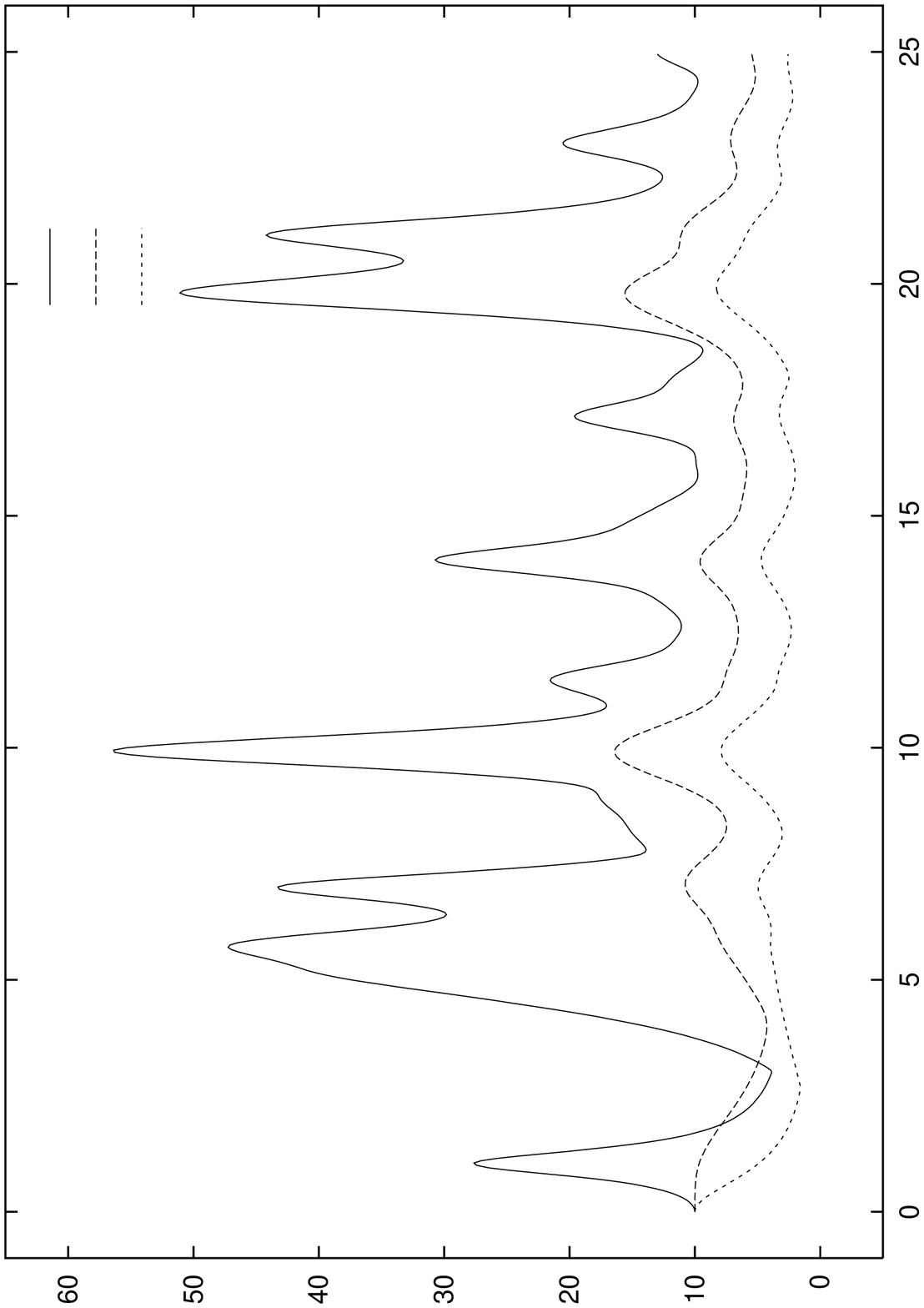}
\put(108,89){$a=1$}
\put(108,85){$a=0.5$}
\put(108,81){$a=0$}
\put(65,8){\small$\theta$}
\put(-10,58){\small$R\xi$}
\put(10,88){\small$^{85}$Rb 63p$_{3/2}\leftrightarrow$ 61d$_{5/2}$}
\put(10,83){\small$R=50~{\rm s}^{-1}$}
\put(10,78){$n_b=0.15$}
\put(10,73){$\gamma=5~{\rm s}^{-1}$}
\end{picture}
\figcap{The correlation length $R\xi$  for the same 
parameters as in Figure  \ref{FigXi} but for $a=0,~0.5,$ and $1$ as a
function of $\theta = g\tau \sqrt{N}$, where $N=R/\gamma$.}
\label{Figinversion}
\end{figure}

\begin{figure}[htb]
\unitlength=1mm
\begin{picture}(140,100)(-10,-5)
\includegraphics{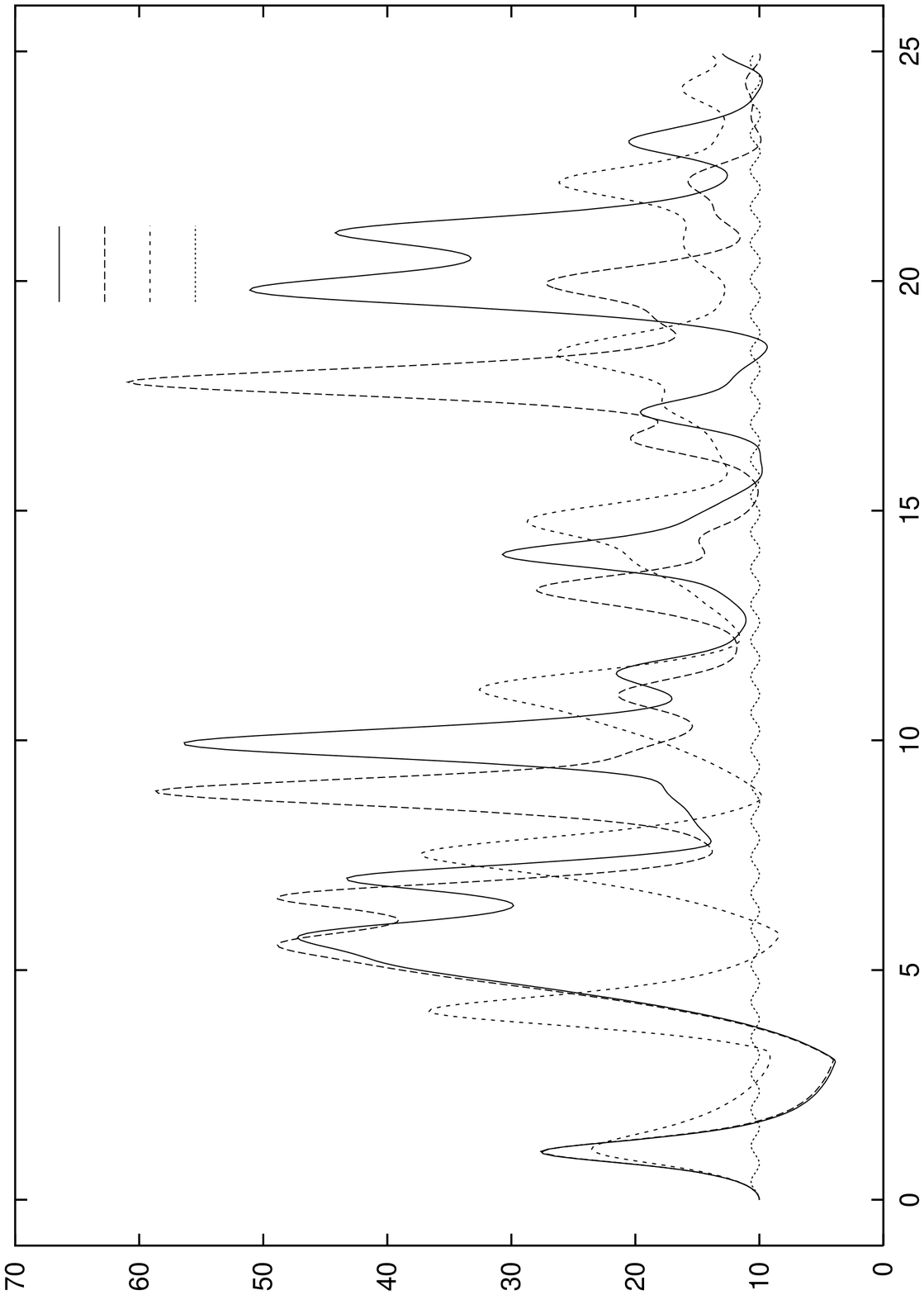}
\put(108,88){$\delta =0$}
\put(108,84){$\delta =1$}
\put(108,79){$\delta =5$}
\put(108,75){$\delta =25$}
\put(60,0){\small$\theta$}
\put(-10,50){\small$R\xi$}
\put(10,88){\small$^{85}$Rb 63p$_{3/2}\leftrightarrow$ 61d$_{5/2}$}
\put(10,83){\small$R=50~{\rm s}^{-1}$}
\put(10,78){$n_b=0.15$}
\put(10,73){$\gamma=5~{\rm s}^{-1}$}
\end{picture}
\figcap{The correlation length $R\xi$  for the same 
parameters as in Figure  \ref{FigXi} but for 
$\delta =\Delta\omega/g = 0,~1,~5$ and $25$ as a
function of $\theta = g\tau \sqrt{N}$, where $N=R/\gamma $.}
\label{Figdetune}
\end{figure}

It is also of importance to notice that the the correlation length is
very sensitive to the inversion to the atomic beam parameter $a$ (see
Figure \ref{Figinversion})
the detuning parameter $\delta = \Delta\omega/g$ (Figure \ref{Figdetune}).
%

Comparison of theory  (solid  curve)  and  MC  data  (dots)  for  the
correlation length $R\xi$ (sample size $10^6$ atoms). The dotted  and  dashed
curves correspond to sub-leading eigenvalues ($\kappa_{2,3}$) of  the  matrix
$S$. The parameters are those of the experiment in Ref.~\cite{Rempe90}.

\section{Analytic Preliminaries}
\label{analytic}
\seqnoll
\begin{flushright}
``{\sl It is futile to employ many principles\\
when it is possible to employ fewer.~}''\\
W.  Ockham
\end{flushright}
In order to tackle the  task  of  determining  the  phase  structure  in  the
micromaser we need to develop  some  mathematical  tools.  The dynamics can be
formulated in two different ways which  are  equivalent  in  the  large  flux
limit. Both are related to Jacobi matrices describing the stochastic process.
Many characteristic features of the correlation length are related to scaling
properties for $N\goto\infty$,  and  require  a  detailed  analysis  of  the
continuum limit. Here we introduce some of the concepts that are used in  the
main analysis in Section \ref{Phasestructure}.

\subsection{Continuous Master Equation}

When the atoms have Poisson distributed  arrival  times  it  is  possible  to
formulate the problem as a differential equation \cite{Lugiato87}. Each  atom
has the same probability $Rdt$ of arriving in an infinitesimal time  interval
$dt$. Provided the interaction with the cavity takes less time  than  this
interval,  \ie\  $\tau\ll  dt$,  we  may  consider  the  transition   to   be
instantaneous and write the transition matrix as $Rdt(M-1)$ so that we get

\begin{formula}{}
\frac{dp}{dt}=-\gamma L_Cp+R(M-1)p\equiv-\gamma Lp~~,
\end{formula}

\noi where $L=L_C-N(M-1)$. This equation obviously has the solution

\begin{formula}{}
p(t)=e^{-\gamma L t}p~~.
\end{formula}

\noi Explicitly we have

\begin{formulas}{}
L_{nm}&=&(n_b+1)(n\delta_{n,m}-(n+1)\delta_{n+1,m})
+n_b((n+1)\delta_{n,m}-n\delta_{n,m+1}) \\
&&+N(
(aq_{n+1}+bq_n)\delta_{n,m}
-aq_n\delta_{n,m+1}
-bq_{n+1}\delta_{n+1,m})
{}~~,
\end{formulas}

\noi and

\begin{formulas}{Maser}
{\displaystyle\frac1\gamma\frac{dp_n}{dt}}
&=&-(n_b+1)(np_n-(n+1)p_{n+1})-n_b((n+1)p_n-np_{n-1})\\
&&-N((aq_{n+1}+bq_n)p_n-aq_np_{n-1}-bq_{n+1}p_{n+1})~~.
\end{formulas}

The equilibrium distribution may be found by the same technique as before,
writing the right-hand side of \eq{Maser} as $J_{n+1}-J_n$ with

\begin{formula}{}
J_n=((n_b+1)n+Nbq_n)p_n-(n_bn+Naq_n)p_{n-1}~~,
\end{formula}

\noi and setting $J_n=0$ for all $n$. The equilibrium distribution is  clearly
given by the same expression (\ref{Equilibrium}) as in the discrete case.


\subsection{Relation to the Discrete Case}

Even if the discrete and continuous  formulation  has  the  same  equilibrium
distribution, there is a difference in the dynamical  behavior  of  the  two
cases. In the discrete case the  basic  propagation  matrix  is  $S^k$, where
$S=(1+L_C/N)^{-1}M$, whereas it is $\exp(-\gamma Lt)$ in the continuous  case.
For high pumping rate $N$ we expect the two formalisms to coincide,  when  we
identify $k\simeq Rt$.  For  the  long-time  behavior  of  the  correlation
functions this implies that the next-to-leading eigenvalues $\kappa_1$ of $S$
and $\lambda_1$ of  $L$  must  be  related  by  $1/\xi=\gamma\lambda_1\simeq
-R\log\kappa_1$.

To prove this, let us compare the two eigenvalue problems. For the continuous
case we have

\begin{formula}{}
(L_C-N(M-1))p=\lambda p~~,
\end{formula}

\noi whereas in the discrete case we may rewrite $Sp=\kappa p$ to become

\begin{formula}{}
\(L_C -\frac{N}{\kappa}(M-1)\)p=N\(\frac1\kappa-1\)p~~.
\end{formula}

\noi Let a  solution  to  the  continuous  case  be  $p(N)$  with  eigenvalue
$\lambda(N)$, making explicit the dependence on $N$. It is then obvious  that
$p(N/\kappa)$ is a solution to the discrete case with eigenvalue $\kappa$
determined by

\begin{formula}{EigenRelation}
\lambda\(\frac N\kappa\)=N\(\frac1\kappa-1\)~~.
\end{formula}

\noi As we shall  see  below,  for  $N\gg1$  the  next-to-leading  eigenvalue
$\lambda_1$ stays finite or goes to zero, and hence $\kappa_1\to1$  at  least
as fast as $1/N$. Using this result it follows that the correlation length is
the same to $\O(1/N)$ in the two formalisms.


\subsection{The Eigenvalue Problem}\label{EigenvalueProblem}

The transition matrix $L$ truncated to size  $(K+1)\times(K+1)$  is  a  special
kind of asymmetric Jacobi matrix

\begin{formula}{LK}
L_K=\left\{\begin{array}{ccccccc}
A_0+B_0&-B_1&0&0&\cdots\\
-A_0&A_1+B_1&-B_2&0&\cdots\\
0&-A_1&A_2+B_2&-B_3\\
\vdots&\vdots&\vdots&\vdots&\vdots\\
&&&&-A_{K-2}&A_{K-1}+B_{K-1}&-B_{K}\\
&&&\cdots&0&-A_{K-1}&A_K+B_K\\
\end{array}\right\}~,\\
\end{formula}

\noi where

\begin{formulas}{}
A_n&=&n_b(n+1)+Naq_{n+1}~~,\\
B_n&=&(n_b+1)n+Nbq_n~~.
\end{formulas}

\noi Notice that the sum over the elements in every column  vanishes,  except
for the first and the last, for which the sums respectively take  the  values
$B_0$ and $A_K$. In our case we have $B_0=0$,  but  $A_K$  is  non-zero.  For
$B_0=0$ it is easy to see  (using  row  manipulation)  that  the  determinant
becomes  $A_0A_1\cdots  A_K$  and  obviously  diverges  in   the   limit   of
$K\to\infty$.  Hence  the  truncation  is  absolutely  necessary.   All   the
coefficients in the characteristic equation diverge, if we do not  truncate.
In order to secure that there is an eigenvalue $\lambda=0$,  we  shall  force
$A_K=0$ instead of the value given above. This means that the matrix  is  not
just truncated but actually changed in the last diagonal element.  Physically
this secures that
there is no external input to the process from cavity occupation
numbers above $K$, a not unreasonable requirement.

An eigenvector to the right satisfies the  equation  $L_Kp=\lambda p$,  which
takes the explicit form

\begin{formula}{}
-A_{n-1}p_{n-1}+(A_n+B_n)p_n -B_{n+1}p_{n+1}=\lambda p_n~~.
\end{formula}

\noi Since we may solve this equation successively for $p_1,  p_2,\ldots,  p_K$
given  $p_0$,  it  follows  that  all  eigenvectors  are  non-degenerate.   The
characteristic polynomial obeys the recursive  equation

\begin{formula}{}
    \det(L_K-\lambda)=(A_K+B_K-\lambda)
    \det(L_{K-1}-\lambda)-A_{K-1}B_K\det(L_{K-2}-\lambda)~~,
\end{formula}

\noi and this is also the characteristic  equation  for  a  symmetric  Jacobi
matrix  with  off-diagonal  elements  $C_n=-\sqrt{A_{n-1}B_n}$.   Hence   the
eigenvalues are the same and therefore all real and, as we  shall  see  below,
non-negative. They may therefore be ordered $0= \lambda_0 < \lambda_1 <\cdots
<\lambda_K$. The equilibrium distribution (\ref{Equilibrium}) corresponds  to
$\lambda=0$ and is given by

\begin{formula}{EquilAB}
p^0_n=p^0_0\prod_{m=1}^n \frac{A_{m-1}}{B_m}=p^0_0
\frac{A_0A_1\cdots A_{n-1}}{B_1B_2\cdots B_n}
{}~~~~\mbox{for $n=1,2,\ldots,K$}~~.
\end{formula}

\noi Notice that this  expression  does  not  involve  the  vanishing  values
$B_0=A_K=0$.

Corresponding to each eigenvector $p$ to the right there  is  an  eigenvector
$u$ to the left, satisfying $u^\top L_K=\lambda u^\top$, which in  components
reads

\begin{formula}{}
A_n(u_n-u_{n-1})+B_n(u_n-u_{n+1})=\lambda u_n~~.
\end{formula}

\noi For $\lambda=0$ we obviously have $u^0_n=1$ for all $n$ and  the  scalar
product $u^0\cdot p^0=1$. The eigenvector to the left is trivially related to
the eigenvector to the right via the equilibrium distribution

\begin{formula}{uRelation}
p_n=p^0_nu_n~~.
\end{formula}

\noi  The  full  set  of  eigenvectors  to  the   left   and   to   the   right
$\{u^\ell,p^\ell~|~\ell=0,1,2,\ldots,K\}$ may now be chosen to  be  orthonormal
$u^\ell\cdot p^{\ell'}=\delta_{\ell,\ell'}$, and is, of course, complete  since
the dimension $K$ is finite.

It is useful to  express  this  formalism  in  terms  of  averages  over  the
equilibrium  distribution  $\<f_n\>_0=\sum_{n=0}^K  f_n  p^0_n$.  Then  using
Eq.~(\ref{uRelation})  we  have,  for  an  eigenvector  with  $\lambda>0$, the
relations

\begin{formulas}{}
\<u_n\>_0&=&0~~,\\
\<u_n^2\>_0&=&1~~,\\
\<u_nu'_n\>_0&=&0&\mbox{for $\lambda\ne\lambda'$}~~.
\end{formulas}

\noi Thus the eigenvectors with $\lambda>0$ may  be  viewed  as  uncorrelated
stochastic functions of $n$ with zero mean and unit variance.

Finally, we rewrite the eigenvalue equation to  the  right  in  the  form  of
$\lambda p_n=J_n-J_{n+1}$ with

\begin{formula}{}
J_n=B_np_n-A_{n-1}p_{n-1}=p^0_nB_n(u_n-u_{n-1})~~.
\end{formula}

\noi Using the orthogonality we then find

\begin{formula}{}
\lambda=\sum_{n=0}^K u_n (J_n-J_{n+1})=\<B_n(u_n-u_{n-1})^2\>_0~~,
\end{formula}

\noi which incidentally proves that all eigenvalues are non-negative.  It  is
also evident that an eigenvalue is built up from the non-constant parts, \ie\
the jumps of $u_n$.

\subsection{Effective Potential}

It is convenient to introduce an effective potential  $V_n$,
first discussed by Filipowicz {\it et al.} \cite{Filipowicz86} in the
continuum limit,   by  writing  the
equilibrium distribution (\ref{Equilibrium}) in the form

\begin{formula}{}
p_n=\frac1Ze^{-NV_n}~~,
\end{formula}

\noi with

\begin{formula}{ExactPot}
V_n=-\frac1N\sum_{m=1}^n\log\frac{n_bm+Naq_m}{(1+n_b)m+Nbq_m}~~,
\end{formula}

\noi for $n\ge1$. The  value  of  the  potential  for  $n=0$  may  be  chosen
arbitrarily, for example $V_0=0$, because of the normalization constant

\begin{formula}{}
Z=\sum_{n=0}^\infty e^{-NV_n}~~.
\end{formula}

\noi It is, of course, completely  equivalent  to  discuss  the  shape  of  the
equilibrium distribution and the shape of the effective potential.
Our definition of $V_n$ differs from the one introduced in 
Refs.~\cite{Filipowicz86,Guzman89}
in the sense that our $V_n$ is {\it exact}
while the one in 
\cite{Filipowicz86,Guzman89}
was derived from a Fokker-Planck equation in the continuum limit.


\subsection{Semicontinuous Formulation}

Another way of making analytical methods, such  as
the Fokker--Planck equation, 
easier to use is to rewrite the formalism (exactly) in  terms  of
the  scaled  photon  number  variable $x$ and the scaled time parameter
$\theta$, defined by 
\cite{Filipowicz86}

\begin{formulas}{Defs}
x&=&\displaystyle{\frac nN}~~,\\
\theta&=&g\tau\sqrt N~~.\\
\end{formulas}

\noi Notice that the variable $x$ and not $n$  is  the  natural  variable  when
observing  the  field  in  the  cavity  by  means  of  the  atomic  beam   (see
Eq.(\ref{Relation})).  
Defining  $\Delta  x=1/N$  and   introducing   the   scaled
probability distribution $p(x)=Np_n$ the conservation of probability takes  the
form

\begin{formula}{}
\sum_{x=0}^\infty \Delta x\; p(x)=1~~,
\end{formula}

\noi where the sum extends over all discrete values of $x$ in the interval.
Similarly the equilibrium distribution  takes the form

\begin{formula}{}
p^0(x)=\frac1Z_x e^{-NV(x)}~~,
\end{formula}

\noi with the effective potential given as an ``integral''

\begin{formula}{Effective}
V(x)=\sum_{x'>0}^x \Delta x'\;D(x')~~,
\end{formula}

\noi with ``integrand''

\begin{formula}{Derivative}
D(x)=-\log\frac{n_bx+aq(x)}{(1+n_b)x+bq(x)}~~.
\end{formula}

\noi The transition probability function is $q(x)=\sin^2\theta\sqrt x$
and the normalization constant is given by

\begin{formula}{}
Z_x=\frac ZN=\sum_{x=0}^\infty\Delta x\;e^{-NV(x)}~~.
\end{formula}

In order to reformulate the master equation (\ref{Maser}) it is convenient
to introduce the discrete derivatives $\Delta_+f(x)=f(x+\Delta x)-f(x)$ and
$\Delta_-f(x)=f(x)-f(x-\Delta x)$. Then we find

\begin{formula}{}
\frac1\gamma\frac{dp(x)}{dt}=\frac{\Delta_+}{\Delta x}J(x)~~,
\end{formula}

\noi with

\begin{formula}{}
J(x)=(x-(a-b)q(x))p(x)+\frac1N(n_bx+aq(x))\frac{\Delta_-}{\Delta x}p(x)~~.
\end{formula}

\noi
For  the  general  eigenvector  we  define  $p(x)=Np_n$   and   write   it   as
$p(x)=p^0(x)u(x)$ with $u(x)=u_n$ and find the equations

\begin{formula}{pEigen}
\lambda p(x)=-\frac{\Delta_+}{\Delta x}J(x)~~,
\end{formula}

\noi and

\begin{formula}{uEq}
J(x)=\frac1N p^0(x) ((1+n_b)x+bq(x))\frac{\Delta_-}{\Delta x}u(x)~~.
\end{formula}

\noi Equivalently the eigenvalue equation for $u(x)$ becomes

\begin{formula}{uEigen1}
\lambda u(x)=(x-(a-b)q(x))\frac{\Delta_-}{\Delta x}u(x)
-\frac1N\frac{\Delta_+}{\Delta x}
\left[(n_bx+aq(x))\frac{\Delta_-}{\Delta x}u(x)\right]~~.
\end{formula}

\noi As before we also have

\begin{formulas}{}
\<u(x)\>_0&=&0~~,\\
\<u(x)^2\>_0&=&1~~,
\end{formulas}

\noi where now the average over $p^0(x)$ is defined as
$\<f(x)\>_0=\sum_x \Delta x f(x) p^0(x)$.
As before we may also express the eigenvalue as an average

\begin{formula}{Eigen1}
\lambda=\frac1N
\left\<((1+n_b)x+bq(x))\(\frac{\Delta_-u(x)}{\Delta x}\)^2\right\>_0~~.
\end{formula}

\noi
Again it should be emphasized that all these formulas are exact rewritings  of
the previous ones,  but  this  formulation  permits  easy  transition  to  the
continuum case, wherever applicable.

\begin{figure}[p]
\unitlength=1mm

\begin{picture}(140,90)(0,0)
\includegraphics{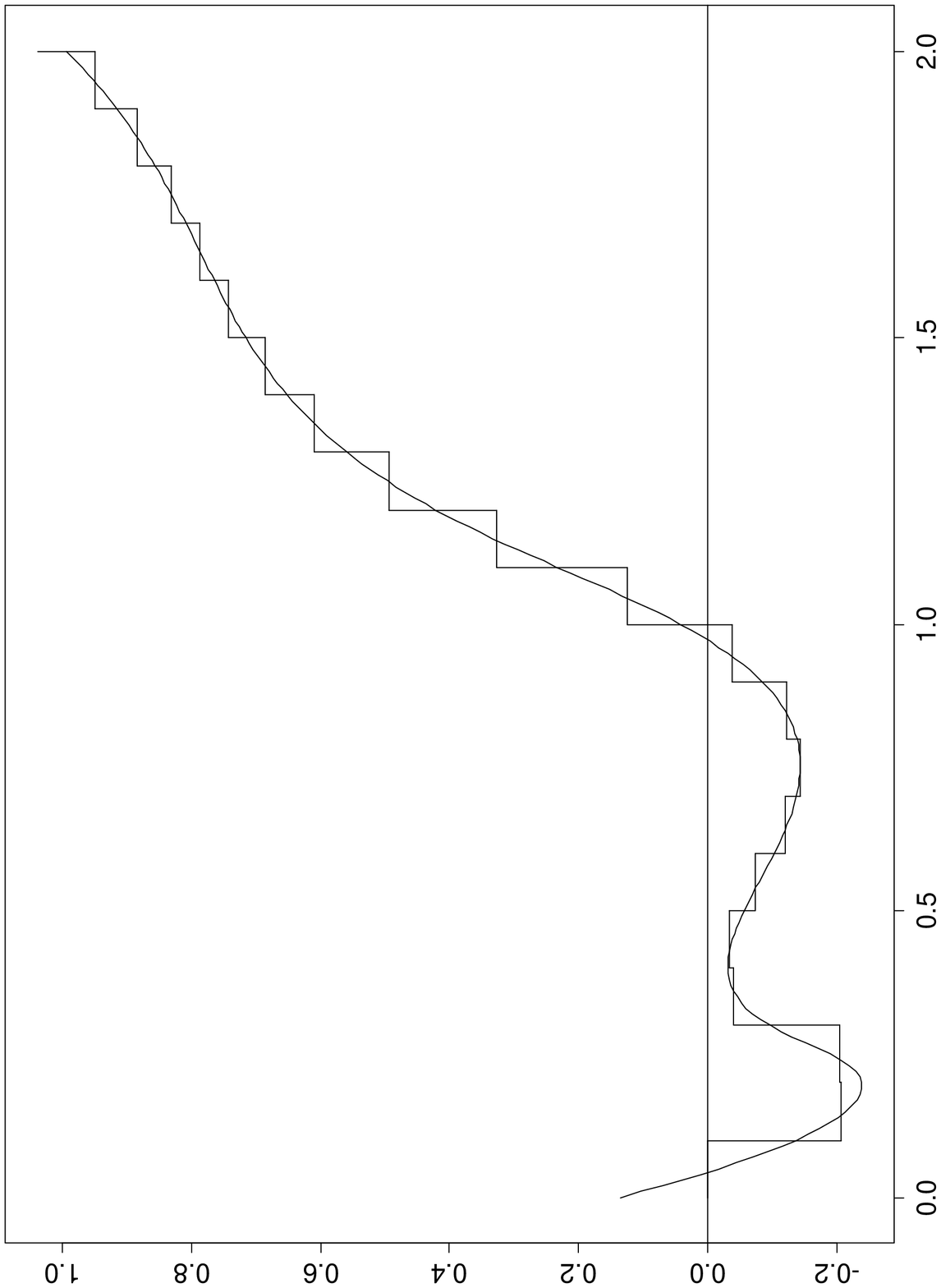}
\put(30,80){\small\parbox{30mm}{$\theta=6.0$\\$N=10$\\$n_b=0.15$\\$a=1,~b=0$}}
\put(34,20){\small$x_0$}
\put(45,32){\small$x_1$}
\put(64,24){\small$x_2$}
\put(0,50){\small$V(x)$}
\end{picture}

\begin{picture}(140,90)(0,0)
\includegraphics{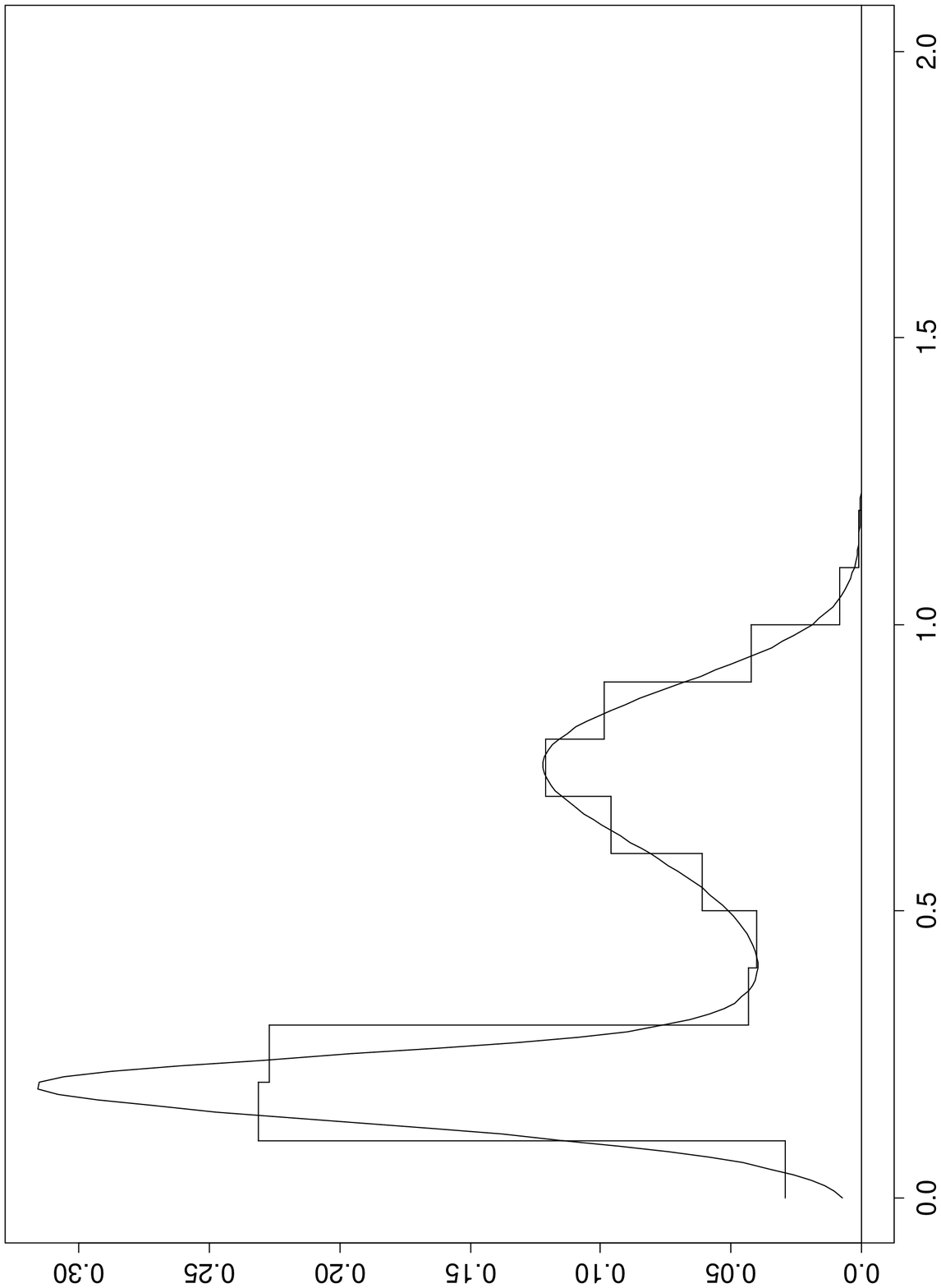}
\put(100,80){\small\parbox{30mm}{$\theta=6.0$\\$N=10$\\$n_b=0.15$\\$a=1,~b=0$}}
\put(0,50){\small$p(x)$}
\put(78,3){\small$x$}
\end{picture}

\figcap{Example of a potential with two minima $x_0,x_2$ and one maximum $x_1$
(upper  graph).  The  rectangular  curve   represents   the   exact   potential
(\ref{ExactPot}), 
whereas the continuous curve is given by Eq. (\ref{Effective})
with the summation replaced  by  an  integral.  The  value  of  the  continuous
potential at $x=0$ has been chosen such as to make the distance minimal between
the two curves. In the lower graph the corresponding  probability  distribution
is shown. }
\label{FigEffPot}
\end{figure}


\subsection{Extrema of the Continuous Potential}
\label{s:Extrema}

The quantity $D(x)$ in Eq.~(\ref{Derivative}) has a natural continuation to all
real values of $x$ as a  smooth  differentiable  function.  The  condition  for
smoothness is that the change in  the  argument  $\theta\sqrt  x$  between  two
neighbouring values, $x$ and $x+\Delta x$ is much  smaller than 1,  or
$\theta\ll
2N\sqrt x$. Hence for $N\to\infty$ the function is smooth everywhere and the
sum in Eq.~(\ref{Effective}) may be replaced by an integral

\begin{formula}{ContPot}
V(x)=\int_0^x dx'\; D(x')~~,
\end{formula}

\noi so that  $D(x)=V'(x)$. In Figure \ref{FigEffPot} we illustrate the typical
behaviour of the potential and the corresponding photon number distribution
in the first critical region (see Section \ref{FirstCritical}). Notice that
the photon-number distribution exhibits Schleich--Wheeler oscillations typical
of a squeezed state 
(see e.g. Refs. \cite{smirnov87}, \cite{Schleich87}-\cite{fabre92}).

 The extrema of this potential are located at
the solutions to $q(x)=x$; they may be parametrized in the form

\begin{formulas}{Extrema}
x&=&(a-b)\sin^2\phi~~,\\
\theta&=&\displaystyle{\frac1{\sqrt{a-b}}~\frac\phi{|\sin\phi|}}~~,\\
\end{formulas}

\noi with $0\le\phi<\infty$. These formulas map out a multibranched function
$x(\theta)$ with critical points where the derivative

\begin{formula}{}
D'(x)=V''(x)=\frac{(a+n_b(a-b))(q(x)-xq'(x))}
{((1+n_b)x+bq(x))(n_bx+aq(x))}
\end{formula}

\noi  vanishes,  which   happens   at   the   values   of   $\phi$   satisfying
$\phi=\tan\phi$.   This   equation    has    an    infinity    of    solutions,
$\phi=\phi_k,~~k=0,1,\ldots$, with $\phi_0=0$ and to a good approximation

\begin{formula}{}
\phi_k=(2k+1)\frac\pi2-\frac1{(2k+1)\frac\pi2}
+\O\(\((2k+1)\frac\pi2\)^{-3}\)
\end{formula}

\noi for $k=1,2,\ldots$, and each of these branches  is  double-valued,
with a sub-branch
corresponding to a minimum ($D'>0$) and another  corresponding
to a maximum ($D'<0$). Since there are always $k+1$ minima and  $k$  maxima, we
denote the minima $x_{2k}(\theta)$ and the maxima $x_{2k+1}(\theta)$. Thus  the
minima have even indices and the maxima have odd indices. They are given  as  a
function of $\theta$ through Eq.~(\ref{Extrema}) when $\phi$ runs  through
certain
intervals. Thus, for the minima of $V(x)$, we have

\begin{formula}{}
\phi_k<\phi<(k+1)\pi,~~\theta_k<\theta<\infty,~~a-b>x_{2k}(\theta)>0,
{}~~~k=0,1,\ldots~~,
\end{formula}

\noi
and for the maxima

\begin{formula}{}
k\pi<\phi<\phi_k,~~\infty>\theta>\theta_k,~0<x_{2k+1}(\theta)<a-b,~~k=1,\ldots
\end{formula}

\noi Here $\theta_k=\phi_k/|\sin\phi_k|\sqrt{a-b}$ is the value of $\theta$ for
which the
$k$'th   branch   comes    into    existence.    Hence    in    the    interval
$\theta_K<\theta<\theta_{K+1}$   there    are    exactly    $2K+1$    branches,
$x_0,x_1,x_2,\ldots,x_{2K-1},x_{2K}$, forming the $K+1$ minima and  $K$  maxima
of $V(x)$. For $0<\theta<\theta_0=1/\sqrt{a-b}$ there are no extrema.

This classification allows us to discuss the different  parameter  regimes
that arise in the limit of $N\to\infty$. Each  regime  is  separated  from  the
others by singularities and are thus equivalent to the phases that arise in the
thermodynamic limit of statistical mechanics.



\section{The Phase Structure of the Micromaser System}
\seqnoll
\label{Phasestructure}
\begin{flushright}
``{\sl Generalization naturally starts from the simplest,\\
 the most transparent particular case.}"\\
G. Polya
\end{flushright}
We shall from now on limit the discussion to the case of  initially  completely
excited  atoms,  $a=1,~b=0$,  which   simplifies   the   following   discussion
considerably. The case of $a \neq 1$ is considered in 
Ref.\cite{rekdal&skagerstam&99}.

\begin{figure}[htb]
\unitlength=1mm
\begin{picture}(140,100)(0,0)
\includegraphics{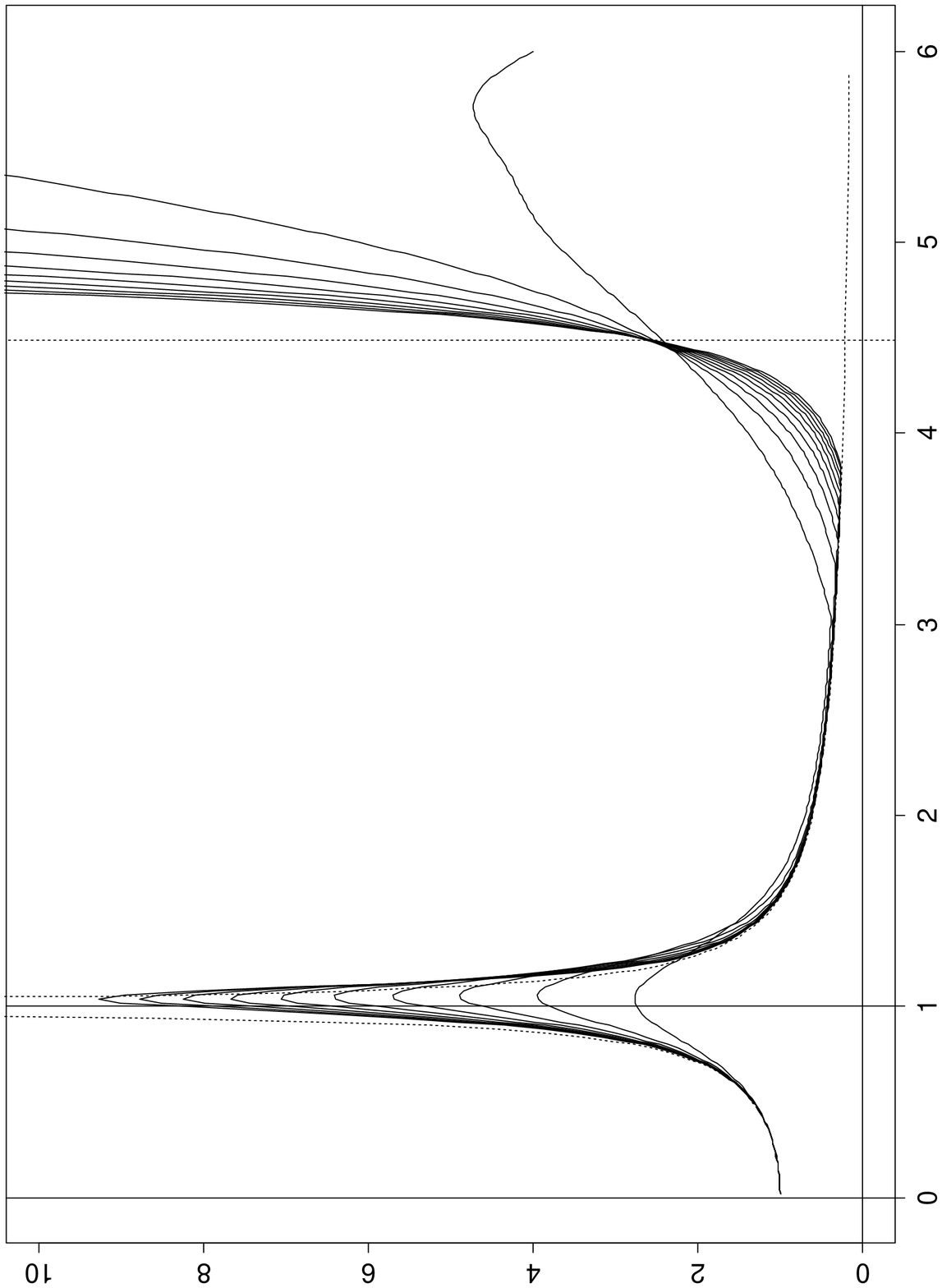}
\put(75,5){\small$\theta$}
\put(5,53){\small$\gamma\xi$}
\put(50,85){\small$N=10,20,\ldots,100$}
\put(50,80){\small$n_b=0.15$}
\end{picture}
\figcap{The correlation  length  in  the  thermal  and  maser
phases as a function of $\theta$ for various values of $N$. The  dotted  curves
are the limiting value for $N=\infty$. The correlation length grows  as  $\sqrt
N$ near $\theta=1$ and exponentially for $\theta>\theta_1\simeq4.603$. }
\label{FigMaser}
\end{figure}

The central issue in these lectures is  the  phase  structure  of  the  correlation
length as a function of the parameter $\theta$. In the limit of infinite atomic
pumping rate, $N\to\infty$, the statistical  system  described  by  the  master
equation (\ref{Master}) has a number of different dynamical  phases,  separated
from each other by singular boundaries in the space of parameters. We shall  in
this section investigate the character of the different  phases,  with  special
emphasis on the limiting behavior of the correlation length. There  turns  out
to be several qualitatively different phases within a range of  $\theta$  close
to experimental values. First, the thermal phase  and  the  transition  to  the
maser phase at $\theta=1$ has previously been discussed  in  terms  of  $\<n\>$
\cite{Filipowicz86,Lugiato87,Guzman89}. The  new  transition  to  the  critical
phase  at  $\theta_1\simeq4.603$  is  not  revealed  by  $\<n\>$  and  the
introduction of the correlation length as an observable is necessary to
describe
it. In the large flux limit $\<n\>$ and $\<(\Delta n)^2\>$ are  only  sensitive
to the probability distribution close to its global  maximum.  The  correlation
length depends crucially also on local  maxima  and  the  phase  transition  at
$\theta_1$ occurs when a new local maximum emerges. At $\theta\simeq 6.3$ there
is a phase transition in $\<n\>$ taking a discrete jump to a higher  value.  It
happens when there are two competing global minima in the  effective  potential
for different values of $n$. At the same point the correlation  length  reaches
its maximum. In Figure \ref{FigMaser} we  show  the  correlation  length  in  the
thermal and maser phases, and in Figure \ref{FigCritic} the critical phases,  for
various values of the pumping rate $N$.

\subsection{Empty Cavity}

When there is no interaction, \ie\ $M=1$, or equivalently $q_n=0$ for all  $n$,
the behavior of the cavity is purely thermal, and then it is possible to  find
the eigenvalues explicitly. Let us in this case write

\begin{formula}{}
L_C=(2n_b+1)L_3-(1+n_b)L_--n_bL_+-\frac12~~,
\end{formula}

\noi
where

\begin{formulas}{}
(L_3)_{nm}&=&\(n+\frac12\)\delta_{nm}~~,\\
(L_+)_{nm}&=&n\delta_{n,m+1}~~,\\
(L_-)_{nm}&=&(n+1)\delta_{n+1,m}~~.\\
\end{formulas}

\noi
These operators form a representation of the Lie algebra of SU(1,1)

\begin{formula}{}
[L_-,L_+]=2L_3~, \quad [L_3,L_\pm]=\pm L_\pm~~.
\end{formula}

\noi It then follows that

\begin{formula}{}
L_C=e^{rL_+}e^{-(1+n_b)L_-}(L_3-\textstyle{\frac12})\,e^{(1+n_b)L_-}e^{-rL+}~~,
\end{formula}

\noi
where $r=n_b/(1+n_b)$. This proves that $L_C$ has the same eigenvalue  spectrum
as the simple number operator $L_3-\textstyle{\frac12}$, \ie\ $\lambda_n=n$ for
$n=0,1,\ldots$ {\it independent} of the $n_b$. Since $M=1$ for $\tau=0$  
this  is  a  limiting  case  for  the
correlation lengths  $\gamma\xi_n=1/\lambda_n=1/n$  for  $\theta=0$.  From  Eq.
(\ref{EigenRelation}) we obtain  $\kappa_n=1/(1+n/N)$  in  the  non-interacting
case. Hence  in  the  discrete  case  $R\xi_n=-1/\log\kappa_n\simeq  N/n$  for
$N\gg n$ and this agrees with the values in Figure \ref{FigXi} for $n=1,2,3$
near
$\tau=0$.


\subsection{Thermal Phase: $0\le\theta<1$}

In this phase  the  natural  variable  is  $n$,  not  $x=n/N$.  The  effective
potential has no extremum for $0<n<\infty$, but is smallest  for  $n=0$.  Hence
for $N\to\infty$ it may be approximated by its leading linear  term  everywhere
in this region

\begin{formula}{}
NV_n=n\log\frac{n_b+1}{n_b+\theta^2}~~.
\end{formula}

\noi Notice that the slope vanishes for $\theta=1$. The higher-order terms play
no  role  as  long  as  $1-\theta^2\gg1/\sqrt{N}$,  and  we  obtain  a   Planck
distribution

\begin{formula}{}
p^0_n=\frac{1-\theta^2}{1+n_b}\(\frac{n_b+\theta^2}{1+n_b}\)^n~~,
\end{formula}

\noi  with photon number average

\begin{formula}{}
\<n\>=\frac{n_b+\theta^2}{1-\theta^2}~~,
\end{formula}

\noi which (for $\theta>0$) corresponds to an increased temperature.  Thus  the
result of pumping the cavity with the  atomic  beam  is  simply  to  raise  its
effective temperature in this region. The mean occupation number  $\<n\>$  does
not depend on the dimensionless pumping rate $N$ (for sufficiently large $N$).

The  variance  is

\begin{formula}{}
\sigma_n^2=\<n^2\>-\<n\>^2=\<n\>(1+\<n\>)=
\frac{(1+n_b)(n_b+\theta^2)}{(1-\theta^2)^2}~~,
\end{formula}

\noi and  the   first   non-leading
eigenvector is easily shown to be

\begin{formula}{BoltzEigen}
u_n=\frac{n-\<n\>}{\sigma_n}~~,
\end{formula}

\noi which indeed has the form of  a  univariate  variable.  The  corresponding
eigenvalue is found from Eq.~(\ref{Eigen1})  $\lambda_1=1-\theta^2$, or

\begin{formula}{}
\gamma\xi=\inv{1-\theta^2}~~.
\end{formula}

\noi Thus the correlation length diverges at $\theta=1$ (for $N\to\infty$).

\subsection{First Critical Point: $\theta=1$}

Around the critical point at $\theta=1$ there is competition between
the linear and quadratic terms in the expansion of the potential for small $x$

\begin{formula}{}
V(x)=x\log\frac{n_b+1}{n_b+\theta^2}
+\frac16x^2\frac{\theta^4}{\theta^2+n_b}+\O(x^3)~~.
\end{formula}

\noi Expanding in $\theta^2-1$ we get

\begin{formula}{}
V(x)=\frac{1-\theta^2}{1+n_b}x+\frac1{6(1+n_b)}x^2+\O\(x^3,(\theta^2-1)^2\)~~.
\end{formula}

\noi  Near  the  critical  point,  \ie\  for  $(1-\theta^2)\sqrt{N}\ll1$,   the
quadratic term dominates, so the average value $\<x\>$ as  well  as  the  width
$\sigma_x$ becomes of $\O(1/\sqrt N)$ instead of $\O(1/N)$.

Let us therefore introduce two scaling variables $r$ and $\alpha$ through

\begin{formula}{Scales}
x=r\sqrt{3(1+n_b)\over N}~,~~
\theta^2-1=\alpha\sqrt{1+n_b\over3N}~~,
\end{formula}

\noi so that the probability distribution in terms of these variables
becomes a Gaussian on the half-line, \ie\

\begin{formula}{}
p^0(r)=\frac1{Z_r}e^{-\frac12(r-\alpha)^2}
\end{formula}

\noi with

\begin{formula}{}
Z_r=\int_0^\infty dr\;e^{-\frac12(r-\alpha)^2}
=\sqrt{\frac\pi2}\(1+\erf\(\frac\alpha{\sqrt2}\)\)~~.
\end{formula}

\noi From this we obtain

\begin{formula}{}
\<r\>=\alpha+\frac{d\log Z_r}{d\alpha}~,~~~
\sigma^2_r=\frac{d\<r\>}{d\alpha}~~.
\end{formula}

\noi For $\alpha=0$ we have explicitly

\begin{formula}{}
\<x\>=\sqrt{12(1+n_b)\over \pi N}~,~~
\sigma^2_x=\frac{6(n_b+1)}{N}\(\frac12-\frac1\pi\)~~.
\end{formula}

This leads to the following equation for $u(r)$

\begin{formula}{Critical}
\rho u=r(r-\alpha)\frac{du}{dr}-\frac{d}{dr}\left[r\frac{du}{dr}\right]
{}~~,
\end{formula}

\noi where

\begin{formula}{}
\rho=\lambda\sqrt{\frac{3N}{1+n_b}}=\left\<r\(\frac{du}{dr}\)^2\right\>_0
{}~~.
\end{formula}

\noi This eigenvalue problem has no simple solution.

We know, however, that $u(r)$ must change sign once, say  at  $r=r_0$.  In  the
neighborhood of the sign change we have $u\simeq r-r_0$ and, inserting  this
into
(\ref{Critical}) we get $r_0=(\alpha+\sqrt{4+\alpha^2})/2$ and
$\rho=\sqrt{4+\alpha^2}$ such that

\begin{formula}{CritXi}
\gamma\xi=\sqrt{\frac{3N}{(1+n_b)(4+\alpha^2)}}~~.
\end{formula}

\subsection{Maser Phase: $1<\theta<\theta_1\simeq4.603$}

In the region above the transition at $\theta=1$  the  mean  occupation  number
$\<n\>$ grows proportionally with the pumping rate $N$, so in this  region  the
cavity acts as a maser. There is a single minimum of  the  effective  potential
described by the branch $x_0(\theta)$, defined by the  region  $0<\phi<\pi$  in
Eq. (\ref{Extrema}). We find  for  $N\gg1$  to  a  good  approximation  in  the
vicinity of the minimum a Gaussian behavior

\begin{formula}{}
p^0(x)=\sqrt{\frac{NV''(x_0)}{2\pi}}e^{-\frac N2 V''(x_0)(x-x_0)^2}~~,
\end{formula}

\noi where

\begin{formula}{}
V''(x_0)=\frac{1-q'(x_0)}{x_0(1+n_b)}~~.
\end{formula}

\noi Hence for $(\theta^2-1)\sqrt N\gg1$ we have a
mean value  $\<x\>_0=x_0$  and
variance $\sigma^2_x=1/NV''(x_0)$. To find the  next-to-leading  eigenvalue  in
this case we introduce the scaling variable  $r=\sqrt{NV''(x_0)}(x-x_0)$, which
has zero mean and unit variance for large $N$. Then  Eq.~(\ref{uEigen1})  takes
the form (in the continuum limit $N\to\infty$)

\begin{formula}{}
\lambda u=(1-q'(x_0))\(r\frac{du}{dr}-\frac{d^2u}{dr^2}\)~~.
\end{formula}

\noi This is the differential equation for Hermite polynomials. The eigenvalues
are  $\lambda_n=n(1-q'(x_0))$,  $n=0,1,\ldots$, and grow linearly with $n$.
This may be observed in Figure \ref{FigXi}. The  correlation
length becomes

\begin{formula}{}
\xi=\inv{1-q'(x_0)}=\inv{1-\phi\cot\phi}\quad\mbox{for $0<\phi<\pi$}~~.
\end{formula}

\noi As in the thermal phase, the correlation length is independent of $N$
(for
large $N$).


\subsection{Mean Field Calculation}

We shall now use a mean field method to get an expression for  the  correlation
length in both the thermal and maser phases and in the critical region. We find
from the time-dependent probability  distribution  (\ref{Maser})  the
following
{\it exact} equation for the average photon occupation number:

\begin{formula}{}
\frac1\gamma\frac{d\<n\>}{dt}=N\<q_{n+1}\>+n_b-\<n\>
{}~~,
\end{formula}

\noi or with $\Delta x=1/N$

\begin{formula}{}
\frac1\gamma\frac{d\<x\>}{dt}=\<q(x+\Delta x)\>+n_b\Delta x-\<x\>
{}~~.
\end{formula}

\noi We shall ignore the fluctuations of $x$ around its mean value and
simply replace this by

\begin{formula}{}
\frac1\gamma\frac{d\<x\>}{dt}=q(\<x\>+\Delta x)+n_b\Delta x-\<x\>
{}~~.
\end{formula}

\noi This is certainly a good approximation in the limit  of  $N\to\infty$  for
the maser phase because the relative fluctuation $\sigma_x/\<x\>$  vanishes  as
$\O(1/\sqrt N)$ here, but it is of dubious validity in the thermal phase, where
the relative  fluctuations  are  independent  of  $N$.  Nevertheless,  we  find
numerically that the mean field description is  rather  precise  in  the  whole
interval $0<\theta<\theta_1$.

The fixed point $x_0$ of the above equation satisfies the mean field equation

\begin{formula}{}
x_0=q(x_0+\Delta x)+n_b\Delta x
{}~~,\end{formula}

\noi which may be solved in parametric form as

\begin{formulas}{}
x_0&=&\sin^2\phi+n_b\Delta x~~,\\
\theta&=&\displaystyle\frac\phi{\sqrt{\sin^2\phi+(1+n_b)\Delta x}}~~.
\end{formulas}

\noi We notice here that there is a maximum region of existence for any
branch of the solution. The maximum is roughly given by $\theta_k^{\max}
=(k+1)\pi\sqrt{N/(1+n_b)}$.

\begin{figure}[htb]
\unitlength=1mm
\begin{picture}(140,90)(0,0)
\includegraphics{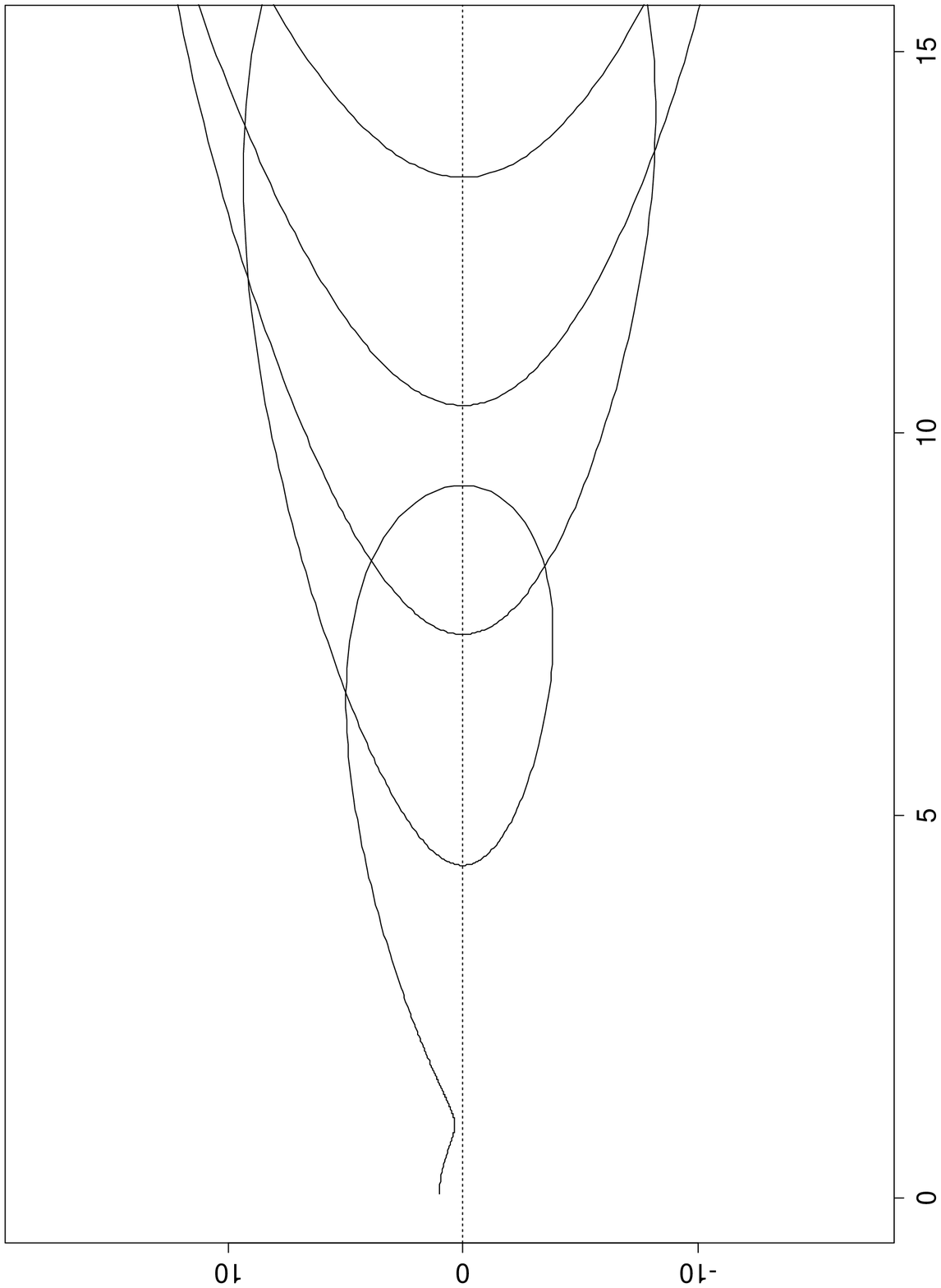}
\put(10,47){\small$\lambda$}
\put(30,80){\small\parbox{30mm}{$N=10$\\$n_b=0.15$}}
\put(80,3){\small$\theta$}
\end{picture}
\figcap{Mean field solution for the sub-leading eigenvalue.}
\label{FigMeanLambda}
\end{figure}

For small perturbations $\<x\>=x_0+\epsilon$ we find the equation of motion

\begin{formula}{}
\frac1\gamma\frac{d\epsilon}{dt}=-(1-q'(x_0+\Delta x))\epsilon
{}~~,\end{formula}

\noi from which we estimate the leading eigenvalue

\begin{formula}{}
\lambda=1-q'(x_0+\Delta x)
=1-\frac{\phi\sin\phi\cos\phi}{\sin^2\phi+(1+n_b)\Delta x}
{}~~.\end{formula}

\noi In Figure \ref{FigMeanLambda} this solution is  plotted  as  a  function  of
$\theta$. Notice that it takes negative  values  in  the  unstable  regions  of
$\phi$. This eigenvalue does not vanish at the critical point $\theta=1$  which
corresponds to

\begin{formula}{}
\phi\simeq\phi_0=\(\frac{3(1+n_b)}N\)^{\frac14}
{}~~,
\end{formula}

\noi but only reaches a small value

\begin{formula}{}
\lambda\simeq2\sqrt{\frac{1+n_b}{3N}}~~,
\end{formula}

\noi which agrees exactly with the previously obtained result (\ref{CritXi}).
Introducing the scaling variable  $\alpha$  from  (\ref{Scales})  and  defining
$\psi=(\phi/\phi_0)^2$    we    easily  get

\begin{formulas}{}
\alpha=(\psi^2-1)/\psi~~,\\
r=\psi~~,\\
\rho=(\psi^2+1)/\psi~~,\\
\end{formulas}

\noi and after eliminating $\psi$

\begin{formulas}{}
r=\frac12(\alpha+\sqrt{\alpha^2+4})~~,\\
\rho=\sqrt{4+\alpha^2}~~,\\
\end{formulas}

\noi which agrees with the previously obtained results.

\begin{figure}[htb]
\unitlength=1mm
\begin{picture}(140,90)(0,0)
\includegraphics{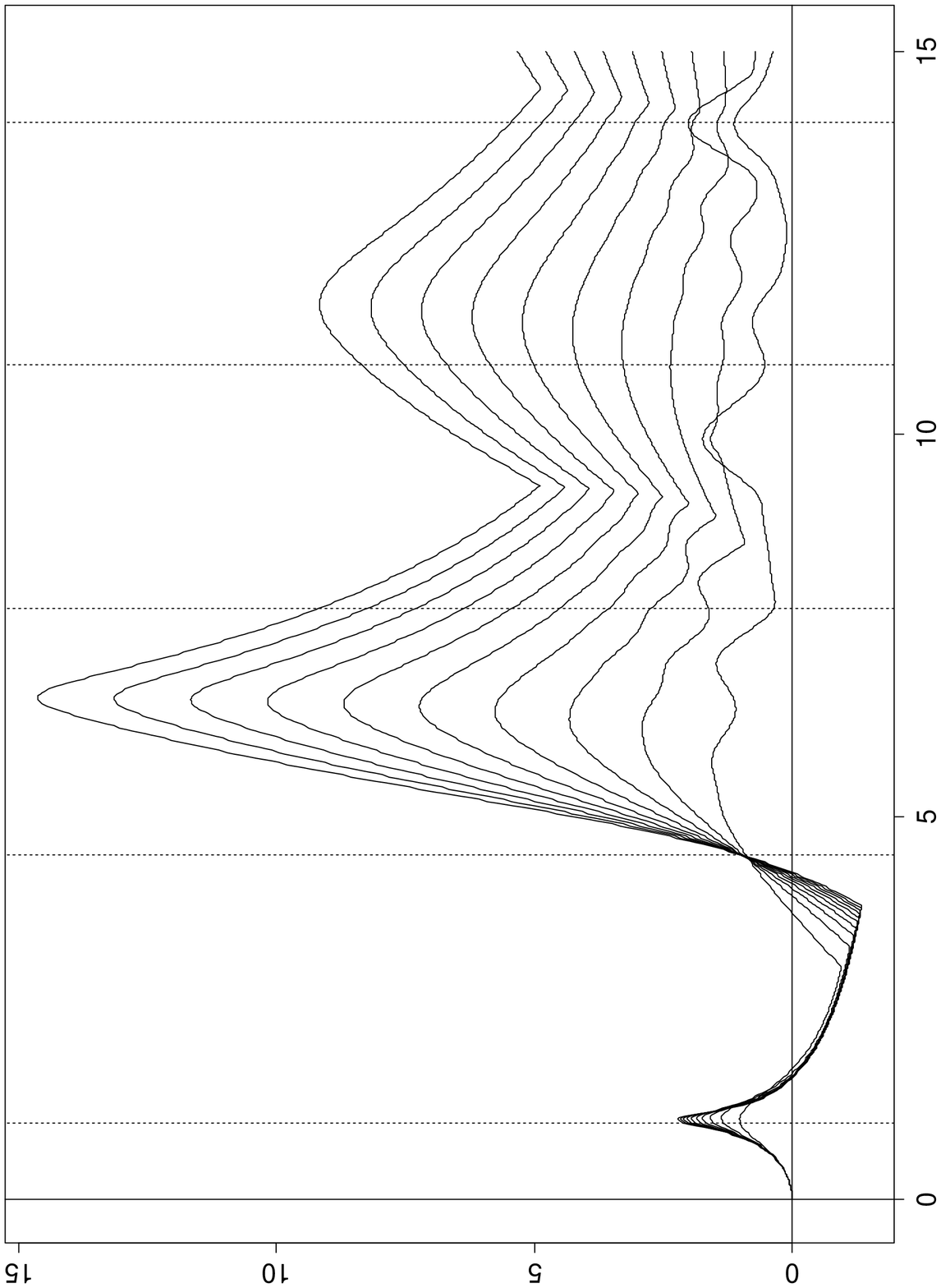}
\put(75,0){\small$\theta$}
\put(5,53){\small$\log(\gamma\xi)$}
\end{picture}
\figcap{The logarithm of the correlation length as a function of  $\theta$  for
various values of $N$
$(10,20,\ldots,100)$. We have $n_b=0.15$  here.  Notice  that
for $\theta>\theta_1$ the logarithm of the correlation  length  grows  linearly
with $N$ for large $N$.  The vertical lines indicate
$\theta_0=1,~\theta_1=4.603,
{}~ \theta_2=7.790, ~\theta_3=10.95$ and $\theta_4=14.10$.
}
\label{FigCritic}
\end{figure}

\subsection{The First Critical Phase:
$4.603\simeq\theta_1<\theta<\theta_2\simeq7.790$}\label{FirstCritical}

We now turn to the first phase in which the effective potential has two  minima
($x_0, x_2$) and a maximum ($x_1$) in between (see Figure \ref{FigEffPot} in
Section~\ref{s:Extrema}).
In this case there is competition between the two  minima
separated by the barrier and for $N\to\infty$ this barrier makes the relaxation
time to equilibrium  exponentially long.  
Hence  we  expect  $\lambda_1$  to  be
exponentially small for large $N$ 
(see Figure \ref{FigCritic})

\begin{formula}{Asymp}
\lambda_1=Ce^{-\eta N}~~,
\end{formula}

\noi where $C$ and $\eta$ are independent of $N$. It is the  extreme  smallness
of the sub-leading  eigenvalue  that  allows  us  to  calculate  it  with  high
precision.

For large $N$ the probability  distribution  consists  of  two  well-separated
narrow maxima, each of  which  is  approximately  a  Gaussian.  We  define  the
\apriori\ probabilities for each of the peaks

\begin{formula}{}
P_0=\sum_{0\le x<x_1}\Delta x\; p^0(x)=\frac{Z_0}{Z}~~,
\end{formula}

\noi and

\begin{formula}{}
P_2=\sum_{x_1\le x<\infty}\Delta x\; p^0(x)=\frac{Z_2}{Z}~~.
\end{formula}

\noi The $Z$-factors are

\begin{formula}{}
Z_0=\sum_{x=0}^{x_1}\Delta x\;e^{-NV(x)}
\simeq e^{-NV_0}\sqrt{\frac{2\pi}{NV''_0}}~~,
\end{formula}

\noi and

\begin{formula}{}
Z_2=\sum_{x=x_1}^\infty\Delta x\;e^{-NV(x)}
\simeq e^{-NV_2}\sqrt{\frac{2\pi}{NV''_2}}~~,
\end{formula}

\noi with $Z=Z_0+Z_2$. The probabilities satisfy of course $P_0+P_2=1$ and
we have

\begin{formula}{}
p^0(x)=P_0p^0_0(x)+P_2p^0_2(x)~~,
\end{formula}

\noi where $p^0_{0,2}$ are individual probability distributions with maximum at
$x_{0,2}$.  The  overlap  error  in  these  expressions  vanishes  rapidly  for
$N\to\infty$, because  the  ratio  $P_0/P_2$  either  converges  towards  0  or
$\infty$ for $V_0\ne V_2$. The transition from one peak being  the  highest  to
the other peak being the highest occurs when the two maxima coincide, \ie\  at
$\theta\simeq7.22$  at  $N=10$,  whereas  for  $N=\infty$   it   happens  at
$\theta\simeq6.66$. At this point the correlation length is also maximal.

Using this formalism, many quantities may be evaluated in the limit
of large $N$. Thus for example

\begin{formula}{}
\<x\>_0=P_0x_0+P_2x_2~~,
\end{formula}

\noi and

\begin{formula}{}
\sigma_x^2=\<(x-\<x\>_0)^2\>_0=\sigma_0^2P_0+\sigma_2^2P_2
+(x_0-x_2)^2P_0P_2 ~~.
\end{formula}

\noi Now there is no direct relation between the variance and  the  correlation
length.

Consider now the expression (\ref{pEigen}),
which shows that since $\lambda_1$
is exponentially small we have an essentially constant $J_n$, except  near  the
maxima of the probability distribution, \ie\ near the minima of the  potential.
Furthermore since  the  right  eigenvector  of  $\lambda_1$  satisfies  $\sum_x
p(x)=0$, we have $0=J(0)=J(\infty)$ so that
\begin{formula}{}
J(x)\simeq
\left\{\begin{array}{ll}
0&0<x<x_0~~,\\
J_1&x_0<x<x_2~~,\\
0&x_2<x<\infty~~.\\
\end{array}\right.
\end{formula}
\noindent  This expression is more accurate away from the minima of the potential,
$x_0$ and $x_2$.

Now it follows  from  Eq.~(\ref{uEq})  that  the  left  eigenvector  $u(x)$  of
$\lambda_1$ must be constant, except near the minimum $x_1$ of the  probability
distribution, where the derivative could be sizeable.
So we conclude that $u(x)$
is constant away from the maximum of the potential. Hence we must approximately
have

\begin{formula}{}
u(x)\simeq
\left\{\begin{array}{ll}
u_0&0<x<x_1~~,\\
u_2&x_1<x<\infty~~.\\
\end{array}\right.
\end{formula}

\noi This expression is more accurate away from the maximum of the potential.

We may now relate the values of $J$ and $u$  by summing Eq.~(\ref{pEigen})
from $x_1$ to infinity

\begin{formula}{}
J_1=J(x_1)=\lambda_1\sum_{x=x_1}^\infty \Delta x\;p^0(x)u(x)
\simeq \lambda_1 P_2u_2~~.
\end{formula}

\noi From Eq.~(\ref{uEq}) we get by summing over the interval between the
minima

\begin{formula}{}
u_2-u_0=\frac{NJ_1}{1+n_b}\sum_{x=x_0}^{x_2}\Delta x  \frac1{xp^0(x)}~~.
\end{formula}

\noi The inverse probability distribution has for $N\to\infty$ a sharp
maximum at the maximum of the potential. Let us define

\begin{formula}{}
Z_1=\sum_{x=x_0}^{x_2} \Delta x\;\frac1xe^{NV(x)}
\simeq \frac1{x_1}e^{NV_1}\sqrt{\frac{2\pi}{N(-V''_1)}}~~.
\end{formula}

\noi Then we find

\begin{formula}{}
u_2-u_0=\frac{NZZ_1J_1}{1+n_b}=\frac{N\lambda_1 Z_1Z_2u_2}{1+n_b}~~.
\end{formula}

\noi But $u(x)$ must be univariate, \ie\

\begin{formulas}{}
u_0P_0+u_2P_2=0~~,\\
u_0^2P_0+u_2^2P_2=1~~,\\
\end{formulas}

\noi from which we get

\begin{formulas}{}
u_0=-\sqrt{\displaystyle{\frac{P_2}{P_0}}}
=-\sqrt{\displaystyle{\frac{Z_2}{Z_0}}}~~,\\\\
u_2=\sqrt{\displaystyle{\frac{P_0}{P_2}}}
=\sqrt{\displaystyle{\frac{Z_0}{Z_2}}}~~.\\
\end{formulas}

\noi Inserting the above solution we may solve for $\lambda_1$

\begin{formula}{LambdaCrit}
\lambda_1=\frac{1+n_b}{N}\frac{Z_0+Z_2}{Z_0Z_1Z_2}~~,
\end{formula}

\noi or more explicitly

\begin{formula}{glorylam}
\lambda_1=\frac{x_1(1+n_b)}{2\pi}\sqrt{-V_1''}
\(\sqrt{V_0''}e^{-N(V_1-V_0)}+\sqrt{V_2''}e^{-N(V_1-V_2)}\)~~.
\end{formula}

\noi
Finally we may read off the coefficients $\eta$ and $C$ from Eq.~(\ref{Asymp}).
We get

\begin{formula}{Eta}
\eta=\left\{\begin{array}{ll}
V_1-V_0&{\rm for}~~ V_0>V_2~~,\\
V_1-V_2&{\rm for}~~ V_2>V_0~~,\\
\end{array}\right.
\end{formula}

\noi and

\begin{formula}{}
C=\frac{x_1(1+n_b)}{2\pi}\sqrt{-V_1''}
\left\{\begin{array}{ll}
\sqrt{V_0''}&{\rm for}~~ V_0>V_2~~,\\
\sqrt{V_2''}&{\rm for}~~ V_2>V_0~~.\\
\end{array}\right.
\end{formula}

\noindent This expression is nothing  but the result of a barrier  penetration  of  a
classical statistical process \cite{Schuss80}. We have derived it in
detail in order to get all the coefficients right.

\begin{figure}[htb]
\unitlength=1mm
\begin{picture}(140,90)(0,0)
\includegraphics{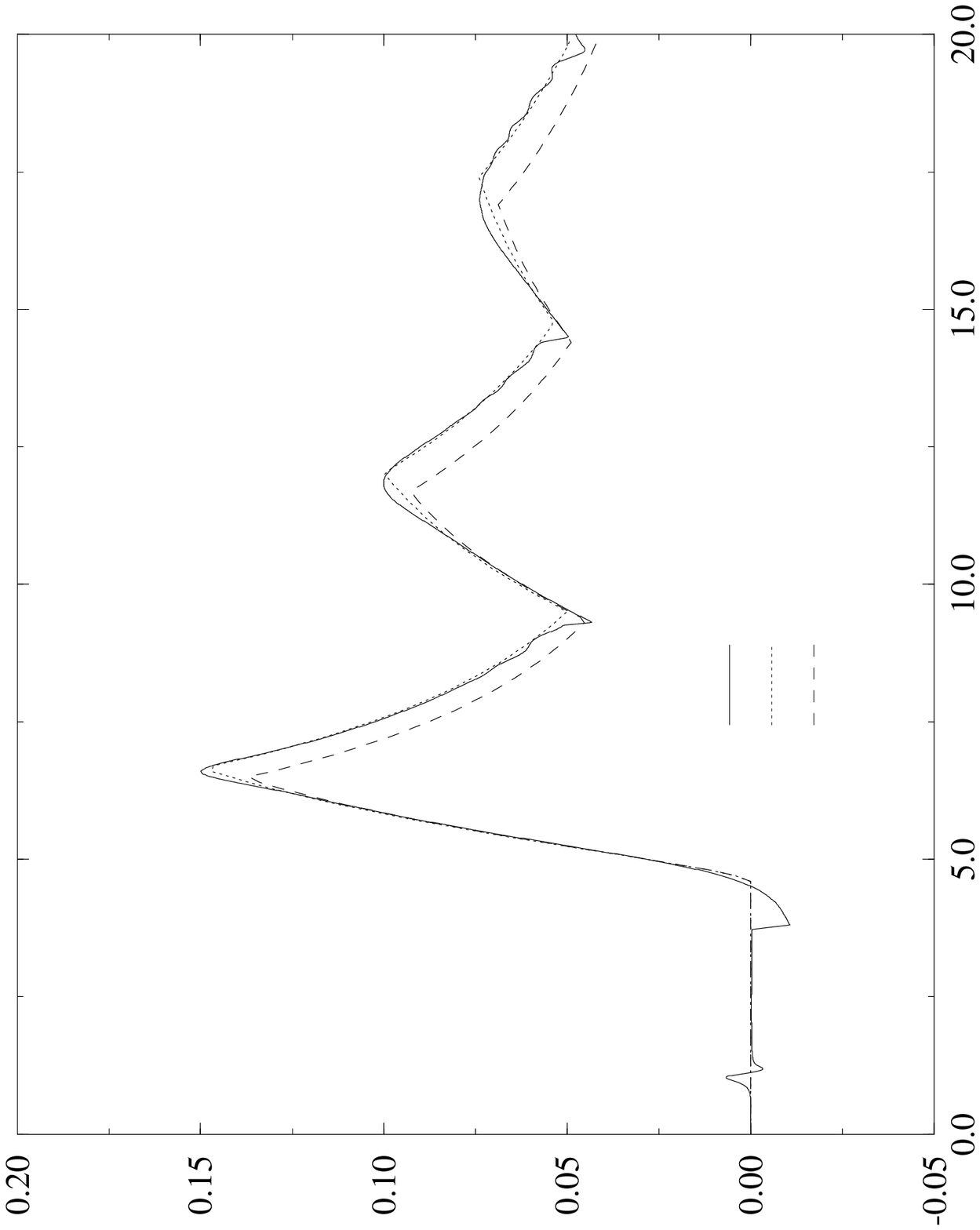}
\put(75,0){\small$\theta$}
\put(10,50){\small$\eta$}
\put(78,25){\small$\log(\xi_{90}/\xi_{70})/20$}
\put(78,20.5){\small Barrier from $V(x)$}
\put(78,16){\small Barrier from Fokker--Planck}
\end{picture}
\figcap{Comparing the barrier height from the potential $V(x)$ with  the  exact
correlation length and the barrier from an approximate Fokker--Planck formula.}
\label{f-barr}
\end{figure}

It  is  interesting  to  check  numerically  how  well  \eq{glorylam}  actually
describes the correlation length.  The  coefficient  $\eta$  is  given
by  Eq.~(\ref{Eta}), and we have numerically computed
the  highest  barrier  from  the
potential $V(x)$ and compared it with an exact calculation in Figure \ref{f-barr}. The
exponent $\eta$ is extracted by comparing two values of the correlation length,
$\xi_{70}$ and $\xi_{90}$, for large values  of  $N$  (70  and  90),
  where  the
difference in the prefactor $C$ should be unimportant.  The  agreement  between
the two calculations is excellent  when  we  use  the  exact  potential.  As  a
comparison  we  also  calculate  the  barrier  height  from  the  approximative
potential     in     the      Fokker--Planck      equation      derived      in
\cite{Filipowicz86,Filipowicz86a}. We find a  substantial  deviation  from  the
exact   value   in    that    case.    It    is    carefully    explained    in
\cite{Filipowicz86,Filipowicz86a} why the Fokker--Planck  potential  cannot  be
expected to give a quantitatively correct  result  for  small  $n_b$.  The
exact
result (solid line) has some  extra  features  at  $\theta=1$  and  just  below
$\theta=\theta_1\simeq 4.603$, due to finite-size effects.

When  the  first  sub-leading  eigenvalue  goes  exponentially   to   zero,   or
equivalently the correlation length grows exponentially, it  becomes  important
to know the density of eigenvalues. If there is an accumulation of  eigenvalues
around 0, the long-time correlation cannot be  determined  by  only  the  first
sub-leading  eigenvalue.  It  is  quite  easy  to  determine  the  density   of
eigenvalues simply by computing them numerically.

\begin{figure}[htb]
\unitlength=1mm
\begin{picture}(140,100)(0,0)
\includegraphics{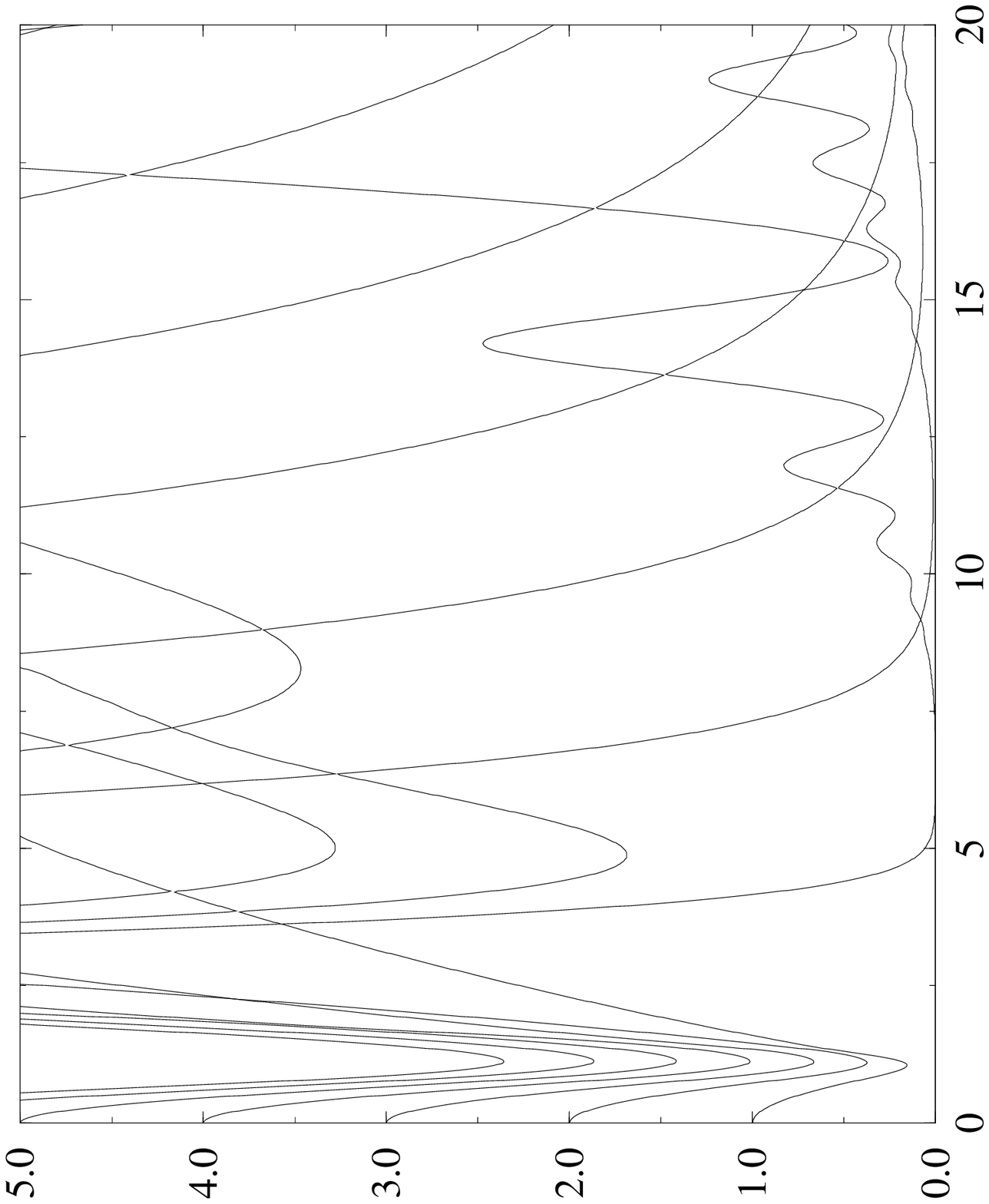}
\put(85,3){\small$\theta$}
\put(10,53){\small$\lambda_{1-7}$}
\end{picture}
\figcap{The first seven sub-leading eigenvalues for $N=50$ and $n_b=0.15$.}
\label{f-lam0-7}
\end{figure}

In Figure \ref{f-lam0-7} we show the first seven sub-leading eigenvalues for $N=50$ and
$n_b=0.15$. It is clear that at the first critical point after the maser  phase
($\theta=\theta_1$) there is only one eigenvalue going to zero.  At  the  next
critical phase $(\theta=\theta_2$) there is one more  eigenvalue  coming  down,
and so on. We find that there is only one exponentially  small  eigenvalue  for
each new minimum in the  potential,  and  thus  there  is  no  accumulation  of
eigenvalues around 0.


\def\st{\sigma_\theta}

\section{Effects of Velocity  Fluctuations}
\seqnoll
\label{Spread}
\begin{flushright}
``{\sl Why, these balls bound; there's noise in it.}"\\
W. Shakespeare
\end{flushright}
The time it takes an atom to  pass  through  the  cavity  is  determined  by  a
velocity filter in front of the cavity. This filter is not perfect  and  it  is
relevant to investigate what a spread  in  flight  time  implies  for  the
statistics of the interaction between cavity  and  beam.  Noisy pump
effects has been discussed before in the literature \cite{schenzle}.
To  be  specific,  we will here
consider the flight time as an independent stochastic variable.  Again,  it  is
more convenient to work with the rescaled variable $\theta$, and we denote  the
corresponding stochastic variable by $\vartheta$.  In  order  to  get  explicit
analytic results we choose the following gamma probability distribution for  positive
$\vartheta$

\begin{formula}{distr}
    f(\vartheta,\alpha,\beta)=\frac{\beta^{\alpha+1}}{\Gamma(\alpha+1)}
    \vartheta^\alpha
    e^{-\beta\vartheta}~~,
\end{formula}

\noi  with  $\beta=\theta/\st^2$   and   $\alpha=\theta^2/\st^2-1$,   so   that
$\<\vartheta\>=\theta$ and $\<(\vartheta-\theta)^2\>=\st^2$. Other choices  are
possible, but are not expected to change the overall qualitative  picture.  The
discrete master equation (\ref{stat1}) for the equilibrium distribution can  be
averaged to yield

\begin{formula}{avemast}
    \< p(t+T)\>=e^{-\gamma L_C T}\< M(\vartheta)\> \< p(t)\>~~,
\end{formula}

\noi The factorization  is  due  to  the  fact  that  $p(t)$  only  depends  on
$\vartheta$ for the preceding atoms,  and  that  all  atoms  are  statistically
independent.      The       effect       is       simply       to       average
$q(\vartheta)=\sin^2(\vartheta\sqrt{x})$ in $M(\vartheta)$, and we get

\begin{formula}{aveq}
    \< q\>=\inv2 \left[1-
      \left(1+\frac{4x\st^4}{\theta^2}\right)^
      {-\frac{\theta^2}{2\st^2}}
      \cos\left(\frac{\theta^2}{\st^2}
        \arctan\left(\frac{2\sqrt{x}\st^2}{\theta}\right)\right)\right]~~.
\end{formula}

\noi This averaged form of $q(\theta)$, which depends on  the  two  independent
variables $\theta/\st$ and $\theta \sqrt{x}$, enters in  the  analysis  of  the
phases in exactly the same way as before. In the limit $\st\goto 0$  we  regain
the original $q(\theta)$, as we should. For  very  large  $\st$  and  fixed
$\theta$, and $\<q\>$ approaches zero.

\subsection{Revivals and Prerevivals}
\label{ss:prerev}
\begin{figure}[p]
\unitlength=1mm
\begin{picture}(140,90)(0,0)
\includegraphics{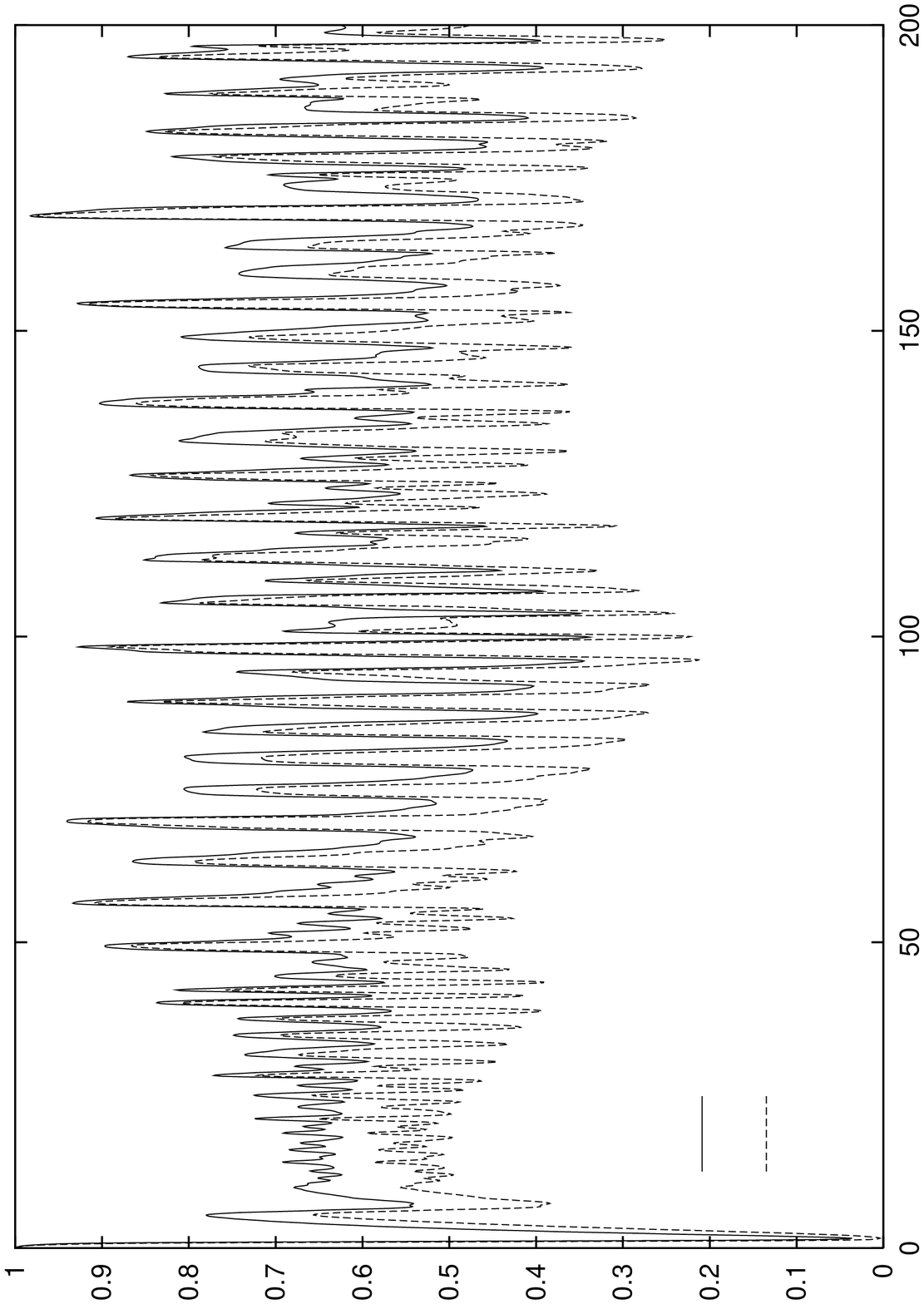}
\put(45,38){\small$\P(+)$}
\put(45,32){\small$\P_0(+,+)$}
\end{picture}

\begin{picture}(140,90)(0,0)
\includegraphics{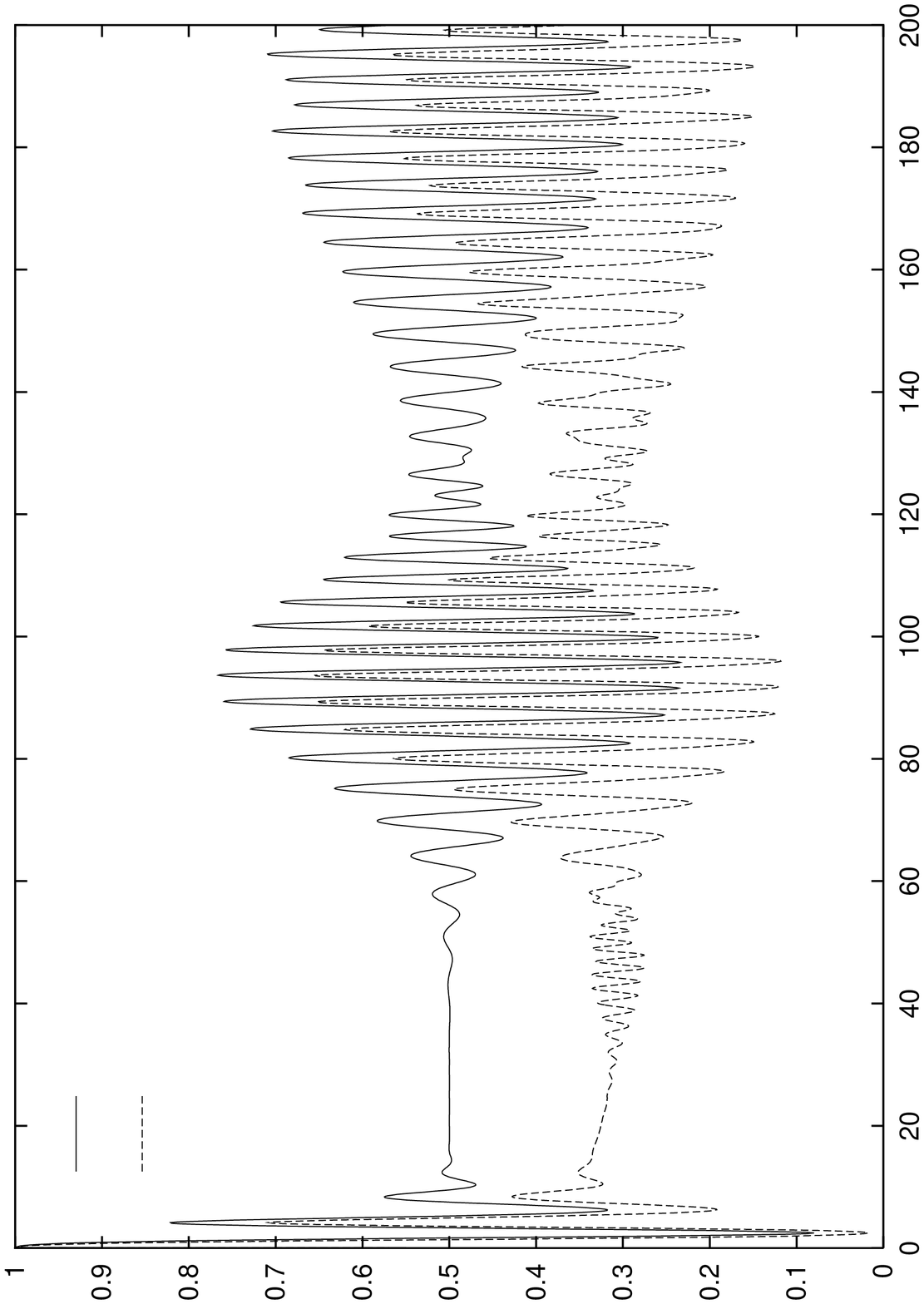}
\put(80,10){\small$\theta$}
\put(45,96){\small$\P(+)$}
\put(45,90){\small$\P_0(+,+)$}
\end{picture}

\figcap{Upper graph: Probabilities of finding one atom, 
or two consecutive ones,  in
the excited state. The flux is given  by  $N=20$  and  the  thermal  occupation
number is $n_b=0.15$. The curves show no evidence for the resonant behavior of
revivals. Lower graph: Presence of  revival  resonances  in  equilibrium  after
averaging the photon distribution over $\theta$.  The  same  parameters  as  in
the upper graph are used but the variance in $\theta$ is now given by $\st^2=10$.}
\label{f:rev0}
\end{figure}

The phenomenon of quantum revival is an essential  feature  of  the  microlaser
system (see e.g. Refs. \cite{Meystre74}--\cite{Filipowicz86b}, and
\cite{Averbukh89a}--\cite{Fleischhauer93}).
The  revivals  are  characterized  by the
reappearance  of  strongly   oscillating
structures in the excitation probability of an outgoing atom which is given  by
Eq.~(\ref{SingleP}):

\begin{formula}{cp}
\P(+)={u^0}^TM(+)p^0=\sum_n (1-q_{n+1}(\theta)) p^0_n~~,
\end{formula}

\noi where $p^0_n$ is the photon distribution (\ref{Equilibrium}) in the cavity
before the atom enters. Here the last equality sign in Eq.(\ref{cp}) is valid only for
$a=1$.
Revivals occur when there is a resonance  between  the
period in $q_n$ and the  discreteness  in  $n$  \cite{Fleischhauer93}.  If  the
photon distribution in the cavity has a sharp peak at $n=n_0$ with  a  position
that does not change appreciably when $\theta$ changes, as for  example  for  a
fixed Poisson distribution, then it is easy  to  see  that  the  first  revival
becomes pronounced in the region of $\theta_{\rm  rev}\simeq  2\pi\sqrt{n_0N}$.
For the equilibrium distribution without any spread in the velocities we do not
expect any dramatic signature of revival, the reason being that  the  peaks  in
the equilibrium distribution $p^0_n(\theta)$ move  rapidly  with  $\theta$.  In
this context it is also natural to study the short-time correlation between two
consecutive atoms, or the probability of finding two consecutive atoms  in  the
excited level \cite{PaulRichter91}. This quantity is given by

\begin{formulas}{}
\P_0(+,+)&=&{u^0}^TM(+)(1+L_C/N)^{-1}M(+)p^0\\
&=&\displaystyle{\sum_{n,m}}(1-q_{n+1}(\theta))(1+L_C/N)^{-1}_{nm}
(1-q_{m+1}(\theta)) p^0_m~~,
\end{formulas}

\noi defined in Eq.~(\ref{kProb}). Here again the last equality sign
is valid only for $a=1$.
 In Appendix \ref{dampingmatrix} we  give  an
analytic  expression  for  the  matrix   elements   of   $(1+L_C/N)^{-1}$.   In
Figure \ref{f:rev0} (upper graph) we present $\P(+)$ and $\P_0(+,+)$ 
for typical values of  $N$  and
$n_b$ \footnote{Notice that we have corrected for a numerical error in
 Figure 10 of Ref.\cite{Elmforsetal95b}.}.

If we on the other hand smear out the equilibrium distribution sufficiently  as
a function of $\theta$, revivals will again appear. The experimental  situation
we envisage is that the  atoms  are  produced  with a
certain  spread  in  their
velocities. The statistically averaged stationary photon  distribution  depends
on the spread. After the  passage  through  the  cavity  we  measure  both  the
excitation level and the speed of the atom. There is thus no averaging  in  the
calculation of $\P(+)$ and $\P_0(+,+)$, but these quantities now also depend on
the actual value $\vartheta$ for each atom. For  definiteness  we  select  only
those atoms that fall in a narrow range around the average value  $\theta$,  in
effect putting in a sharp velocity filter {\it after} the interaction. The result for
an averaged photon distribution is presented  in  Figure \ref{f:rev0}  (lower  graph),
where clear signs of revival are found. We also  observe  that  in  $\P_0(+,+)$
there are {\em prerevivals}, occurring
for a value of $\theta$ half as large  as
for the usual revivals. Its origin is obvious since in  $\P_0(+,+)$  there  are
terms containing $q_n^2$ that vary with the double of the frequency of
$q_n$.
It is clear from  Figure \ref{f:rev0}  (lower  graph) that the
addition of noise to the system can enhance the signal. This
observation suggest a connection to noise synchronization in
non-linear systems \cite{maritan}. The micromaser system can also be
used to study the phenomena of stochastic resonance (see
Ref.\cite{buchl} and references therein) which, however, corresponds to a
different mechanism for signal-noise amplification due to the presence
of additional noise in a physical system.

\subsection{Phase Diagram}
\label{ss:avephd}

The different phases discussed in Section \ref{Phasestructure} depend  strongly
on the structure of the effective potential. Averaging over $\theta$ can easily
change this structure and the phases. For instance, averaging with large  $\st$
would typically wash out some of the minima and lead to  a  different  critical
behavior. We shall determine a two-dimensional phase diagram in the parameters
$\theta$ and $\st$ by finding the lines where new minima occur  and  disappear.
They are determined by the equations

\begin{formulas}{phdeq}
    \< q\>=x~~,\\
    {\displaystyle\frac{d\< q\>}{dx}}=1~~.
\end{formulas}

\begin{figure}[htb]
\unitlength=1mm
\begin{picture}(140,110)(0,0)
\includegraphics{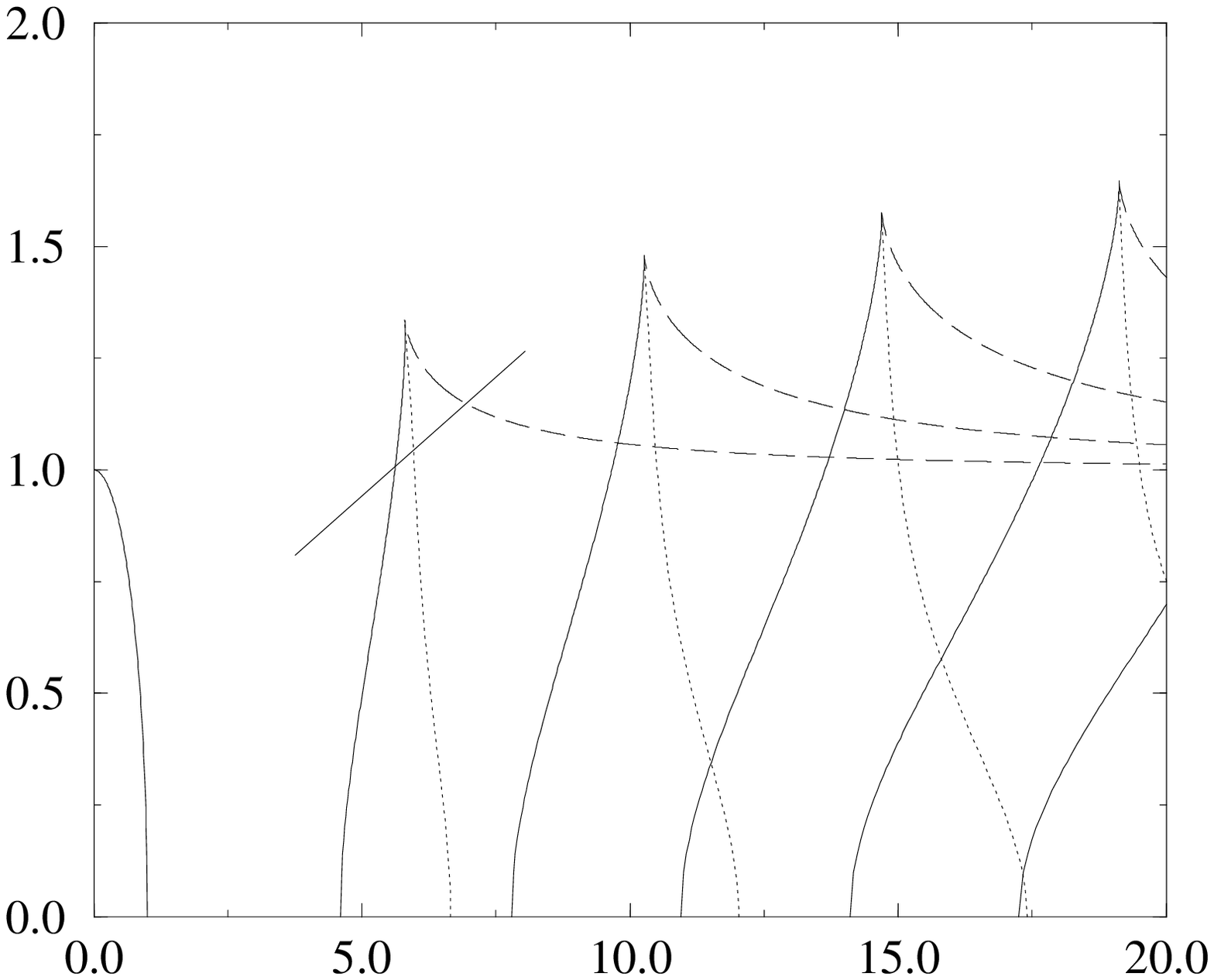}
\end{picture}
\put(-60,5){$\theta$}
\put(-130,70){$\st$}
\put(-85,80){\small $a$}
\put(-100,50){\small $b$}
\put(-71,76){\small $c$}
\figcap{Phase diagram in the $\theta$--$\st$ plane. The  solid  lines  indicate
where new minima in the effective potential emerge. In the lower  left  corner
there is only one minimum at $n=0$, this is the  thermal  phase.  Outside  that
region there is always a minimum for non-zero $n$ implying that the cavity acts
as a maser. To the right of the solid line starting at  $\theta\simeq4.6$,  and
for not too large $\st$, there are two or more minima and thus the  correlation
length grows exponentially with the flux. For increasing $\st$ minima disappear
across the dashed lines, starting with those at small  $n$.  The  dotted  lines
show where the two lowest minima are equally deep.}
\label{f:avephd}
\end{figure}

The phase boundary between the thermal and the maser phase
 is determined by the effective potential  for
small  $x$.  The   condition   $\theta^2=1$   is   now   simply   replaced   by
$\<\vartheta^2\>=\theta^2+\st^2=1$, which also follows from the  explicit  form
of $\<q\>$ in \eq{aveq}. The transitions from the maser phase to  the
critical phases are determined numerically and presented in Figure \ref{f:avephd}. The
first line starting from $\theta\simeq4.6$ shows where the  second  minimum  is
about to form, but exactly on this line it is only an inflection point. At  the
point $a$ about $\st\simeq1.3$ it disappears,  which  occurs  when  the  second
minimum fuses with the first minimum. From the cusp at point $a$ there is a new
line (dashed) showing where the first  minimum  becomes  an  inflection  point.
Above the cusp at point $a$ there is only one minimum.  Going  along  the  line
from point $b$ to $c$ we thus first have one minimum,  then  a  second  minimum
emerges, and finally the first minimum disappears before we  reach  point  $c$.
Similar things happens at the other cusps,
 which represent the fusing points for
other minima. Thus, solid lines show where a new minimum emerges for large  $n$
($\sim N$) as $\theta$ increases, while  dashed  lines  show  where  a  minimum
disappears for small $n$ as $\st$ increases. We have also indicated (by  dotted
lines) the first-order maser transitions where  the  two  dominant  minima  are
equally deep. These are the lines where $\xi$ and  $Q_f$  have  peaks  and
$\<n\>$ makes a discontinuous jump.

\section{Finite-Flux Effects}
\label{finite}
\seqnoll
\begin{flushright}
``{\sl In this age people are experiencing a delight, the
tremendous\\
 delight that you can guess how nature will work \\
in a new situation never seen before.}"\\
R. Feynman
\end{flushright}
So far, we have mainly discussed characteristics of the large flux limit. These
are  the  defining   properties   for   the   different   phases   in   Section
\ref{Phasestructure}. The parameter that controls finite flux  effects  is  the
ratio between the period of oscillations in the potential and the size  of  the
discrete steps in $x$. If $q=\sin^2(\theta\sqrt{x})$ varies slowly over $\Delta
x=1/N$, the continuum limit is usually a good approximation, while  it  can  be
very poor in the opposite case. In the discrete case there exist,  for  certain
values of $\theta$, states that cannot be pumped above  a  certain  occupation
number since $q_n=0$ for that level. This effect is not seen in  the  continuum
approximation.   These   states    are    called    {\em    trapping    states}
\cite{Filipowicz86c} and  we  discuss  them  and  their  consequences  in  this
section.

The continuum approximation starts breaking down for small photon numbers  when
$\theta\simgeq  2\pi\sqrt{N}$,  and  is  completely  inappropriate   when   the
discreteness is manifest for  all  photon  numbers lower  than  $N$,  \ie\  for
$\theta\simgeq    2\pi    N$.    In    that    case     our     analysis     in
Section~\ref{Phasestructure} breaks down and the system  may  occasionally,
  for
certain values of $\theta$, return to a non-critical phase.

\begin{figure}[htb]
\unitlength=1mm
\begin{picture}(140,90)(0,0)
\includegraphics{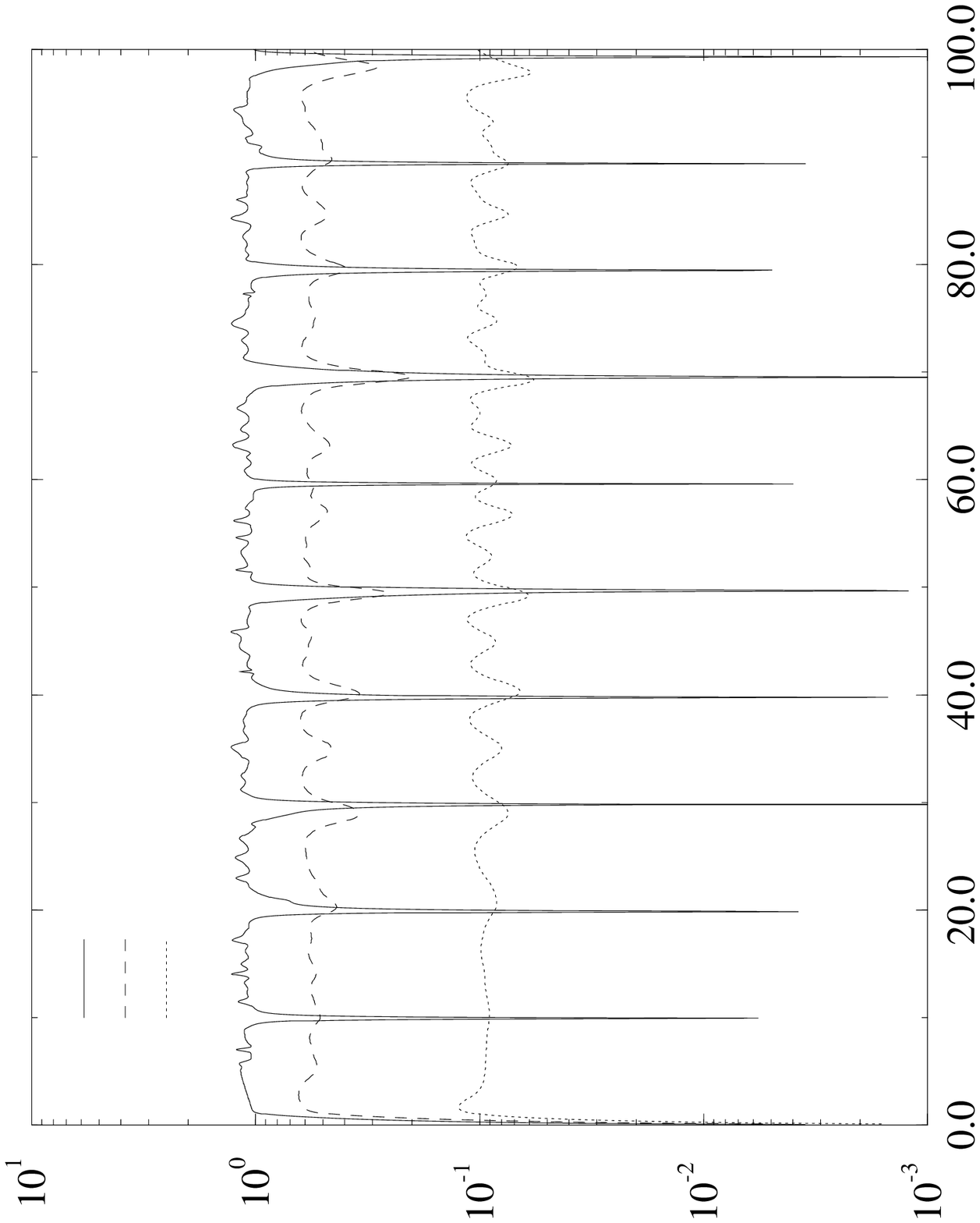}
\put(80,-1){$\theta$}
\put(9,53){$d_{L^2}$}
\put(48,83){\footnotesize$n_b=0.0001$}
\put(48,79){\footnotesize$n_b=1.0$}
\put(48,75){\footnotesize$n_b=10.0$}
\end{picture}
\figcap{Distance between the initial probability
distribution $p_n(0)$ and $p_n(\theta)$ measured by $d_{L^2}(\theta)$
in \eq{dist}.}
\label{f-dist}
\end{figure}

\subsection{Trapping States}\label{Trapping}

The  equilibrium  distribution  in  \eq{Equilibrium}  has  peculiar  properties
whenever $q_m=0$ for some value of $m$, in particular when
$n_b$ is  small,  and
dramatically    so    when    $n_b=0$.    This    phenomenon    occurs     when
$\theta=k\pi\sqrt{N/m}$ and is called a trapping state.  When  it  happens,  we
have $p_n=0$ for all $n\geq m$ (for $n_b=0$). The physics behind  this  can  be
found in \eq{Pump},
 where $M(-)$ determines the pumping of  the  cavity  by  the
atoms. If $q_m=0$ the cavity cannot be pumped above  $m$  photons  by  emission
from the  passing  atoms.  For  any
non-zero  value  of  $n_b$  there  is  still  a
possibility for thermal fluctuation above $m$ photons and $p_n\neq 0$ even  for
$n\geq m$. The effect of trapping is lost in  the  continuum  limit  where  the
potential  is  approximated   by   Eq.~(\ref{ContPot}).   Some   experimental
consequences of trapping states  were  studied  for  very  low  temperature  in
\cite{Meystre88} and it was stated that  in  the  range  $n_b=0.1$--1.0  no
experimentally measurable effects were present. Recently trapping states have actually been
observed in the micromaser system \cite{trapping99}. Below we show that 
there are  clear signals of trapping states in the correlation length  even  for
$n_b=1.0$.

\begin{figure}[htb]
\unitlength=1mm
\begin{picture}(140,60)(0,0)
\includegraphics{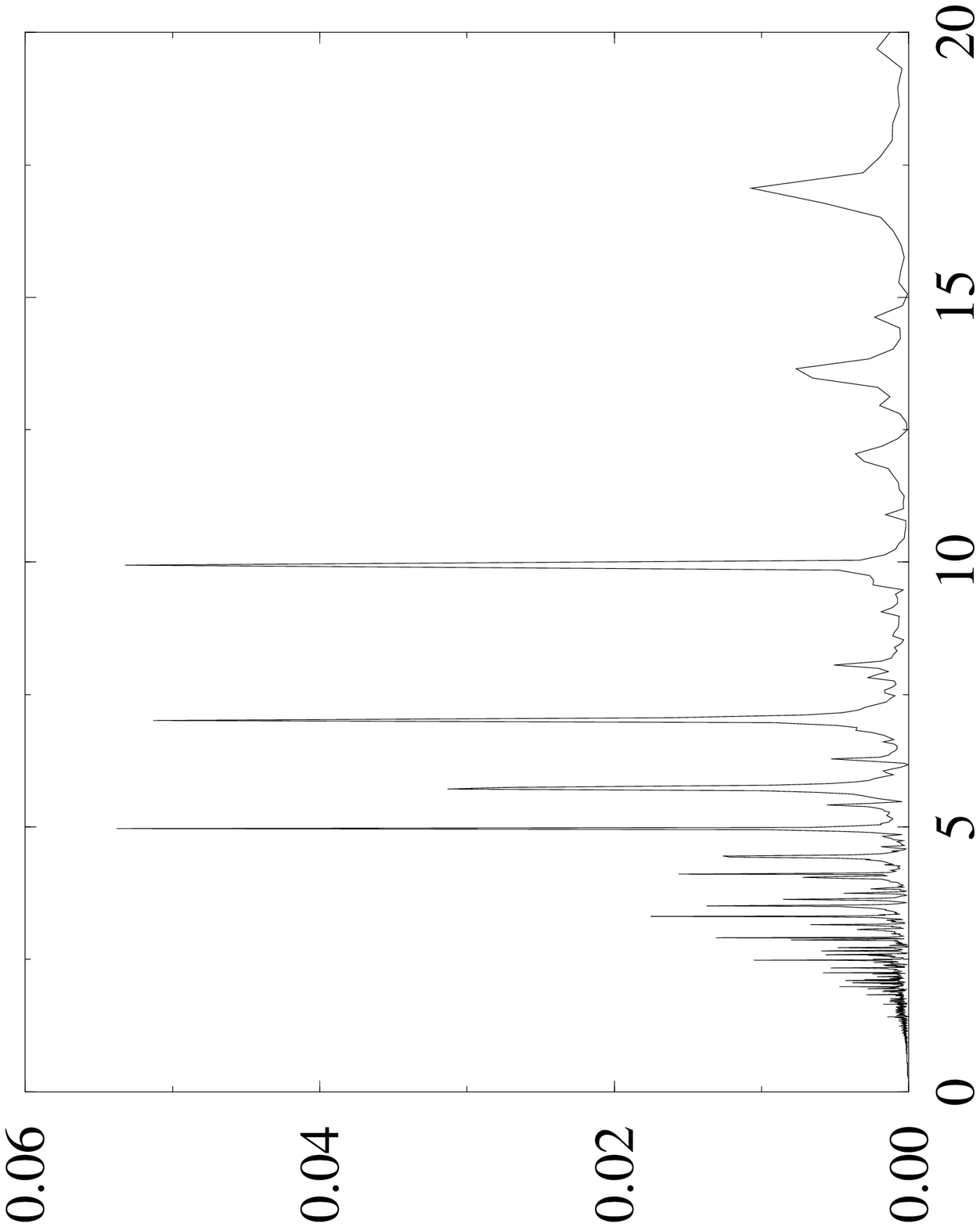}
\put(9,30){\small$\abs{\cF d_{L^2}(\theta)}$}
\end{picture}

\begin{picture}(140,70)(0,0)
\includegraphics{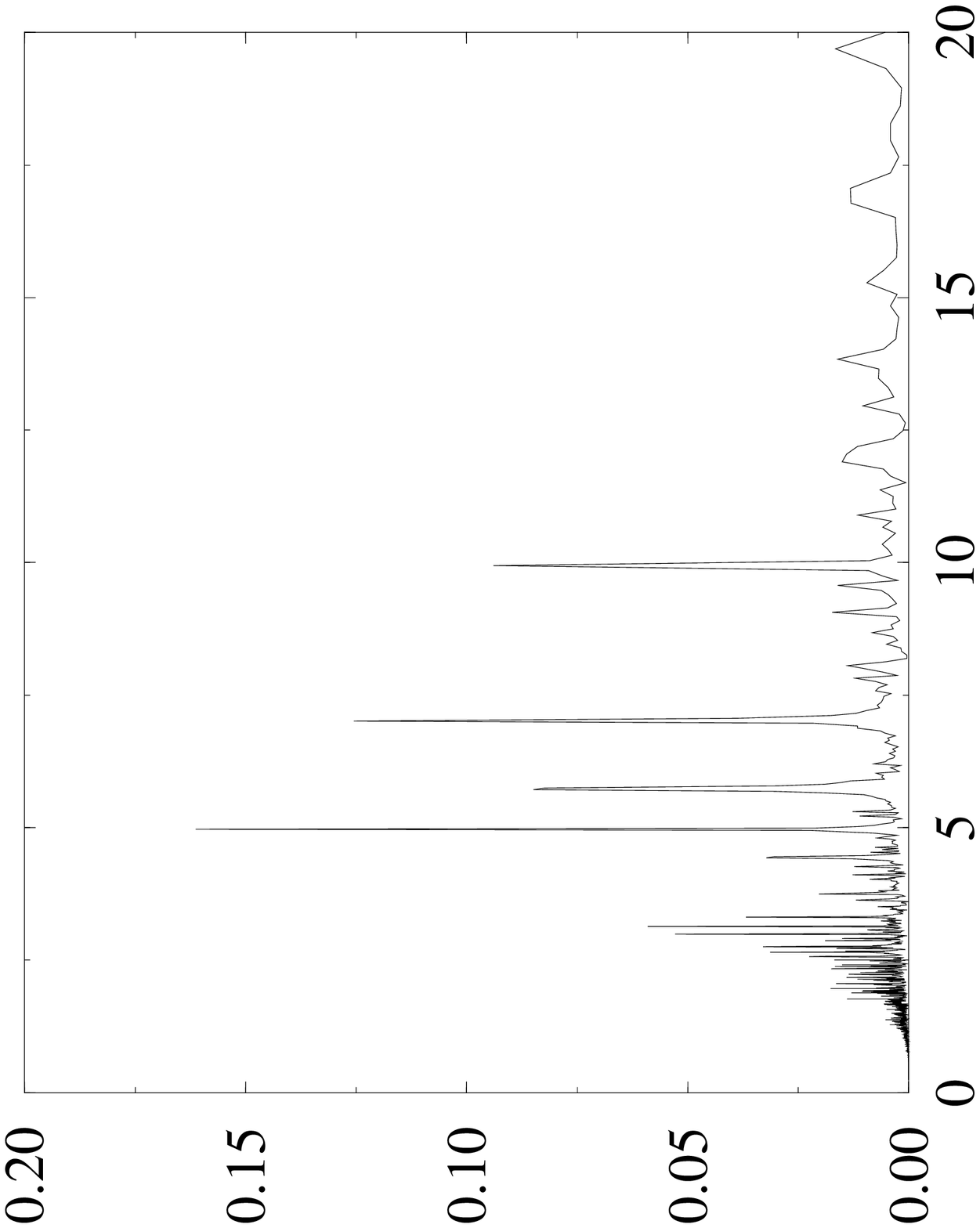}
\put(80,5){\small$\Delta\theta$}
\put(9,48){\small$\abs{\cF \xi(\theta)}$}
\end{picture}
\figcap{Amplitudes of Fourier modes  of  $d_{L^2}(\theta)$  (upper  graph)  and
$\xi(\theta)$ (lower graph) as functions of periods using $N=10$, $n_b=1.0$
and
scanning $0<\theta<1024$. There are pronounced peaks at the values of  trapping
states: $\Delta\theta=\pi \sqrt{N/n}$.}
\label{f-FTFxid}
\end{figure}

\subsection{Thermal Cavity Revivals}
\label{cavrev}

Due to the trapping states, the cavity  may  revert  to  a  statistical  state,
resembling the thermal state at $\theta=0$,  even  if  $\theta>0$.  By  thermal
revival we mean that the state of the cavity returns to the $\theta=0$  thermal
state for other values of $\theta$. Even if the equilibrium state for  non-zero
$\theta$ can resemble a thermal state,
 it does not at all mean that the dynamics
at that value of $\theta$ is similar to what it is  at  $\theta=0$,  since  the
deviations  from  equilibrium  can  have  completely  different  properties.  A
straightforward measure of the deviation  from  the  $\theta=0$  state  is  the
distance in the $L^2$ norm

\begin{formula}{dist}
    d_{L^2}(\theta)=\left(\sum_{n=0}^\infty
    [p_n(0)-p_n(\theta)]^2\right)^{1/2}~~.
\end{formula}

\noi In Figure \ref{f-dist} we exhibit $d_{L^2}(\theta)$ for $N=10$ and several
values of $n_b$.

For small  values  of  $n_b$  we  find  cavity revivals  at  all  multiples  of
$\sqrt{10}\pi$,
  which  can  be  explained  by  the   fact   that   $\sin(\theta
\sqrt{n/N})$ vanishes for $n=1$ and $N=10$ at those points, \ie\ the cavity  is
in a trapping state. That implies  that  $p_n$  vanishes  for  $n\geq  1$  (for
$n_b=0$) and thus there are no photons in the  cavity.  For  larger  values  of
$n_b$ the trapping is less efficient and the  thermal  revivals  go  away.

Going to much larger values of $\theta$ we can start to look for  periodicities
in the fluctuations in $d_{L^2}(\theta)$. In Figure \ref{f-FTFxid}  (upper  graph)  we
present the spectrum of periods occurring in $d_{L^2}(\theta)$ over  the  range
$0<\theta<1024$.

Standard  revivals  should  occur  with   a   periodicity   of   $\Delta\theta=
2\pi\sqrt{\<n\>N}$, which is typically between 15 and 20,
 but there  are  hardly  any  peaks  at
these
values. On the other hand, for periodicities corresponding to trapping  states,
\ie\ $\Delta\theta=\pi\sqrt{10/n}$, there are very  clear  peaks,  even  though
$n_b=1.0$, which is a relatively large value.

\begin{figure}[htb]
\unitlength=1mm
\begin{picture}(140,100)(0,0)
\includegraphics{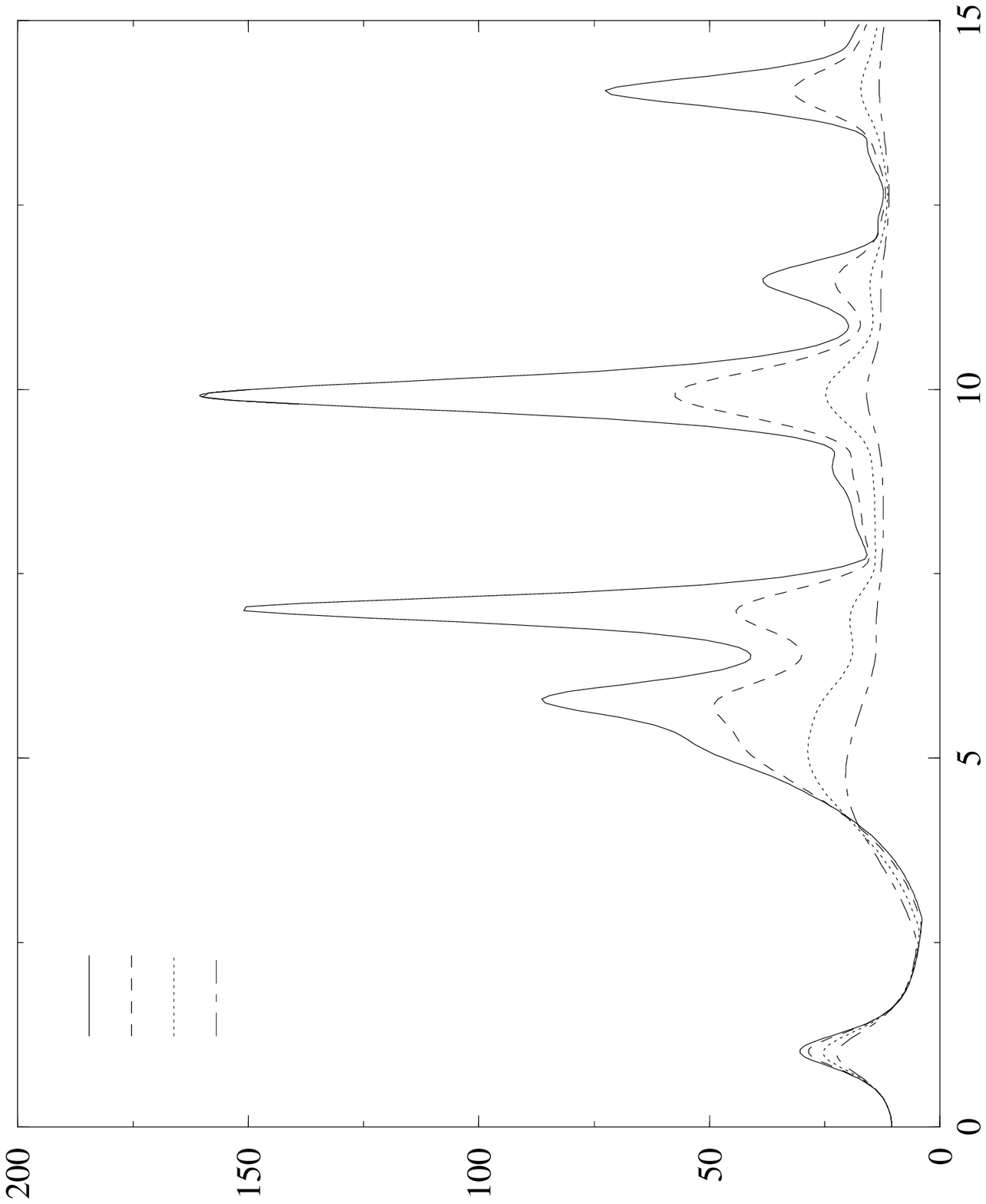}
\put(80,0){\small$\theta$}
\put(9,53){\small$\xi$}
\put(43,82){\small$n_b=0.0$}
\put(43,78){\small$n_b=0.1$}
\put(43,74){\small$n_b=1.0$}
\put(43,70){\small$n_b=10.0$}
\put(30,60){\small$N=10.0$}
\end{picture}
\figcap{Correlation lengths for different values of $n_b$.
The high peaks occur for trapping states and go away as
$n_b$ increases.}
\label{f-xinb}
\end{figure}

In order to see whether trapping states influence  the  correlation  length  we
pre\-sent 
in Figure \ref{f-FTFxid} a similar  spectral  decomposition  of  $\xi(\theta)$
(lower graph) and we find the same peaks. A more direct way of seeing  the
effect of trapping states is to study the correlation length for  small  $n_b$.
In Figure \ref{f-xinb} we see some very pronounced peaks for small $n_b$ which rapidly
go away when $n_b$ increases. They are located at $\theta=\pi k \sqrt{N/n}$ for
every integer $k$ and $n$. The effect is most dramatic when $k$  is  small.  In
Figure \ref{f-xinb}  there  are  conspicuous  peaks  at  $\theta=\pi  \sqrt{10}  \cdot
\{1/\sqrt{3}, ~1/\sqrt{2}, ~1, ~2/\sqrt{3}, ~2/\sqrt{2}\}$, agreeing well with
the formula for  trapping  states.
Notice how sensitive the correlation length is to the temperature when $n_b$ is
small \cite{Meystre88}.


\section{Conclusions}
\label{conclusions}
\seqnoll
\begin{flushright}
``{\sl The more the island of knowledge expands in the sea of\\
ignorance, 
the larger its boundary to the unknown.}"\\
V. F. Weisskopf
\end{flushright}

In the first set of lectures we have discussed the notion of a photon
in quantum physics. We have observed that single photon states can be
generated in the laboratory and that the physics of such quantum
states can be studied under various experimental conditions. A
relativistic quantum-mechanical description of single photon states
has been outlined which constitutes an explicit realization of a representation
(irreducible or reducible) of the Poincar\'{e}
group. The Berry phase for a {\it single } photon has been derived
within such a framework.
 
In the second set of lectures we have outlined the physics of the
micromaser system. 
We have thoroughly discussed various aspects of long-time correlations  in  the
micromaser. It is truly remarkable that this simple dynamical system  can  show
such a rich structure of different phases. The  two  basic
parameters, for $a=1$, in  the
theory are the time the atom spends in  the  cavity,  $\tau$,  and  the  ratio
$N=R/\gamma$ between the rate at which atoms arrive and the decay constant  of
the cavity. We have also argued that the population probability of
the excited state of in the atoms entering into the cavity is of
importance.
 The natural observables  are  related  to  the  statistics  of  the
outgoing atom beam, the average excitation being the simplest one. In
Refs.\cite{Elmforsetal95,Elmforsetal95b}
it is  proposed
to use the long-time correlation  length  as  a  second  observable  describing
different aspects of the photon statistics in the cavity. The  phase  structure
we have investigated is defined  in  the  limit  of  large  flux,  and  can  be
summarized as follows:

\begin{itemize}

\item
{\bf Thermal phase, $0\simleq\theta<1$.}\\
The mean number of photons $\<n\>$
is  low  (finite  in  the  limit  $N\goto\infty$),  and  so  is  the  variance
$\sigma_n$ and the correlation length $\xi$.

\item
{\bf Transition to maser phase, $\theta\simeq 1$.}\\
The maser is starting to get pumped up and $\xi$, $\<n\>$, and $\sigma_n$  grow
like $\sqrt{N}$.

\item
{\bf Maser phase, $1<\theta<\theta_1\simeq4.603$.}\\ The maser is pumped up  to
$\<n\>\sim N$, but fluctuations remain  smaller,  $\sigma_n\sim  \sqrt{N}$,
whereas
$\xi$ is finite.

\item
{\bf  First  critical  phase,  $\theta_1<\theta<\theta_2\simeq  7.790$.}\\  The
correlation length increases exponentially  with  $N$, but  nothing  particular
happens with $\<n\>$ and $\sigma_n$ at $\theta_1$.

\item
{\bf Second maser transition, $\theta\simeq 6.6$}\\ As the  correlation  length
reaches its maximum, $\<n\>$ makes a discontinuous  jump  to  a  higher  value,
though in both phases it is of the order of $N$. The fluctuations grow like $N$
at this critical point.
\end{itemize}

At higher values of $\theta$ there  are  more  maser  transitions  in  $\<n\>$,
accompanied by critical growth of $\sigma_n$, each time the photon distribution
has two competing maxima. The correlation length remains exponentially large as
a function of $N$, as long as there are several maxima, though the  exponential
factor depends on the details of the photon distribution.

No quantum interference effects have been important in  our  analysis
of the {\it phase structure} of the micromaser system,
apart from field quantization in the cavity, and  the
statistical aspects are therefore purely classical. The reason is that we only study  one
atomic observable, the excitation level, which  can  take  the  values  $\pm1$.
Making an analogy with a spin system,
 we can say that we only measure  the  spin
along one direction. It would be  very  interesting  to  measure  non-commuting
variables, \ie\ the spin in different directions or linear superpositions of an
excited and decayed atom, and see how the phase transitions can be described in
terms of such observables \cite{Krause86,Zaheer89}. Most effective
descriptions
of phase transitions in quantum field theory rely on classical concepts,  such
as the free energy and the expectation value of some field, and do not describe
coherent effects. Since linear superpositions of excited and decayed atoms  can
be injected into the cavity,  it is therefore  possible  to  study
coherent phenomena in phase transitions both theoretically and  experimentally,
using resonant micro cavities (see e.g. Refs.\cite{haroche&etal&96,haroche&gimo}). 
The long-time correlation effects we have
discussed in great detail in these lectures have actually recently
been observed in the laboratory \cite{benson}.

\section{Acknowledgment}
\begin{flushright}
``{\sl The faculty of being acquainted with things other then\\
itself is the main characteristic of a mind.}"\\
B. Russell
\end{flushright}
We are very grateful to Professor Choonkyu Lee and
the organizers of this wonderful meeting and  for providing this
opportunity to present various ideas in the field of modern quantum
optics.
The generosity shown to us during the meeting is ever memorable. 
The work presented in these lectures is  based on fruitful
collaborations with many of our friends.  The
work done on resonant cavities and the micromaser system
 has been done in collaboration with Per
Elmfors, Benny Lautrup and also, recently, with Per Kristian Rekdal. 
Most of the other work has been done in collaboration
with A. P. Balachandran, G. Marmo and A. Stern. We are grateful to all of our
collaborators for allowing us to present joint results in the form of these
lectures at
the Seoul 1998 meeting. I am grateful to
John R. Klauder for his encouragement and enthusiastic support
 over the years and for all
these interesting things I have learned from him on the 
notion of coherent states.
 We are also very grateful for many useful comments, discussions and communications 
 with  M. Berry, R. Y. Chiao, P. L. Knight, N. Gisin, R. Glauber, W. E. Lamb Jr.,
 D. Leibfried, E. Lieb ,
 Y. H. Shih, C. R. Stroud Jr.,  
A. Zeilinger, and in particular H. Walther. The friendly and spiritual support of
Johannes M. Hansteen, University of Bergen, is deeply acknowledged.
Selected parts of the material presented in these
 lectures
 have also been 
 presented in lectures/seminars at e.g.: the {\it 1997 Nordic Meeting on Basic
 Problems in Quantum Mechanics}, Rosendal Barony; the 1997 ESF
Research Conference on {\it Quantum Optics}, Castelvecchio Pascoli; 
 the {\it 1998 Symposium on the Foundations
 of Quantum Theory}, Uppsala University; the {\it 1998 NorFA Iceland Meeting on 
Laser-Atom Interactions}; the University of Alabama, Tuscaloosa;
the University of Arizona, Tucson; the University of
Bergen; Chalmers University of Technology, Ericssons Components,
Kista; Link\"{o}ping University; the Max-Planck
Institute of Quantum Optics, Garching; the University of Oslo, the
University of Uppsala,  and at 
the Norwegian University of Science and Technology.  
We finally thank the participants
at these lectures/seminars for their 
interest and all their intriguing, stimulating  questions and remarks
on the topics discussed.

\newpage
\appendix


\section{Jaynes--Cummings With Damping}
\seqnoll
\label{AppDamping}

In most experimental situations the time the atom spends in the cavity is small
compared to the average time between the  atoms  and  the  decay  time  of  the
cavity. Then it is a good  approximation  to  neglect  the  damping  term  when
calculating the  transition probabilities from the cavity--atom  interaction.
In order to establish the range of validity of the approximation we  shall  now
study the full interaction governed by the JC Hamiltonian in \eq{JCH}  and  the
damping in \eq{Damping}. The density matrix for the cavity and one atom can  be
written as

\begin{formula}{dmca}
   \rho=\rho^0\otimes\id +\rho^z\otimes\sigma_z
   +\rho^+\otimes\sigma\_+\rho^-\otimes\sigma_+~~,
\end{formula}

\noi   where   $\rho^\pm=\rho^x\pm   i\rho^y$   and    $\sigma_\pm=(\sigma_x\pm
i\sigma_y)/2$. We want to restrict the cavity part of the density matrix to  be
diagonal, at least the $\rho_0$ part,
 which is the only part of  importance  for
the following atoms, provided that  the  first  one  is  left  unobserved  (see
discussion in Section~\ref{mixed}). Introducing the notation

\begin{eqnarray}\label{rhonot}
   \rho^0_n&=&\< n|\rho_0|n\>~~,\nn
   \rho^z_n&=&\< n|\rho_z|n\>~~,\\
   \rho^\pm_n&=&\< n|\rho_+|n-1\>-
   \< n-1|\rho_-|n\>~~,\non
\end{eqnarray}
\noi the equations of motion can be written as

\begin{eqnarray}\label{rhoeqom}
    \frac{d\rho^0_n}{dt}&=&
    \frac{ig}{2}(\sqrt{n}\rho^\pm_n-\sqrt{n+1}\rho^\pm_{n+1})
    -\gamma \sum_m L^C_{nm}\rho^0_m~~,\nn
    \frac{d\rho^z_n}{dt}&=&
    -\frac{ig}{2}(\sqrt{n}\rho^\pm_n+\sqrt{n+1}\rho^\pm_{n+1})
    -\gamma\sum_m L^C_{nm}\rho^z_m~~,\\
    \frac{d\rho^\pm_n}{dt}&=&
    i2g\sqrt{n}(\rho^0_n-\rho^0_{n-1}-\rho^z_n-\rho^z_{n-1})
    -\gamma \sum_m L^\pm_{nm}\rho^\pm_m~~,\non
\end{eqnarray}

\noi where

\begin{formulas}{LCLpm}
   L^C_{nm}&=&[(n_b+1)n+n_b(n+1)]\,\delta_{n,m}
           -(n_b+1)(n+1)\,\delta_{n,m-1}
           -n_b n \,\delta_{n,m+1}~~,\nn
   L^\pm_{mn}&=&[n_b(2n+1)-{\textstyle\inv{2}}]\,\delta_{n,m}
           -(n_b+1)\sqrt{n(n+1)}\,\delta_{n,m-1}
           -n_b\sqrt{n(n-1)}\,\delta_{n,m+1}~~.
\end{formulas}

\noi
It is thus consistent to study the particular form  of  the  cavity  density
matrix,
which has only one non-zero diagonal or sub-diagonal for each  component,
even when damping is included. Our strategy shall be  to  calculate  the
first-order
correction  in  $\gamma$  in  the   interaction   picture,   using   the
JC Hamiltonian as the free part. The JC part of \eq{rhoeqom} can be drastically
simplified using the variables

\begin{eqnarray}\label{newrho}
    \rho_n^s&=&\rho^n_0+\rho^{n-1}_0-\rho_z^n+\rho_z^{n-1}~~,\nn
    \rho_n^a&=&\rho^n_0-\rho^{n-1}_0-\rho_z^n-\rho_z^{n-1}~~.
\end{eqnarray}

\noi
The equations of motion then take the form

\begin{eqnarray}\label{newreq}
   \frac{d\rho^s_n}{dt}&=&-\frac{\gamma}{2}\sum_m
               \left[(L^C_{nm}+L^C_{n-1,m-1})\rho^s_m+
               (L^C_{nm}-L^C_{n-1,m-1})\rho^a_m\right]~~,\nn
   \frac{d\rho_a^n}{dt}&=&2ig\sqrt{n}\rho^\pm_{n}-\frac{\gamma}{2} \sum_m
               \left[(L^C_{nm}-L^C_{n-1,m-1})\rho^s_m+
                            (L^C_{nm}+L^C_{n-1,m-1})\rho^a_m\right],\nn
   \frac{d\rho^\pm_n}{dt}&=&2ig\sqrt{n} \rho^a_n
           -\gamma \sum_m L^\pm_{nm}\rho^\pm_m~~.\nn
\end{eqnarray}

\noi
The  initial  conditions  $\rho^s_n(0)=p_{n-1}$,   $\rho^a_n(0)=-p_{n-1}$   and
$\rho^\pm_n(0)=0$ are obtained from

\begin{eqnarray}\label{beg}
    \Tr\(\rho(0) |n\>\< n|\otimes\id\)&=&2 \rho_0^n(0)=p_n~~,\nn
    \Tr\(\rho(0) |n\>\< n|\otimes\inv{2}(\id-\sigma_z)\)&=&
    \rho_0^n(0)-\rho_z^n(0)=0~~,\\
    \Tr\(\rho(0) |n\>\< n|\otimes\sigma_x\)&=&
    \Tr\(\rho(0) |n\>\< n|\otimes\sigma_y\)=0~~.\non
\end{eqnarray}

\noi
In the limit $\gamma\goto 0$ it is easy to solve \eq{newreq} and  we  get  back
the standard solution of the JC equations, which is

\begin{eqnarray}\label{JCsol}
   \rho_s^n(t)&=&p_{n-1}~~,\nn
   \rho_a^n(t)&=&-p_{n-1} \cos(2gt\sqrt{n})~~,\\
   \rho_\pm^n(t)&=&-i p_{n-1}  \sin(2gt\sqrt{n})~~.\non
\end{eqnarray}

\noi
Equation   (\ref{newreq})    is    a    matrix    equation    of    the    form
$\dot{\rho}=(C_0-\gamma C_1)\rho$. When $C_0$ and $C_1$ commute the solution
can
be written as $\rho(t)=\exp(\gamma C_1  t)\exp(C_0  t)\rho(0)$,  which  is  the
expression used in \eq{stat1}. In our case $C_0$ and $C_1$ do not  commute  and
we have to solve the equations perturbatively in $\gamma$.  Let  us  write  the
solution  as  $\rho(t)=\exp(C_0  t)\rho_1(t)$  since  $\exp(C_0  t)$   can   be
calculated explicitly. The equation for $\rho_1(t)$ becomes

\begin{formula}{r1eq}
    \frac{d\rho_1}{dt}=-\gamma e^{-C_0 t}C_1e^{C_0t}\rho_1(t)~~,
\end{formula}

\noi
which to lowest order in $\gamma$ can be integrated as

\begin{formula}{r1int}
    \rho_1(\tau)=-\gamma \int_0^\tau dt\, e^{-C_0 t}C_1e^{C_0t}\rho(0)
    +\rho(0)~~.
\end{formula}

\noi
The explicit expression for $\exp( C_0 t)$ is

\begin{formula}{C0}
        e^{C_0 t}=\delta_{nm}
        \(\begin{array}{ccc} 1 & 0 & 0\\
        0 & \cos(2gt\sqrt{n}) & i\sin(2gt\sqrt{n}) \\
        0 & i \sin(2gt\sqrt{n}) & \cos(2gt\sqrt{n}) \end{array}\)~~,
\end{formula}

\noi
and, therefore, $\exp(-C_0t)C_1\exp(C_0t)$ is a bounded  function  of  $t$.
The
elements  of  $C_1$  are  given  by  various  combinations  of  $L^C_{nm}$  and
$L^\pm_{nm}$ in \eq{LCLpm} and they grow  at  most  linearly  with  the  photon
number. Thus the integrand of \eq{r1int} is of the order of  $\<n\>$  up  to an
$n_b$-dependent factor. We conclude that the damping is negligible as  long  as
$\gamma\tau\<n\>\ll 1$,
unless $n_b$ is very large. When  the  cavity  is  in a
maser phase, $\<n\>$  is  of  the  same  order  of  magnitude  as
$N=R/\gamma$, so the condition becomes $\tau R\ll 1$.

Even though this equation can be integrated explicitly it  only  results  in  a
very complicated expression which does not really tell us directly anything about  the
approximation. It is more useful to estimate the size by recognizing that $C_0$
only has one real eigenvalue which is equal to zero, and two imaginary ones, so
the norm of $\exp(-C_0 t)$ is 1. The condition for neglecting the damping while
the atom is in the cavity is that $\gamma C_1 \tau  \rho_1$  should  be  small.
Since $C_1$ is essentially linear in $n$ the condition reads  $\gamma  \tau  \<
n\> \ll 1$. In the maser phase and above we have that $\< n\>$ is of  the  same
order of magnitude as $N=R/\gamma$, so the condition becomes $\tau R\ll 1$.

\section{Sum Rule for the Correlation Lengths}\label{AppSumRule}
\seqnoll

In this appendix we derive the sum rule quoted in Eq.~(\ref{SubSumRule}) and
use the notation of Section \ref{EigenvalueProblem}. No assumptions
are made for the parameters in the problem.

For $A_K=0$ the determinant $\det L_K$ becomes $B_0B_1\cdots  B_K$  as  may  be
easily derived by row manipulation. Since $A_K$ only occurs  linearly  in  the
determinant it must obey the recursion relation $\det  L_K=B_0\cdots  B_K  +A_K
\det L_{K-1}$. Repeated application of this relation leads to the expression

\begin{formula}{Determinant}
\det L_K=\sum_{k=0}^{K+1} B_0\cdots B_{k-1}A_k\cdots A_K~~.
\end{formula}

\noi This is valid for arbitrary values of $B_0$ and $A_K$. Notice that here we
define $B_0\cdots B_{k-1}$ $=1$ 
for $k=0$  and  similarly  $A_k\cdots  A_K=1$  for
$k=K+1$.

In the actual case  we  have  $B_0=A_K=0$,  so  that  the  determinant
vanishes. The characteristic polynomial consequently takes the form

\begin{formula}{}
\det(L_K-\lambda)=(-\lambda)   (\lambda_1-\lambda)   \cdots
(\lambda_K- \lambda)=-D_1\lambda+D_2\lambda^2+\O(\lambda^3)~~,
\end{formula}

\noi where the last expression is valid for $\lambda\to0$. The coefficients
are

\begin{formula}{}
D_1=\lambda_1\cdots\lambda_K~~,
\end{formula}

\noi and

\begin{formula}{}
D_2=D_1\sum_{k=1}^K \frac1{\lambda_k}~~.
\end{formula}

To calculate $D_1$ we note that it is the sum of the $K$ subdeterminants  along
the diagonal. The subdeterminant obtained by removing the $k$'th row and column
takes the form

\begin{formula}{}
\left|\begin{array}{cccccccc}
A_0+B_0&-B_1&\\
&&\vdots\\
&-A_{k-2}&A_{k-1}+B_{k-1}&0\\
&&0&A_{k+1}+B_{k+1}&-B_{k+2}\\
&&&\vdots\\
&&&&-A_{K-1}&A_K+B_K\\
\end{array}\right|~~,
\end{formula}

\noi which decomposes into the product of two  smaller  determinants which
may be calculated using Eq.~(\ref{Determinant}). Using that $B_0=A_K=0$ we get

\begin{formula}{}
D_1=\sum_{k=0}^K A_0\cdots A_{k-1} B_{k+1}\cdots B_K~~.
\end{formula}

\noi Repeating this procedure for $D_2$ which is a sum of all possible diagonal
subdeterminants with two rows and columns removed ($0\le k<l\le K$) , we find

\begin{formula}{}
D_2=\sum_{k=0}^{K-1}\sum_{l=k+1}^K \sum_{m=k+1}^l
A_0\cdots A_{k-1} B_{k+1}\cdots B_{m-1}A_m\cdots A_{l-1}B_{l+1}\cdots B_K~~.
\end{formula}

\noi Finally, making use of Eq.~(\ref{EquilAB}) we find

\begin{formula}{}
D_1=\frac{B_1\cdots B_K}{p^0_0}\sum_{k=0}^K p^0_k~~,
\end{formula}

\noi and

\begin{formula}{}
D_2=\frac{B_1\cdots B_K}{p^0_0}
\sum_{k=0}^{K-1}\sum_{l=k+1}^K \sum_{m=k+1}^l \frac{p^0_kp^0_l}{B_mp^0_m}~~.
\end{formula}

\noi Introducing the cumulative probability

\begin{formula}{}
P^0_n=\sum_{m=0}^{n-1} p^0_m~~,
\end{formula}

\noi and interchanging the sums, we get the correlation sum rule

\begin{formula}{SumRule}
\sum_{n=1}^K
\frac1{\lambda_n}=\sum_{n=1}^K\frac{P^0_n(1-P^0_n/P^0_{K+1})}{B_np^0_n}~~.
\end{formula}

This sum rule is valid for finite $K$ but diverges  for  $K\to\infty$,  because
the equilibrium distribution $p^0_n$  approaches  a  thermal  distribution  for
$n\gg N$. Hence the right-hand side diverges logarithmically in that limit. The
left-hand side also diverges logarithmically with the truncation  size  because
we have $\lambda^0_n=n$ for the untruncated thermal  distribution.  We  do  not
know the thermal eigenvalues for the truncated case, but expect that  they
will
be of the form $\lambda^0_n=n+\O(n^2/K)$ since they should vanish for $n=0$ and
become progressively worse as $n$ approaches $K$. Such a correction leads to  a
finite correction to $\sum_n 1/\lambda_n$. In fact, evaluating  the
right-hand
side of Eq.~(\ref{SumRule}), we get for large $K$

\begin{formula}{}
\sum_{n=1}^K \frac1{\lambda_n^0}
\simeq\sum_{n=1}^K \frac{1-[n_b/(1+n_b)]^n}{n}
\simeq\sum_{n=1}^K \frac1n -\log(1+n_b)~~.
\end{formula}

Subtracting the thermal case  from  Eq.~(\ref{SumRule})
we  get  in  the  limit
of
$K\to\infty$

\begin{formula}{}
\sum_{n=1}^\infty \(\frac1{\lambda_n}-\frac1{\lambda_n^0}\)
=\sum_{n=1}^\infty\(\frac{P^0_n(1-P^0_n)}{B_np^0_n}
-\frac{1-[n_b/(1+n_b)]^n}n\)~~.
\end{formula}

\noi Here we have extended the summation to infinity under the assumption  that
for large $n$ we have $\lambda_n\simeq\lambda^0_n$. The left-hand side  can  be
approximated by $\xi-1$ in regions where the leading correlation length is much
greater than the others. A comparison of the exact eigenvalue and the  sum-rule
prediction is made in Figure \ref{FigSumRule}.

\section{Damping Matrix}
\label{dampingmatrix}
\seqnoll

In this appendix we find an integral representation for the matrix elements  of
$(x+L_C)^{-1}$, where $L_C$ is given by Eq.~(\ref{LC}). Let

\begin{equation}
v_n = \sum_{m=0}^{\infty} (x\delta _{nm} +( L_{C})_{nm})w_m~~~,
\end{equation}

\noi and introduce generating functionals $v(z)$ and  $w(z)$  for complex $z$
defined by

\begin{equation}
v(z) = \sum_{n=0}^{\infty} z^n v_n~~~,\quad
w(z) = \sum_{n=0}^{\infty} z^n w_n~~~.
\end{equation}

\noi By making use of

\begin{equation}
v(z) = \sum_{n,m=0}^{\infty} (x+L_C )_{nm}z^n w_m~~~,
\end{equation}

\noi one can derive a first-order differential equation for $w(z)$,

\begin{equation}
(x+n_{b}(1-z))w(z) + (1+n_{b}(1-z))(z-1)\frac{dw(z)}{dz}=v(z)~~~,
\end{equation}

\noi which can  be  solved  with  the  initial  condition  $v(1)=1$, \ie\
$w(1)=1/x$. If we consider the monomial $v(z) = v_mz^m$ and  the  corresponding
$w(z)=w_{m}(z)$, we find that

\begin{equation}
\label{matrixelements}
w_m (z) = \int_{0}^{1} dt(1-t)^{x-1}\frac{[z(1-t(1+n_{b}))
+t(1+n_{b})]^m}{[1+n_{b}t(1 -z)]^{m+1}}~~~.
\end{equation}

\noi Therefore $(x+L_{c})^{-1}_{nm}$ is given by the coefficient of $z^n$ in
the
series expansion of $w_{m}(z)$. In particular, we obtain for $n_b = 0$ the
result

\begin{equation}
(x+L_{C})^{-1}_{nm} = \left(
\begin{array}{c} m\\n \end{array}
\right)
\frac{\Gamma (x+n)\Gamma (m-n+1)}{\Gamma (x+m+1)}~~~,
\end{equation}

\noi where $m\geq n$. We then find that

\begin{equation}
\label{above}
\P_0(+,+) = \sum_{n=0}^{\infty} \cos^{2}(g\tau\sqrt{n+1})\sum_{m=n}^{\infty}
\frac{m!}{n!}\frac{N\Gamma (N+n)}{\Gamma
(N+m+1)}\cos^{2}(g\tau\sqrt{m+1})p_{m}^{0}~~~,
\end{equation}

\noi  where   $p_{m}^{0}$   is   the   equilibrium   distribution   given   by
\eq{Equilibrium}, and where $x=N=R/\gamma $. Equation
(\ref{above})  can  also  be
derived
from  the  known  solution  of  the  master  equation  
in Eq.~(\ref{Damping})  for
$n_b   =   0$
\cite{Agarwal73}. For small $n_b$ and/or large $x$, \eq{matrixelements}  can
be
used to a find a series expansion in $n_b$.

\newpage
For the convenience of the readers we have attached a capital 
$\fet{R}$ to those
references which are reprinted in Ref.\cite{klauderskagerstam85}.
We apologize to the many authors whose work  we may have overlooked
or to those feel that their work should have been referred to.
\def\pub#1#2#3#4#5{\bibitem{#2}#3, ``{\sl #4~}'', #5.}

\end{document}